\documentclass[apj,square,numbers]{emulateapj}

\usepackage{graphicx}
\usepackage{latexsym}
\usepackage{epsfig}
\usepackage{color}
\usepackage[colorlinks, linkcolor=blue, citecolor=blue, urlcolor=blue]{hyperref}
\usepackage{apjfonts}
\usepackage{amsmath,amssymb,bm,color}
\usepackage{ulem}
\usepackage{url}
\usepackage{breakurl}

\hypersetup{colorlinks=true,citecolor=blue,urlcolor=magenta,linkcolor=red}

\newcommand{\WISC}{WI$\times$SC}
\newcommand{\Xia}{\textcolor{blue}{X15}}

\shorttitle{Tomographic imaging of the $\gamma$-ray sky}
\shortauthors{Cuoco et al.}

\begin{document}

\title{Tomographic imaging of the {\it Fermi}-LAT $\gamma$-ray sky through  cross-correlations:\\A wider and deeper look}

\author{Alessandro Cuoco$^{2}$}
\author{Maciej Bilicki$^{3,4,5}$}
\author{Jun-Qing Xia$^{1}$}
\author{Enzo Branchini$^{6,7,8}$}

\affiliation{$^1$ Department of Astronomy, Beijing Normal University, Beijing 100875, P. R. China}

\affiliation{$^2$ Institute for Theoretical Particle Physics and Cosmology, RWTH Aachen University, Otto-Blumenthal-Strasse, 52057, Aachen, Germany}

\affiliation{$^3$ Leiden Observatory, Leiden University, P.O. Box 9513,  NL-2300 RA Leiden, The Netherlands}
\affiliation{$^4$ National Center for Nuclear Research, Astrophysics Division, P.O.Box 447, 90-950 \L{}\'{o}d\'{z}, Poland}
\affiliation{$^5$ Janusz Gil Institute of Astronomy, University of Zielona G\'ora, ul. Lubuska 2, 65-265 Zielona G\'ora, Poland}

\affiliation{$^6$ Dipartimento di Matematica e  Fisica, Universit\`a degli Studi ``Roma Tre'', via della Vasca Navale 84, I-00146 Roma, Italy}
\affiliation{$^7$ INFN, Sezione di Roma Tre, via della Vasca Navale 84, I-00146 Roma, Italy}
\affiliation{$^8$ INAF Osservatorio Astronomico di Roma, INAF, Osservatorio Astronomico di Roma, Monte Porzio Catone, Italy}

\email{cuoco@physik.rwth-aachen.de,  bilicki@strw.leidenuniv.nl,}
\email{xiajq@bnu.edu.cn,  branchin@fis.uniroma3.it}


\begin{abstract}

We investigate the nature of the extragalactic unresolved  $\gamma$-ray background (UGRB)
by cross-correlating several galaxy catalogs with sky-maps of the UGRB built from
78 months of Pass 8 {\it Fermi}-Large Area Telescope  data.
This study updates and improves similar previous analyses  in several aspects.
Firstly,  the use of
a larger  $\gamma$-ray  dataset allows us to investigate the
energy dependence of the cross-correlation in more detail, using up to 8 energy bins
over a wide energy range of [0.25-500] GeV. Secondly, we consider larger and deeper catalogs
(2MASS Photometric-Redshift catalog, 2MPZ;  WISE~$\times$~SuperCOSMOS,  \WISC; and SDSS-DR12 photometric-redshift dataset)
in addition to the ones employed in the previous studies (NVSS and SDSS-QSOs).
Thirdly, we exploit the redshift information available for the above catalogs to divide them into
redshift bins and perform the cross-correlation separately in each of them.

Our results confirm, with higher statistical significance, the detection of cross-correlation signal between the UGRB maps and \textit{all} the catalogs
considered, on angular scales smaller than $1^{\circ}$.
Significances  range from
$16.3\;\sigma$ for NVSS, $7\;\sigma$ for SDSS-DR12 and \WISC, $5\;\sigma$ for 2MPZ and $4\;\sigma$ for SDSS-QSOs.
Furthermore, including redshift tomography, the significance of the SDSS-DR12 signal
strikingly rises up to $\sim 12\;\sigma$ and  the one of \WISC\ to  $\sim 10.6\;\sigma$.
We offer a simple interpretation of the signal in the framework of the halo-model.
The precise redshift and energy information allows us to clearly detect a change over redshift in the spectral
and clustering behavior of the $\gamma$-ray sources contributing to the UGRB.

\end{abstract}



\section{Introduction}
\label{sec:intro}

The extragalactic $\gamma$-ray background  (EGB) is the gamma-ray emission
observed at high galactic latitudes after subtraction of the diffuse emission from our Galaxy.
It is mainly contributed by various classes of astrophysical sources, like common star-forming galaxies (SFGs) and active
galactic nuclei (AGNs) such as blazars. Contributions from purely diffuse processes,
for example cascades from ultra-high-energy cosmic-rays, are also possible,
as well as exotic scenarios like $\gamma$-rays from dark matter (DM) annihilation or decay (see \citealt{Fornasa:2015qua} for a review).
In the era of the \textit{Fermi} Large Area Telescope  \citep[LAT,][]{2009ApJ...697.1071A}, with  its strong sensitivity to point sources,
a sizable fraction of the EGB has been resolved into sources. Indeed, the third \textit{Fermi} $\gamma$-ray catalog of sources \citep[3FGL,][]{Acero:2015hja}
contains $\sim$3000 sources.  The  resolved sources constitute typically 10-20\% of the EGB for energies below $\sim$10 GeV,
while above  this energy the fraction rises up to 50\% or more \citep{Ackermann:2014usa,TheFermi-LAT:2015ykq}.
This large number of detected sources has been fundamental to study in detail the different populations of
emitters, and to infer their properties in the so-far unresolved regime 
\citep{Ackermann:2011bg,Inoue:2011bm,Achermann12,AjelloFSRQs,AjelloBLLs,DiMauro:2013xta,DiMauro:2013zfa}.
The still-unresolved EGB emission is typically indicated with the name of unresolved (or isotropic) gamma-ray background \citep[UGRB,][]{Ackermann:2014usa} and   
  is the subject
of the present analysis.

Together with population studies of resolved sources,
in recent years a number of different and complementary techniques have been developed to
study the UGRB in a more direct way, exploiting the information contained in the spatial as well as in the energy properties of the UGRB maps.
Among these we can list anisotropy analyses
\citep{2006PhRvD..73b3521A,andokomatsu,2009PhRvD..80b3520A,FermiAPS_12,cuoco12,Harding:2012gk,Fornasa:2012gu,DiMauro:2014wha,Ando2017, FermiAPS_16},
pixel statistic analyses
\citep{Dodelson:2009ih,Malyshev:2011zi,Feyereisen:2015cea,Lisanti:2016jub,Zechlin:2015wdz,Zechlin:2016pme}, and cross-correlations with tracers
of the large-scale structure    of the Universe
\citep{Ando:2014aoa,Ando:2013xwa,Fornengo:2013rga,Shirasaki:2014noa,Camera:2014rja,Cuoco:2015rfa,Fornengo:2014cya,Regis:2015zka,xia15,Feng:2016fkl,Shirasaki:2016kol,Troester2016},
which we will investigate in the following.

In \cite{xia15} (herafter \Xia), \cite{Cuoco:2015rfa} and \cite{Regis:2015zka}
5-years $\gamma$-ray maps of the UGRB from \textit{Fermi}-LAT were cross-correlated with
different catalogs of galaxies, i.e.,  SDSS-DR6 quasars
\citep{richards09}, SDSS-DR8 Luminous Red Galaxies
\citep{2008arXiv0812.3831A}, NVSS radiogalaxies
\citep{1998AJ....115.1693C}, 2MASS galaxies
\citep{jarrett2000}, and SDSS DR8
main sample galaxies \citep{dr8}. Significant correlation (at the level of 3-5 $\sigma$)
was observed at small angular scales, $\lesssim 1^\circ$, for all the catalogs except the Luminous Red Galaxies,
 and the results interpreted in terms
of constraints on the composition of the UGRB.
This work updates these  analyses in several aspects  :
  {\it i)} we use a larger amount of \textit{Fermi} data, almost 7 years compared to the 5 years.
In doing so, we employ the new \textit{Fermi}-LAT Pass 8 data selection \citep{Atwood:2013rka}, based on
improved event reconstruction algorithm, and providing a $\sim$30\% larger effective area.
The full Pass 8 dataset is roughly two times larger than the 5 years Pass 7 dataset.
With such large dataset, we can perform our cross-correlation analysis in more energy bins. We now consider up to
eight energy bins instead of the three ones used in \Xia.
{\it ii)} we use updated versions of the original galaxy catalogs.
For example, we now use the 2MPZ catalog instead of 2MASS. 2MPZ extends the 2MASS dataset by adding precise
photometric redshifts which were not available before (but see \citealt{Jarrett2004}).
Thanks to this we can perform cross-correlation analysis subdividing the sample into a number of different $z$-bins.
Similarly, instead of the SDSS main sample galaxies, we now consider
the latest SDSS DR12 photometric galaxy catalog.
As for the NVSS catalog and the QSO sample we consider the same datasets used in the
previous analyses.
 {\it iii)} we consider a new dataset:
the WISE~$\times$~SuperCOSMOS photometric redshift catalog \citep[\WISC,][]{bilicki16}.
This is a natural extension of 2MPZ providing coverage of $\sim75\%$ of sky
and reaching in redshift up to almost $z\sim 0.5$.

In our analysis, we will use the same methodology as in \Xia\ and estimate
the angular two-point cross-correlation function (CCF) and the cross-angular power spectrum
(CAPS) of the UGRB maps and discrete objects catalogs.
The rationale for computing two quantities, CCF and CAPS, which contain the same information
is that  they are largely complementary since their estimates
 are affected by different types
of biases and, which is probably more important, the properties of the error covariance are different in
the two cases.

The layout of the paper is as follows: in Section~\ref{sec:fermimaps}
we present the {\it Fermi}-LAT maps, their accompanying masks
and discuss the procedure adopted to remove potential spurious contributions
to the extragalactic signal.
In Section~\ref{sec:maps} we present the catalogs of different types of extragalactic sources
that we cross-correlate with the {\it Fermi} UGRB maps. In Section~\ref{sec:corranalysis}
we briefly describe the CCF and CAPS estimators and their uncertainties.
In Section~\ref{sec:chi2} we propose a simple, yet physically motivated model for the cross-correlation
signal and introduce the $\chi^2$ analysis used to perform the comparison  with the data.
The results of the cross-correlation analysis are described in
Section~\ref{sec:results}  and discussed in Section~\ref{sec:discussion}
in which we also summarize our main conclusions.
An extended discussion of the
systematic errors is presented in Appendix \ref{apdx:validation},
where we describe the results of a series of tests to assess the robustness of our results.
Appendix \ref{apdx:moreplots} contains additional plots that show results of the cross correlation 
analysis not included in the main text.

To model the expected angular cross-correlations
we assume a flat Cold Dark Matter model with a
cosmological constant ($\Lambda$CDM) with  cosmological parameters
$\Omega_{\rm b} h^2 = 0.022161$,
$\Omega_{\rm c} h^2 = 0.11889$, $\tau= 0.0952$, $h = 0.6777$, $\ln{10^{10}A_{\rm s}} = 3.0973$ at $k_0=0.05$ Mpc$^{-1}$,
and $n_{\rm s} =0.9611$, in accordance with the most recent Planck results \citep{Ade:2015xua}.

The data files containing the results of our cross correlation analysis  are publicly available
at \url{https://www-glast.stanford.edu/pub_data/}.


\section{{\it Fermi}-LAT maps}
\label{sec:fermimaps}

In this section we describe the EGB maps obtained from 7  years of
{\it Fermi}-LAT observations and the masks and procedures
 used to subtract  contributions from  {\it i)} $\gamma$--ray resolved sources,
{\it ii)} Galactic diffuse emission due to interaction of cosmic rays with the interstellar medium and
{\it iii)} additional Galactic emission located high above the Galactic plane in prominent structures
such as the Fermi Bubbles  \citep{2010arXiv1005.5480S} and Loop~I  \citep{2009arXiv0912.3478C}.

{\it Fermi}-LAT is a pair-conversion telescope onboard the {\it Fermi} Gamma-ray Space
Telescope  \citep{2009ApJ...697.1071A}. It covers the
energy range between 20 MeV and $\sim 1$ TeV, most of which will be used in our
analysis (E$=[0.25,500]$ GeV), and has an excellent angular resolution ($\sim 0.1^{\circ}$)
above 10 GeV over a large  field of view ($\sim 2.4$ sr).

For our study we have used 78 months of data from August
4, 2008 to January 31, 2015 (\textit{Fermi} Mission Elapsed
Time 239557418 s - 444441067 s), considering the Pass 8 event selection \citep{Atwood:2013rka}
 and excluding photons detected
with measured zenith angle larger than $100^\circ$
 to reduce the contamination from
the bright Earth limb emission.
 We used both back-converting and front-converting events.
The corresponding exposure maps were produced
using the standard routines from the LAT
{\it Science Tools}\footnote{\url{http://fermi.gsfc.nasa.gov/ssc/data/analysis/documentation/Cicerone/}}
 version 10-01-01, and the \texttt{P8R2\_CLEAN\_V6}
instrument response functions (IRFs).
We have also used for a cross-check the \texttt{P8R2\_ULTRACLEANVETO\_V6} IRFs,
which provide a data selection where residual contamination of the $\gamma$-ray sample from charged cosmic rays
is substantially reduced, at the price of a decrease in effective area of $\sim 30\%$.
To pixelize the photon counts we have used the GaRDiAn package \citep{Ackermann:2009zz,FermiLAT:2012aa}.
The count maps were generated in   HEALPix\footnote{\url{http://healpix.jpl.nasa.gov/}} format \citep{2005ApJ...622..759G}
 containing
$N_{\rm pix} = 12, 582, 912 $
 pixels with mean spacing of $0.06^{\circ}$  corresponding to the
HEALPix resolution parameter $N_{\rm side}=1024$.

Thanks to the large event statistics  we consider
eight bins with energy edges $E=$ 0.25, 0.5, 1, 2, 5, 10, 50, 200,
500 GeV. In several cases  we have grouped the events in three wider
intervals in order to have better statistics and  higher signal-to-noise: $0.5 < E < 1$ GeV, $1 < E < 10 $ GeV, and
$10 < E < 200$ GeV.


The masking, the cleaning procedure and the tests aimed at assessing our ability to remove
contributions from the Galactic foreground and resolved sources have been described in detail in  \cite{xia11}
and  in  \Xia. Here we summarize the main steps.

{\it i)}~The geometry mask excludes the Galactic Plane \mbox{$|b|<  30^{\circ}$},
the region associated with the Fermi Bubbles and the Loop~I structure,
and two circles of  $5^{\circ}$ and $3^{\circ}$ radius at the position of the Large and Small Magellanic Clouds, respectively.
The 500 brightest point sources (in terms of the integrated photon flux in the 0.1-100 GeV energy range) from the 3FGL catalog are masked
with a disk of radius $2^\circ$, and the remaining ones with a disk of $1^\circ$ radius.
We notice that in several of the cross-correlation analyses (in particular the ones involving SDSS-related catalogs) presented below, the mask of the catalog
largely overlaps and sometimes includes the \textit{Fermi} one so that the effective geometry mask used
is more conservative than the one described here.

{\it ii)}~The Galactic diffuse emission in the unmasked region has been removed
by subtracting the model \verb"gll_iem_v05_rev1.fit"\footnote{\url{http://fermi.gsfc.nasa.gov/ssc/data/access/lat/BackgroundModels.html}} \citep{Acero:2016qlg}
from the observed emission.
More precisely, in the unmasked region we performed,  separately in each energy bin,  a two-component fit
including  the Galactic emission from the model above and a purely isotropic emission. Convolution of the two template maps with the IRFs  
and subsequent fit to the observed counts were then performed with GaRDiAn.
The best-fit  isotropic plus Galactic emission, in count units, was then subtracted off from the
$\gamma$-ray count maps, and finally divided by the exposure map  in the considered energy range to obtain the
residual flux maps used for the analysis.
The robustness of this cleaning procedure has been tested against a different Galactic diffuse emission model,
 \verb"ring_2year_P76_v0.fits"$^3$. We have found that
 the two models are very similar in our
region of interest.
 As a result, their residuals agree at the percent level.
We stress, nonetheless, that the Galactic foregrounds are not expected to correlate with
extragalactic structures, and thus it is not crucial to achieve a {\it perfect} cleaning.
Indeed, in \Xia, we did show that even without foreground removal 
the recovered cross-correlations were unbiased, while  the main impact  of foreground removal
was to suppress the background and thus to reduce the size of the random errors. 
Similar conclusions were reached in the recent cross-correlation analysis of weak lensing catalogs with {\it Fermi}-LAT performed by \cite{Troester2016}.
{Analogous considerations apply to the point sources.
Especially at low energies, some leakage of the point sources outside the mask
is expected, since the point-spread function of the instrument becomes large
and the tails lie outside the mask.
Nonetheless, bright point sources  should not  correlate with extragalactic sources,
so the leakage is expected to increase the random noise but not to introduce biases.
In Appendix~\ref{apdx:validation}, we test the validity of this assumption estimating the correlation using different point source masks 
and find that the results is insensitive to the choice of the mask.}

{\it iii)}~An imperfect cleaning procedure may induce spurious features in the diffuse $\gamma$-ray signal on large angular scales.
These should not significantly affect our cross-correlation analysis since they are not expected to correlate with the
sources responsible for the UGRB.
 Nevertheless, to minimize the chance that
spurious large-scale correlation power may leak into the genuine signal,
we performed an additional cleaning step (dubbed \textit{$\ell10$ cleaning}) and removed
contributions up to multipoles   $\ell=10$,  including the monopole and dipole, from
all the maps, using standard  HEALPix tools.  This cleaning procedure was also adopted  in \cite{xia11}.

\begin{figure}
\centering
\includegraphics[width=0.33\textwidth,angle=90]{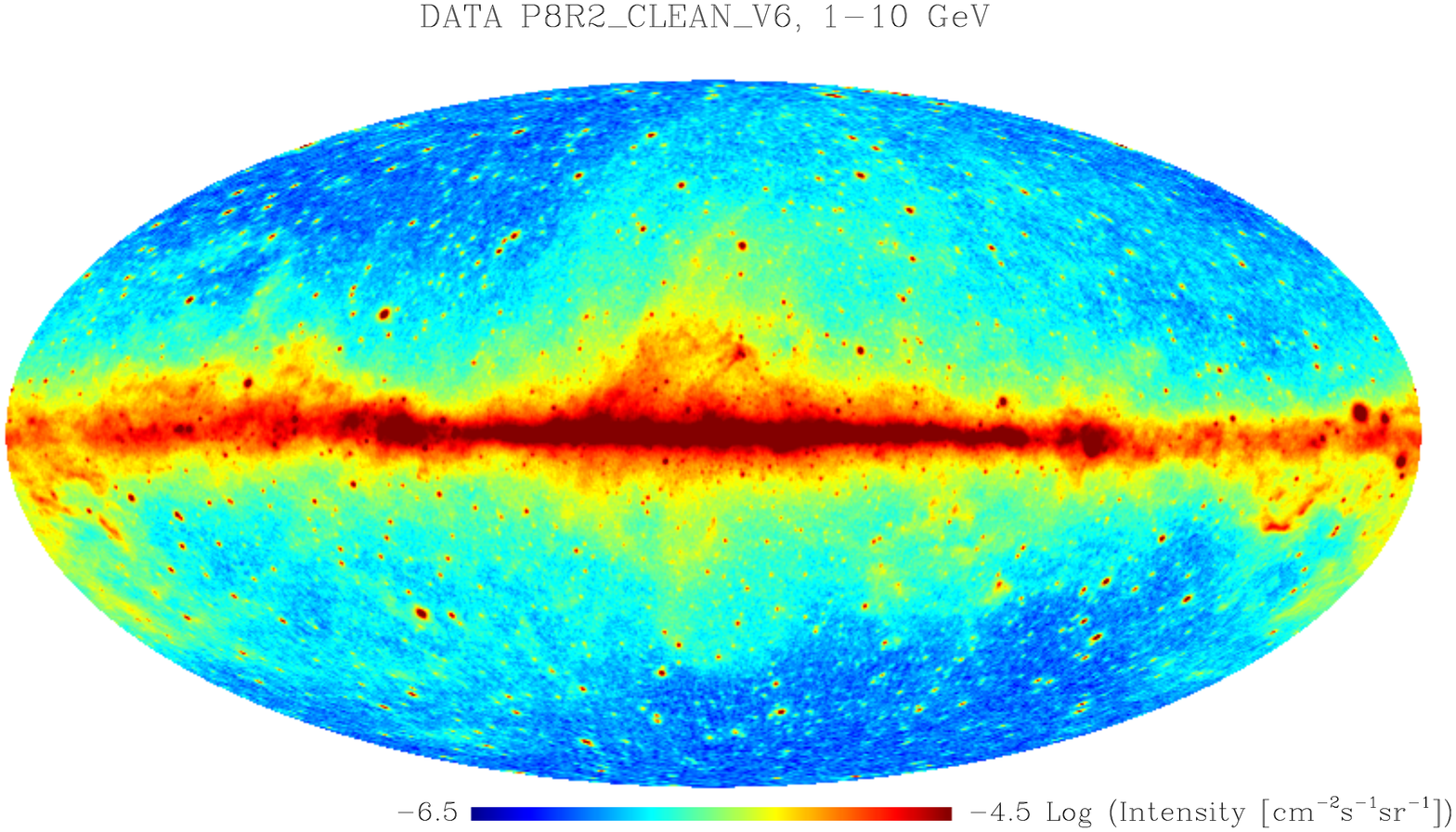}
\includegraphics[width=0.33\textwidth,angle=90]{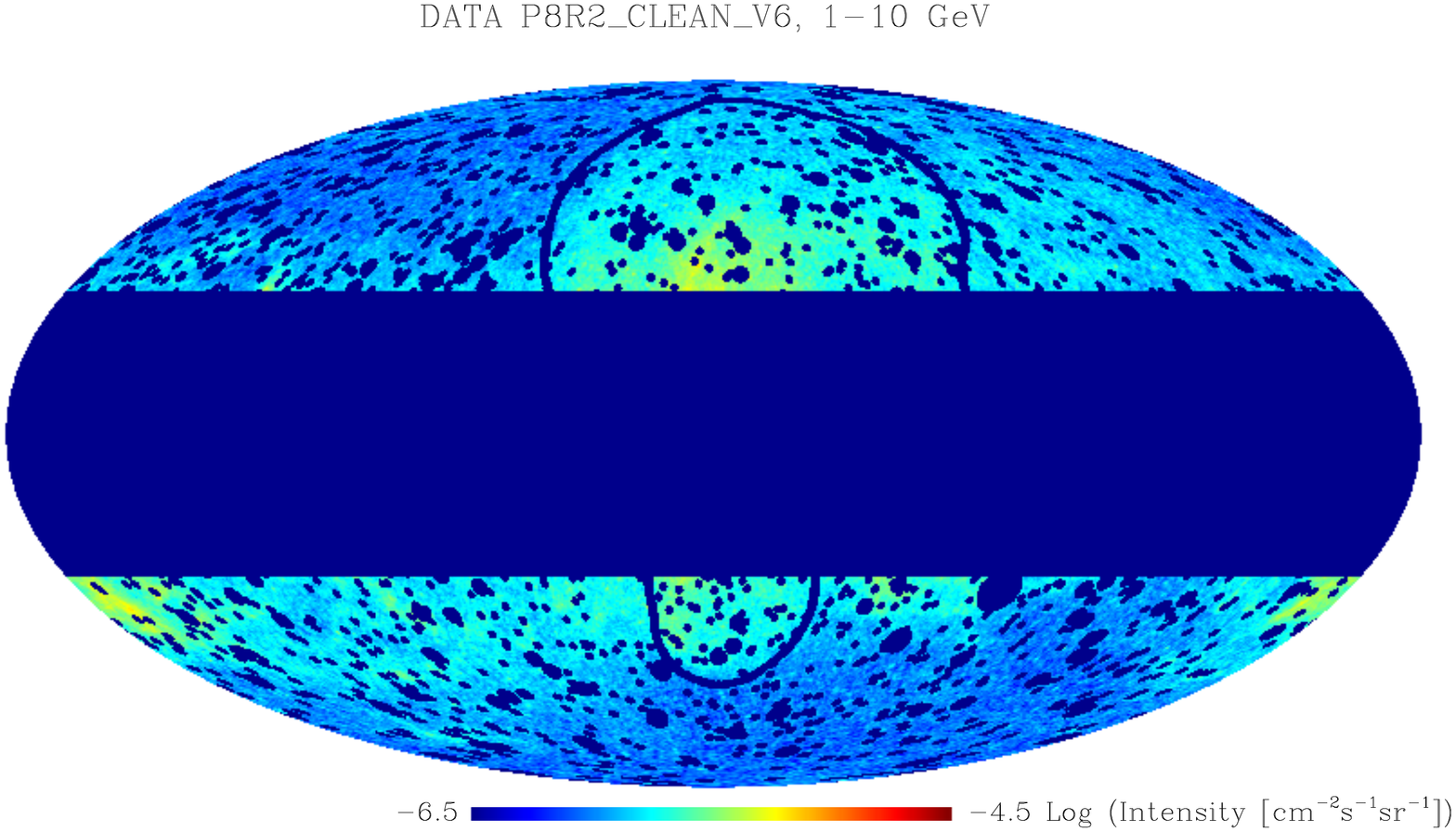}
\includegraphics[width=0.33\textwidth,angle=90]{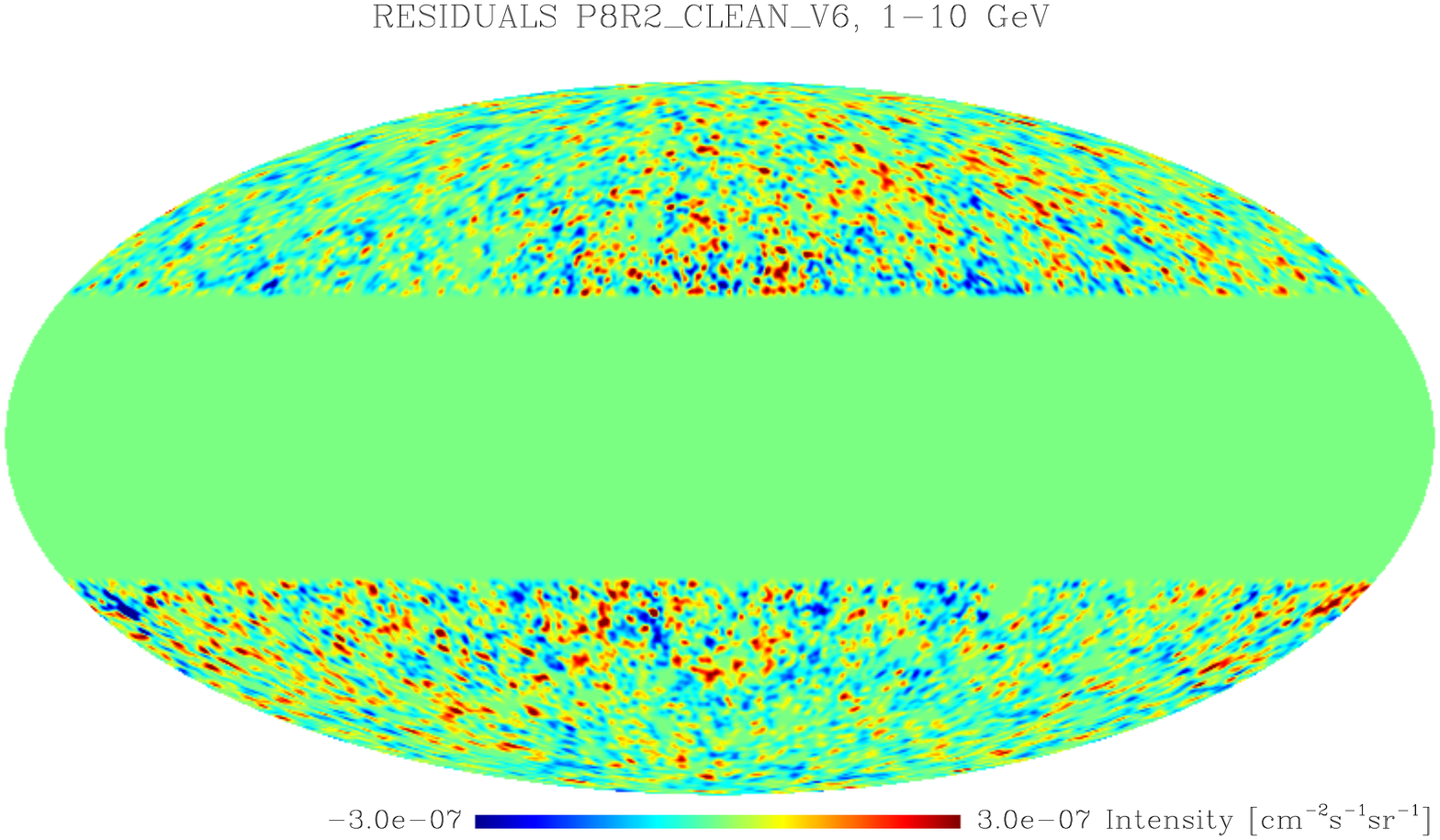}
\caption{All-sky Mollweide projections of {\it Fermi}-LAT   total flux maps in the energy range 1-10 GeV.
{\it Upper panel:} Flux map without mask.
{\it Middle panel:} Flux map together with the fiducial mask, covering 3FGL point sources and the Galactic $|b|<30^\circ$ region.
The two further visible lines enclose the region covering the {\it Fermi} Bubbles and the Loop I area.
{\it Lower panel:} residual flux maps after Galactic foreground subtraction and $\ell10$ cleaning.
For better visualization the upper two maps have been smoothed with a Gaussian beam of $\mathrm{FWHM}=0.5^\circ$, and the lower
one with $\mathrm{FWHM}=1^\circ$.}
\label{fig:maps-Fermi}
\end{figure}

Example  maps are shown in Fig.~\ref{fig:maps-Fermi}
for the energy range \mbox{1-10 GeV}, with and without the fiducial mask,
and after the foreground subtraction and $\ell10$ cleaning.
In the bottom panel, the residuals are shown without the Bubbles/Loop I mask
in order to show that the cleaning works well nonetheless also in this region.

\section{Catalogs of Discrete Sources}
\label{sec:maps}

\begin{figure}
\includegraphics[width = 0.48\textwidth]{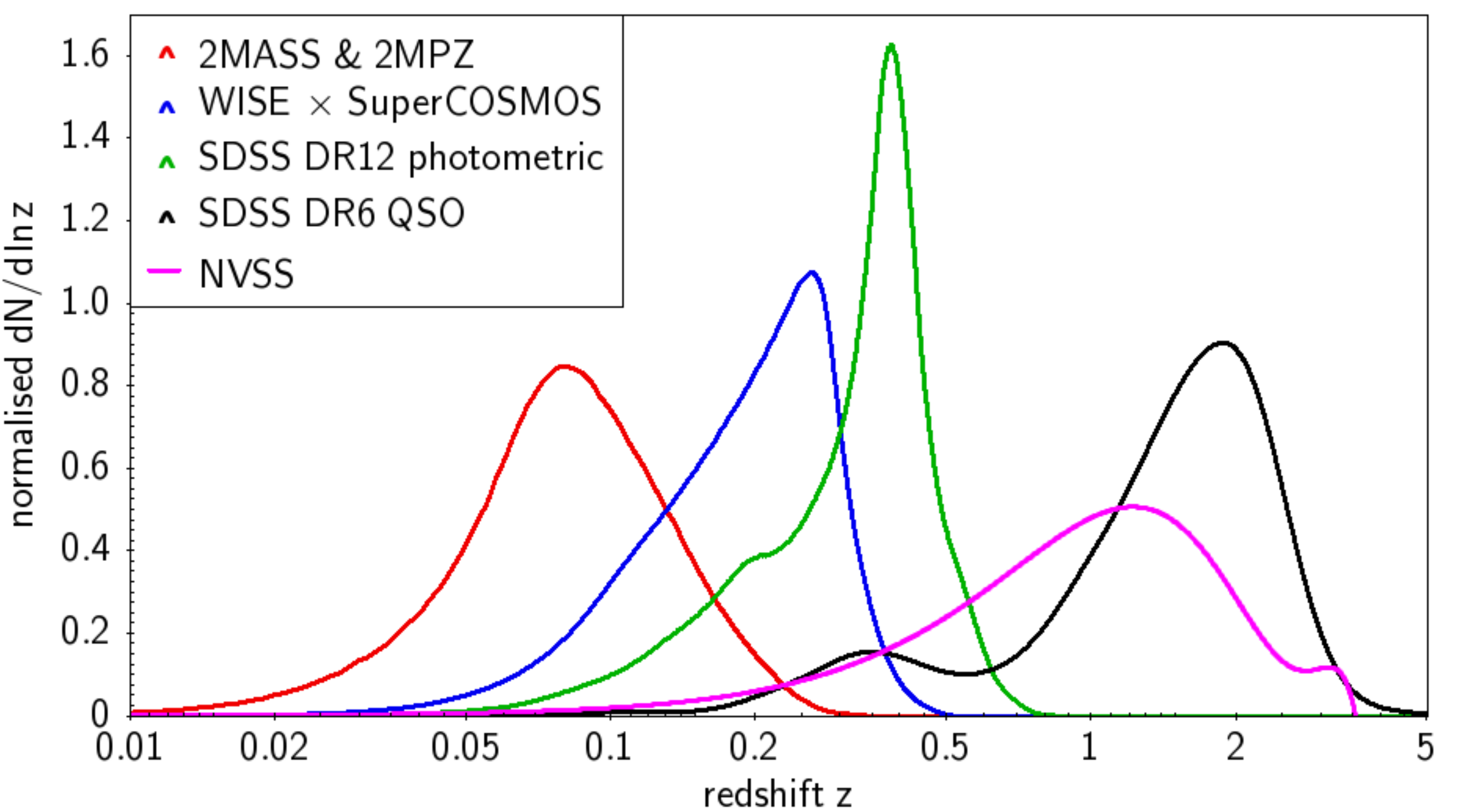}
\caption{Redshift distributions of the five datasets used in our analysis. The curves show their normalized
 $dN/dz$ distributions, based on photometric redshifts of the sources. An exception
is the NVSS where no redshift estimates are available, hence the analytical approximation
described in the text  was assumed.}
\label{fig:dNdzs}
\end{figure}

In this work we use five different catalogs of extragalactic objects for the cross-correlation analysis.
They span wide, overlapping redshift ranges,
contain different types of objects (galaxies, quasars) detected at several wavelengths (UV, optical, near- and mid-IR, radio)
whose distances, when available, are inferred from photometric redshifts.
They all share two important characteristics: large angular coverage to maximize the number of Fourier modes
available to the cross-correlation analysis, and a large number of objects to minimize shot noise errors.
The redshift distributions of the sources in the various catalogs are shown in Fig.~\ref{fig:dNdzs}.
Overall, they span an extended range of redshift, from $z=0$ out to $z\sim5$. Such a wide redshift coverage
is of paramount importance to identify the nature of the UGRB that could be generated both by nearby
(star forming galaxies and DM annihilation processes in halos)  and high redshift sources (e.g. blazars). 
In Table \ref{tab:catalogs} we summarise the basic properties of the source catalogs used in our analysis, such as their sky coverage, source number and mean surface density of the objects  in the region of sky effectively used for the analysis, i.e., after applying  both the catalog and $\gamma$-ray masks.
{In the following sections, instead, when describing a given catalog, we will report
numbers referred to the nominal mask of the catalog itself.}

\begin{table}
\begin{tabular}{lcrc}
\hline 
source & sky & number & mean surface \\
catalog & coverage & of sources & density [deg$^{-2}$] \\ 
\hline 
NVSS    & 25.5\% & 177,084 & 16.8 \\ 
2MPZ    & 28.8\% & 293,424 & 24.7 \\ 
WISE$\times$SCOS   & 28.7\% & 7,544,862 & 638 \\ 
SDSS DR12   & 12.3\% & 15,194,640 & 2980 \\ 
SDSS DR6 QSO      & 11.7\% & 340,162 & 70.3 \\ 
\hline 
\end{tabular} 
\caption{\label{tab:catalogs} Statistics of the source catalogs used for the cross-correlation. 
The sky coverage indicates the 
area effectively used in the analysis, i.e., after applying both the catalog and $\gamma$-ray masks. 
The numbers refer to the objects contained in the selected regions.}
\end{table}

\subsection{NVSS}
\label{sec:nvss}

The NRAO VLA catalog \citep[NVSS,][]{1998AJ....115.1693C} is the largest catalog of radio sources currently available.
The sample considered in our analysis contains  $\sim5.7\times10^5$
objects with a flux $> 10$ mJy,
located at declinations $\delta\gtrsim -40^\circ$ and outside a relatively narrow Zone of Avoidance ($|b| > 5^{\circ}$). The mean surface density of sources is
$\sim 16.9$~deg$^{-2}$.
This is the same NVSS dataset  used in the cross-correlation analysis of \Xia.
The map showing the sky coverage and angular positions of  the objects can be found in  \cite[][fig.~9]{xia11}.

The main reason for repeating the cross-correlation analysis
using  the new Pass-8 \textit{Fermi} data is to check the robustness of
 the strong correlation signal at small angular separation found by
 \Xia\ and interpreted as contributed by the same NVSS galaxies emitting in gamma rays.

Radio sources in the NVSS catalog do not come with an estimate of their redshift. 
{We use the redshift distribution  determined
by~\cite{2008MNRAS.385.1297B}. Their sample, 
contained 110 sources with $S > 10$ mJy, 
of which 78(i.e. 71 \% of the total) had spectroscopic redshifts, 23 had
redshift estimates from the $K - z$ relation for radio sources,
and 9 were not detected in the $K$-band and therefore had a
lower limit to $z$.
We adopt the smooth parametrization of this distribution
given in \cite{dezotti10}, shown in Fig.~\ref{fig:dNdzs} with the magenta line.}

\subsection{SDSS DR6 QSO}
\label{sec:QSO}

\begin{figure*}
\includegraphics[width=0.33\textwidth]{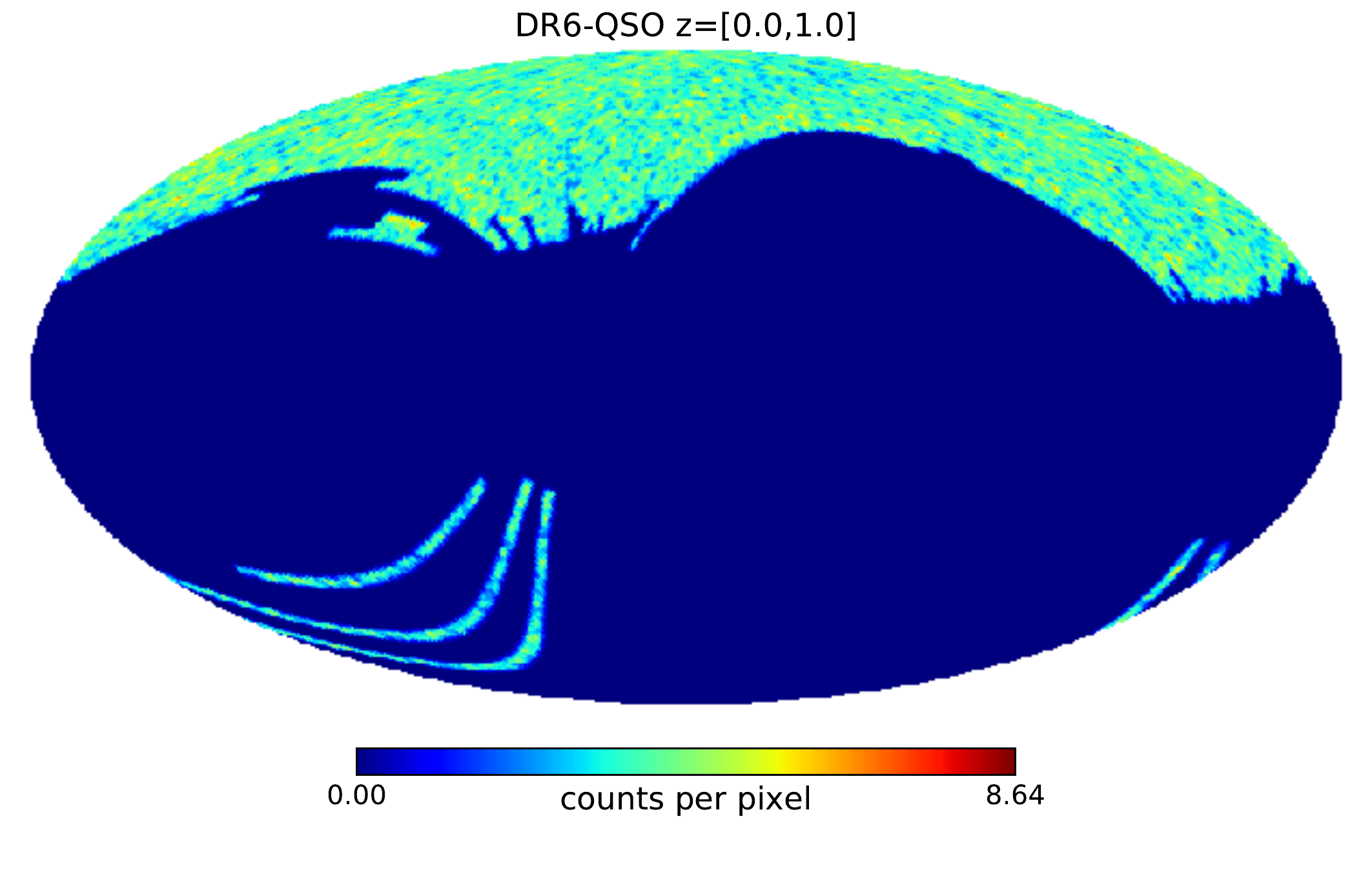}
\includegraphics[width=0.33\textwidth]{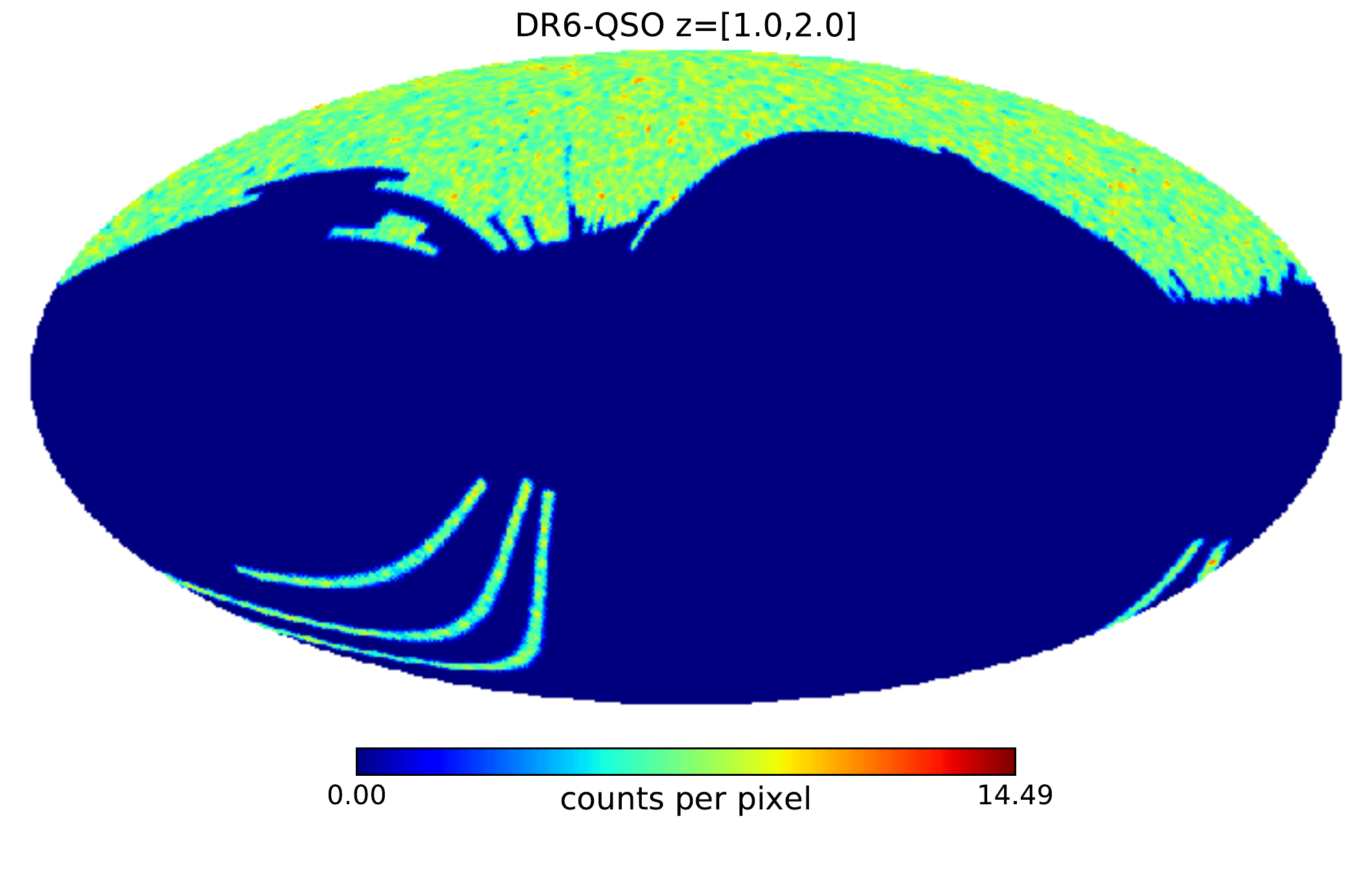}
\includegraphics[width=0.33\textwidth]{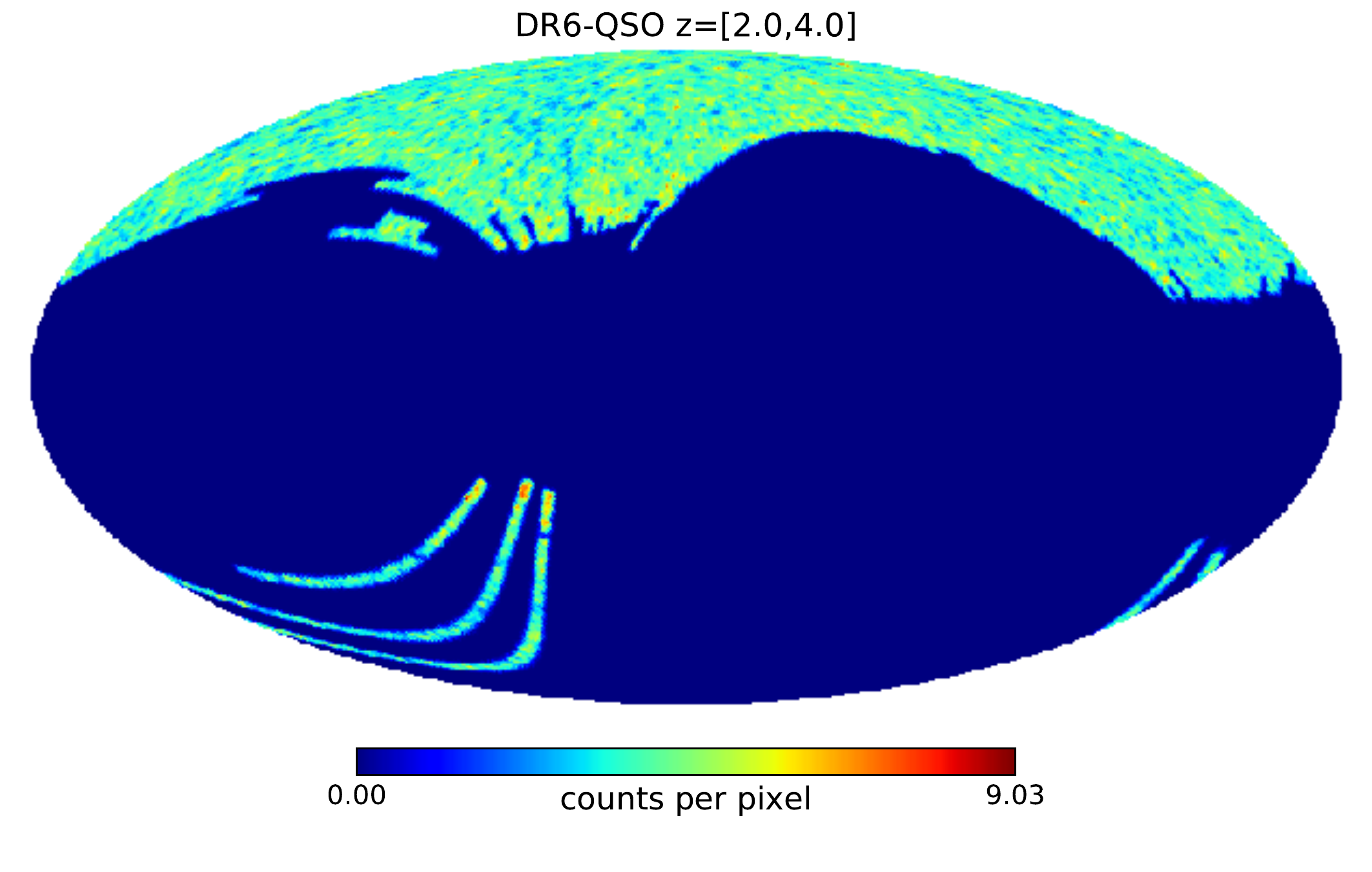}
\caption{All-sky projections of the SDSS DR6 QSO  distribution in the three redshift shells adopted in the analysis.
The maps have HEALPix resolution $N_\mathrm{side}=128$ and include additional Gaussian smoothing of $\mathrm{FWHM}=1^\circ$  for better visualization.}
\label{fig:maps-DR6-QSO}
\end{figure*}

In recent years several quasar catalogs have been obtained based on the SDSS dataset,  complemented in some cases with additional
information, most notably from  the Wide-field Infrared Survey Explorer (WISE). They all are meant to supersede the SDSS DR6 QSO catalog
\citep[][hereafter DR6-QSO]{richards09} used in the previous cross-correlation analyses by \cite{xia11,xia15}.
We checked the adequacy of these new samples using two criteria: the surface number density of objects, that has to be large to minimize
the shot noise error, and
the uniformity in the selection function of the catalog across the sky to ensure  a uniform calibration of the catalog.
Our tests have shown that none of the newer datasets satisfy these requirements better than the original DR6-QSO one
since in all  the new samples we detected large variations in the number density of sources across the sky.
Neither aggressive cleaning procedures  nor  geometry cuts were able to  guarantee angular homogeneity without heavily compromising the surface density of sources.  

For these reasons, we decided to rely on the original DR6-QSO catalog.
We applied the same preselection procedures as in \cite{xia11,xia15}. In particular, we
considered only the sources with an UV excess flag $\mathrm{uvxts}=1$, since this criterion provides a uniform selection. There are about $6\times10^5$ sources in the sample selected this way,  covering $\sim25\%$ of the sky, with photometric redshifts $0<z<5.75$ ($\langle z \rangle \simeq 1.5$) of typical accuracy $\sigma_z \sim 0.24$. Fig.\ \ref{fig:dNdzs} shows smoothed $dN/dz$ of this dataset (black line). We note however that the original histogram as derived from the \cite{richards09} data is very non-uniform, exhibiting multiple peaks (see, e.g.,\ fig.\ 1 in \citealt{xia09}), probably an artifact of the photo-$z$ assignment method. Nevertheless, this is of minor importance for the present paper, as for the cross-correlation we use very broad redshift shells. In particular, we split the DR6-QSO dataset into three bins of $z\in [0.0,1.0]$, $[1.0,2.0]$, and $[2.0, 4.0]$, selected in a way to have similar number of objects in each bin. Usage of redshift shells is, together with the updated \textit{Fermi} data and binning in energy, a novel element of the QSO -- $\gamma$-ray cross-correlation analysis in comparison to \cite{xia11,xia15}, where the same quasar sample was considered as one broad bin encompassing all the data. Fig.\ \ref{fig:maps-DR6-QSO} shows all-sky projections of the three redshift shells of the DR6-QSO catalog in HEALPix format.
We have excluded from the analysis the three narrow stripes present in the south Galactic sky and use only the northern region.

\subsection{2MPZ}
\label{sec:2MPZ}

\begin{figure*}
\includegraphics[width=0.33\textwidth]{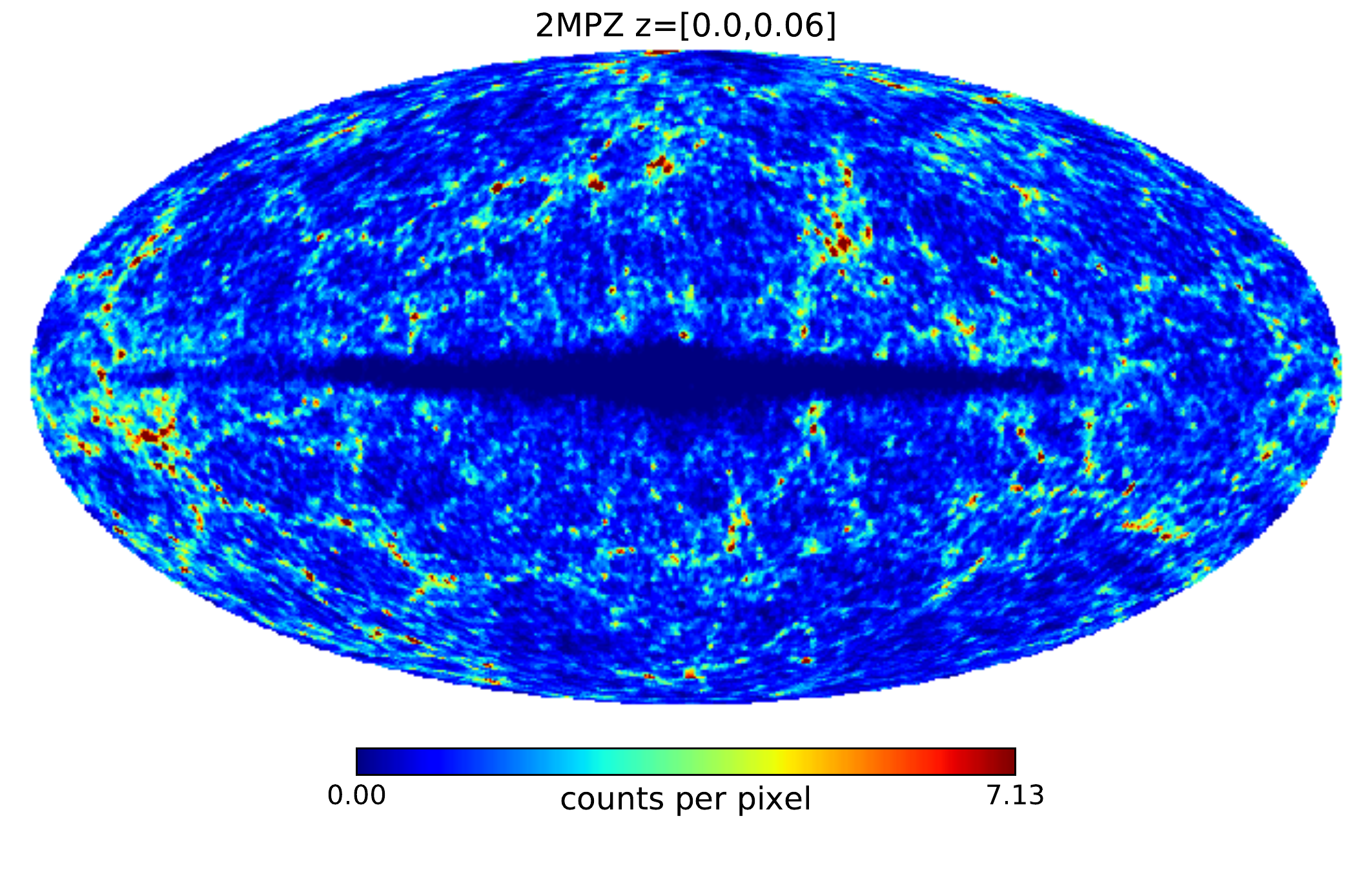}
\includegraphics[width=0.33\textwidth]{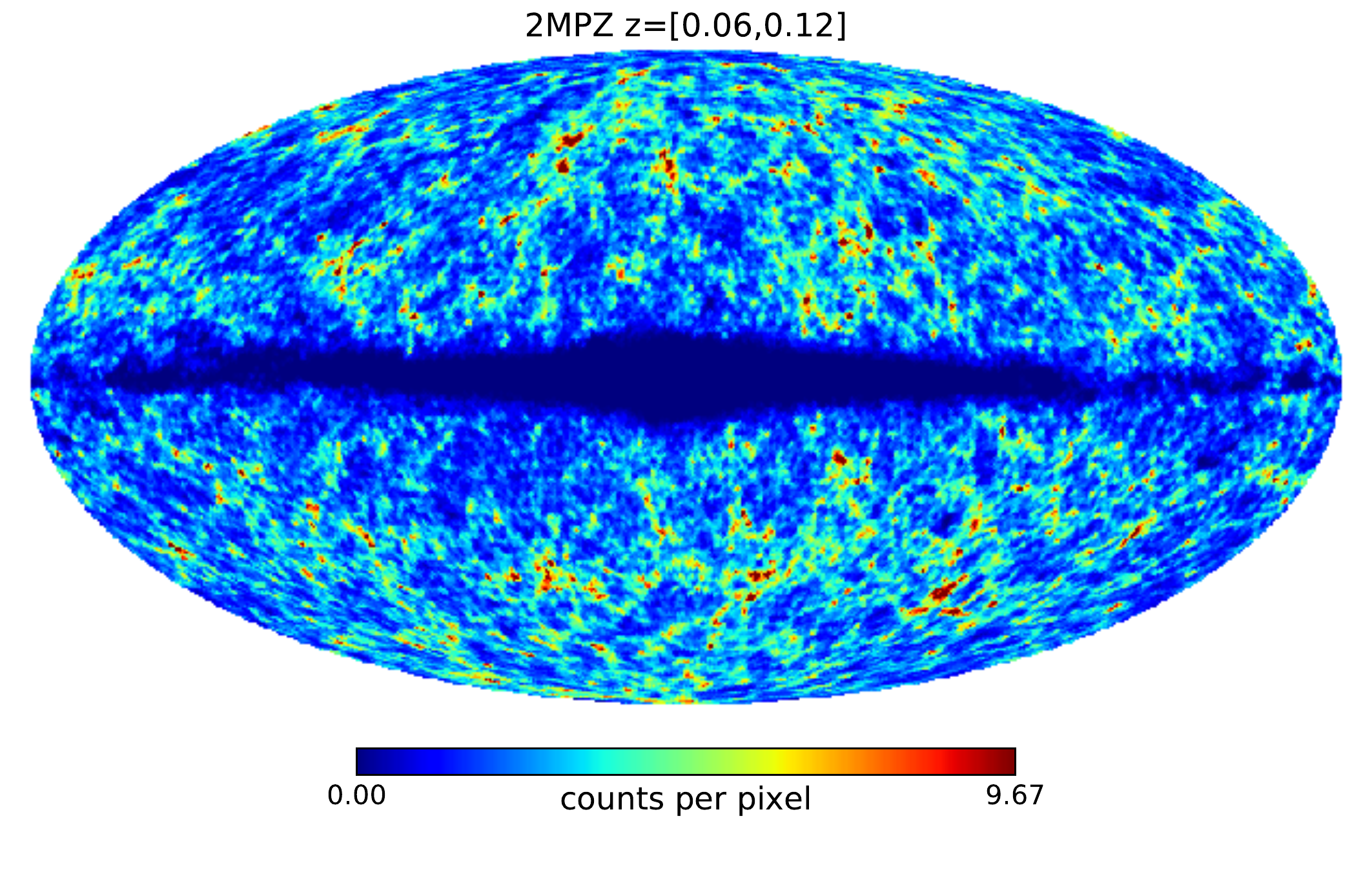}
\includegraphics[width=0.33\textwidth]{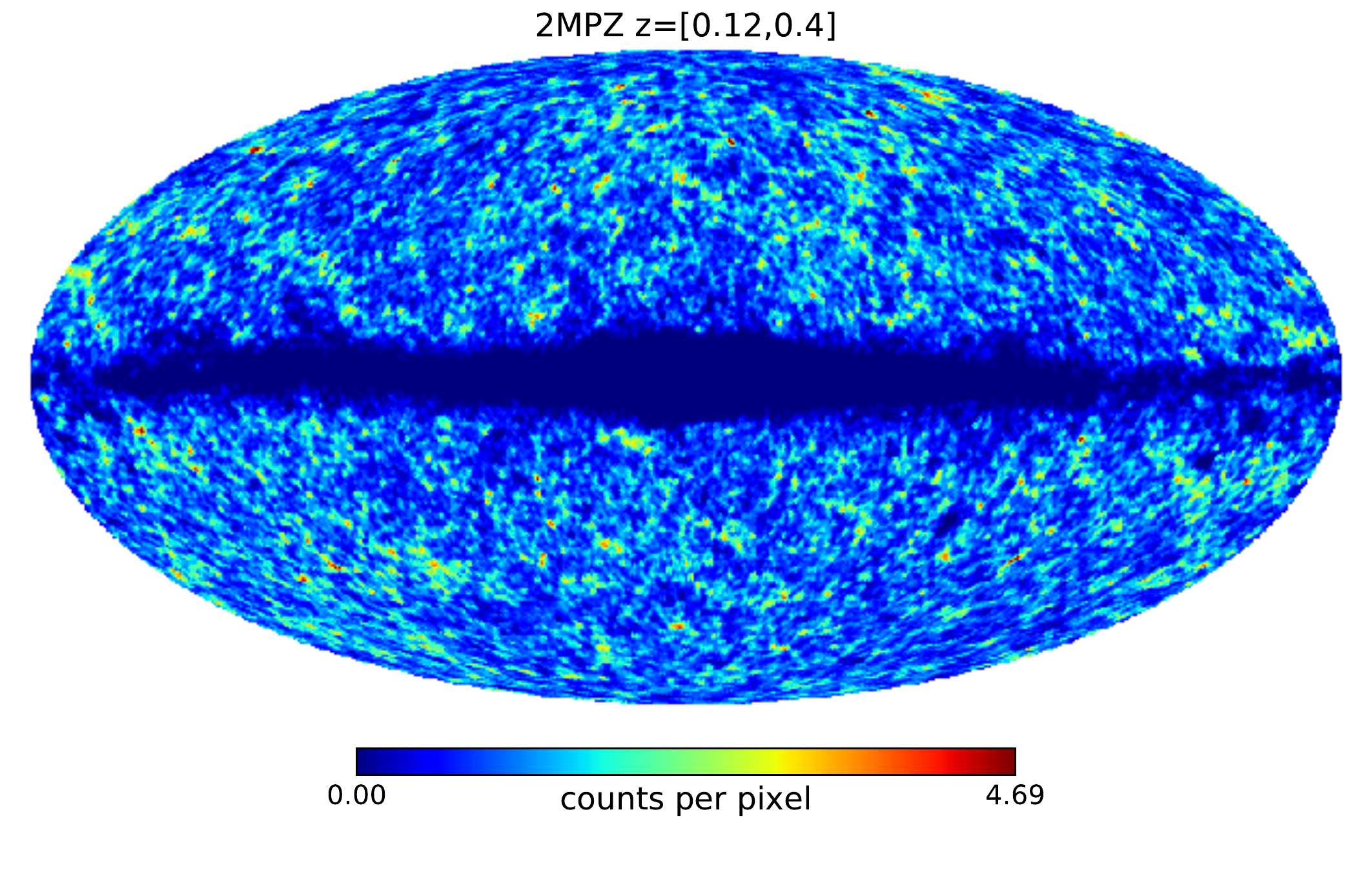}
\caption{All-sky projections of the 2MPZ galaxy distribution in the three redshift shells adopted in the analysis. 
The maps have HEALPix resolution $N_\mathrm{side}=128$ and include additional Gaussian smoothing of $\mathrm{FWHM}=1^\circ$  for better visualization.}
\label{fig:maps-2MPZ}
\end{figure*}

The 2MASS Photometric Redshift catalog \citep[2MPZ,][]{bilicki14} is a dataset of galaxies with
measured photometric redshifts  constructed by
cross-matching three all-sky datasets covering different energy bands:
2MASS-XSC \citep[near-infrared,][]{jarrett2000},  WISE \citep[mid-infrared,][]{wright10}
and SuperCOSMOS  scans of
UKST/POSS-II photographic plates \citep[optical,][]{peacock16}.
2MPZ is flux limited at $K_s < 13.9$ and contains
$\sim$ 935,000 galaxies over most of the sky.
However, since the strip at  $| b | < 10^{\circ}$ is undersampled, in
our analysis we masked out this region
as well as other incompleteness areas, using a mask similar to the one shown in \cite{alonso15}.

The 2MPZ photo-$z$s are generally unbiased ($\langle \delta z \rangle \sim 0$). Their random errors are
almost distance-independent, their distribution has an {\it rms} scatter $\sigma_z=0.015$
with 1\% of outliers beyond $3\sigma_z$.
The  redshift distribution of 2MPZ galaxies is  shown in Fig.~\ref{fig:dNdzs} (red line). It peaks at $z\sim 0.06$ and has    $\langle z \rangle \sim 0.08$.
The surface density of objects is $\sim30$ sources per square degree.
2MPZ is the only wide catalog that comprehensively probes the nearby Universe ($z\lesssim 0.2$) all-sky and has reliable redshift estimates. This feature and the possibility
of dividing the sample in different redshift shells are crucial to constrain the composition of the UGRB.
For our analysis we split the catalog in three redshift bins: $z \in [0.00,0.06]$, $[0.06,0.12]$ and $[0.12, 0.40]$.
The binning was designed to bracket the mean redshift in the second bin  and to guarantee a reasonably large number of objects in
the two other bins. Moreover, this binning has a good overlap with that adopted to slice the SDSS-DR12 sample  (Section~\ref{sec:DR12}).
In Section \ref{sec:ccf2mpz} we shall also use the full 2MPZ sample for the cross-correlation analysis (the case dubbed  ``ZA'') so that the results
can be directly compared with those of \Xia, obtained using the  2MASS catalog.

The all-sky distribution of 2MPZ  galaxies in each of the three redshift bins is shown in Fig.~\ref{fig:maps-2MPZ}.

\subsection{WISE $\times$ SuperCOSMOS}
\label{sec:WIXSC}

\begin{figure*}
\includegraphics[width=0.33\textwidth]{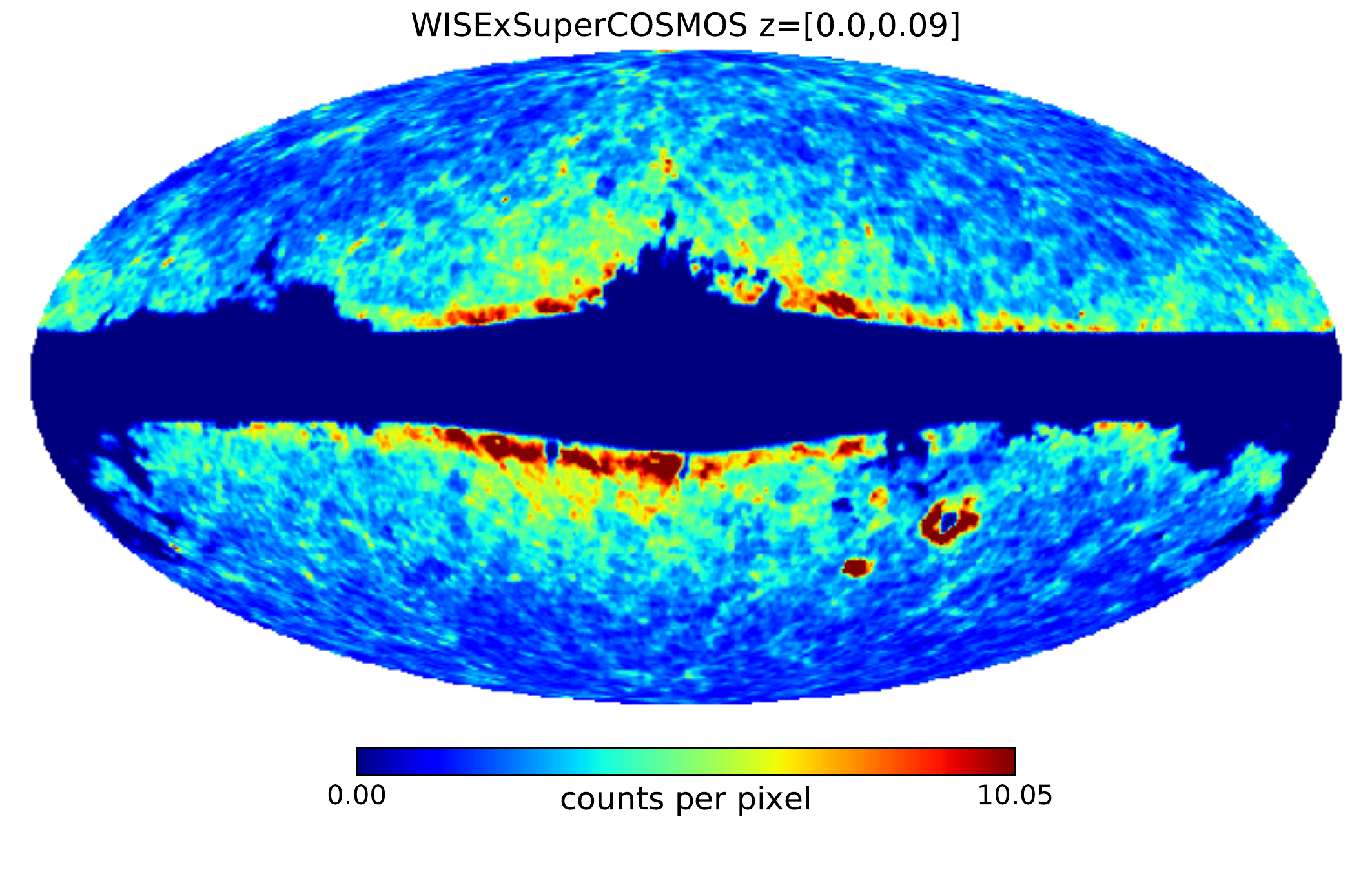}
\includegraphics[width=0.33\textwidth]{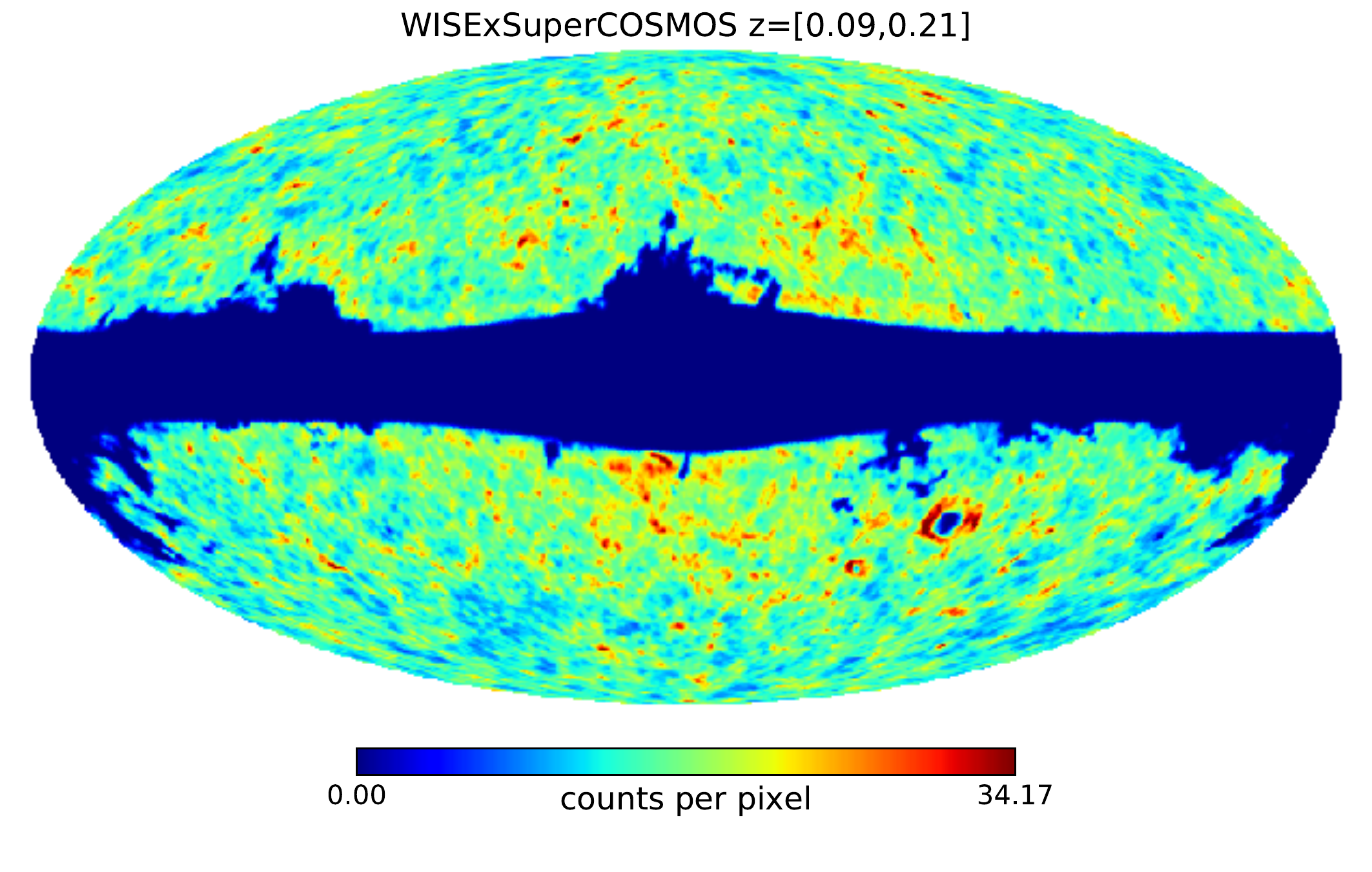}
\includegraphics[width=0.33\textwidth]{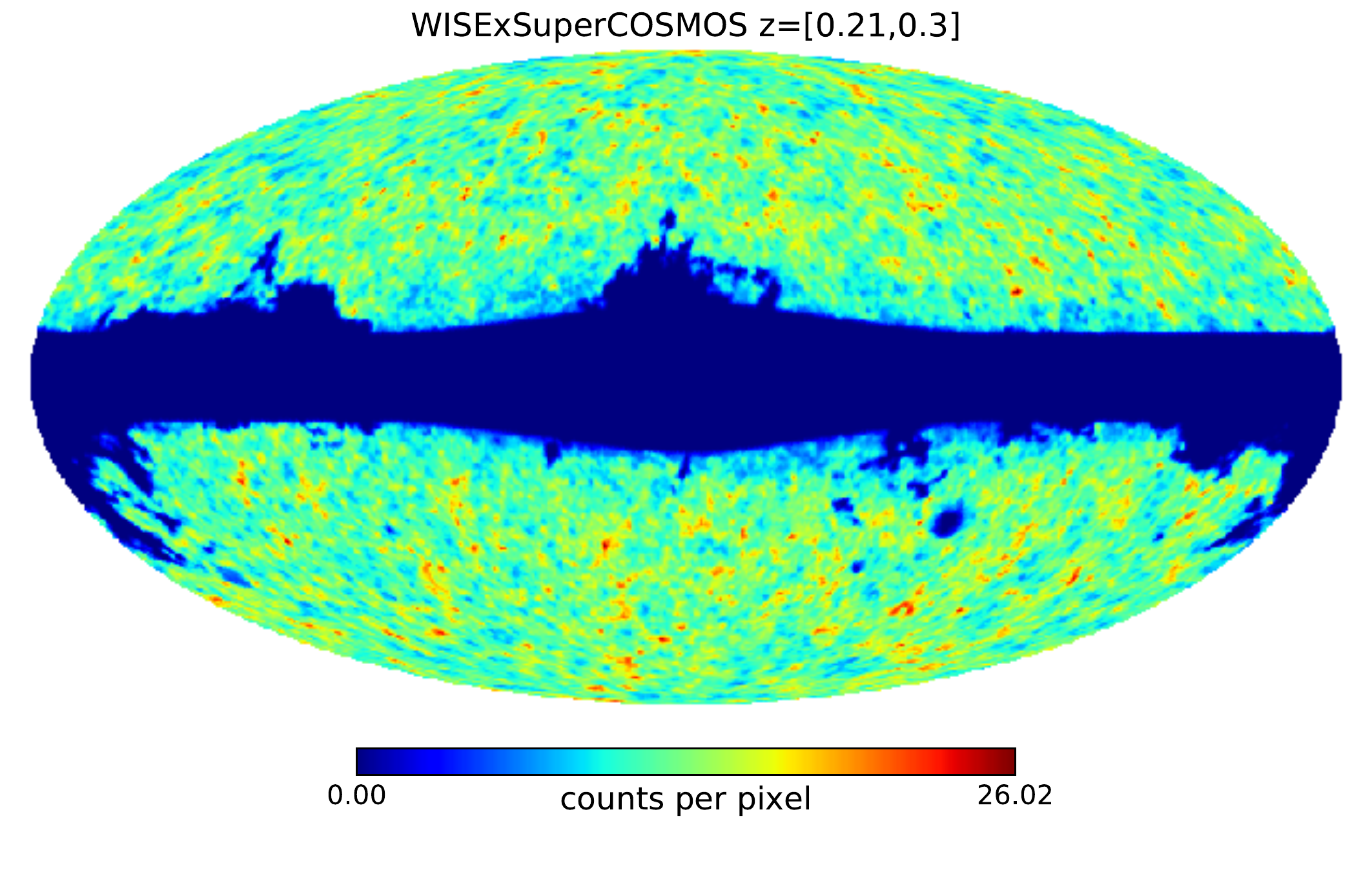}
\includegraphics[width=0.33\textwidth]{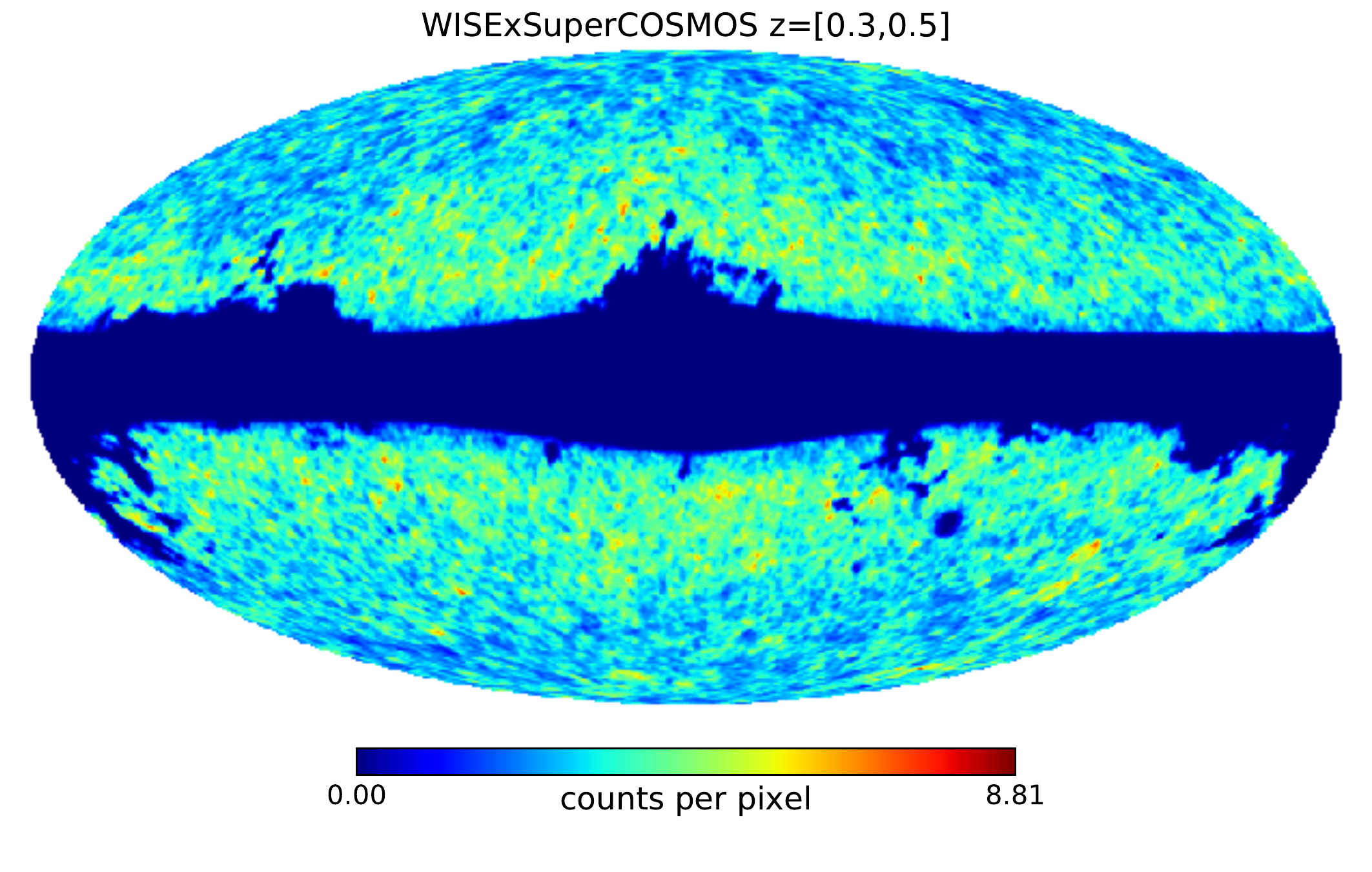}
\includegraphics[width=0.33\textwidth]{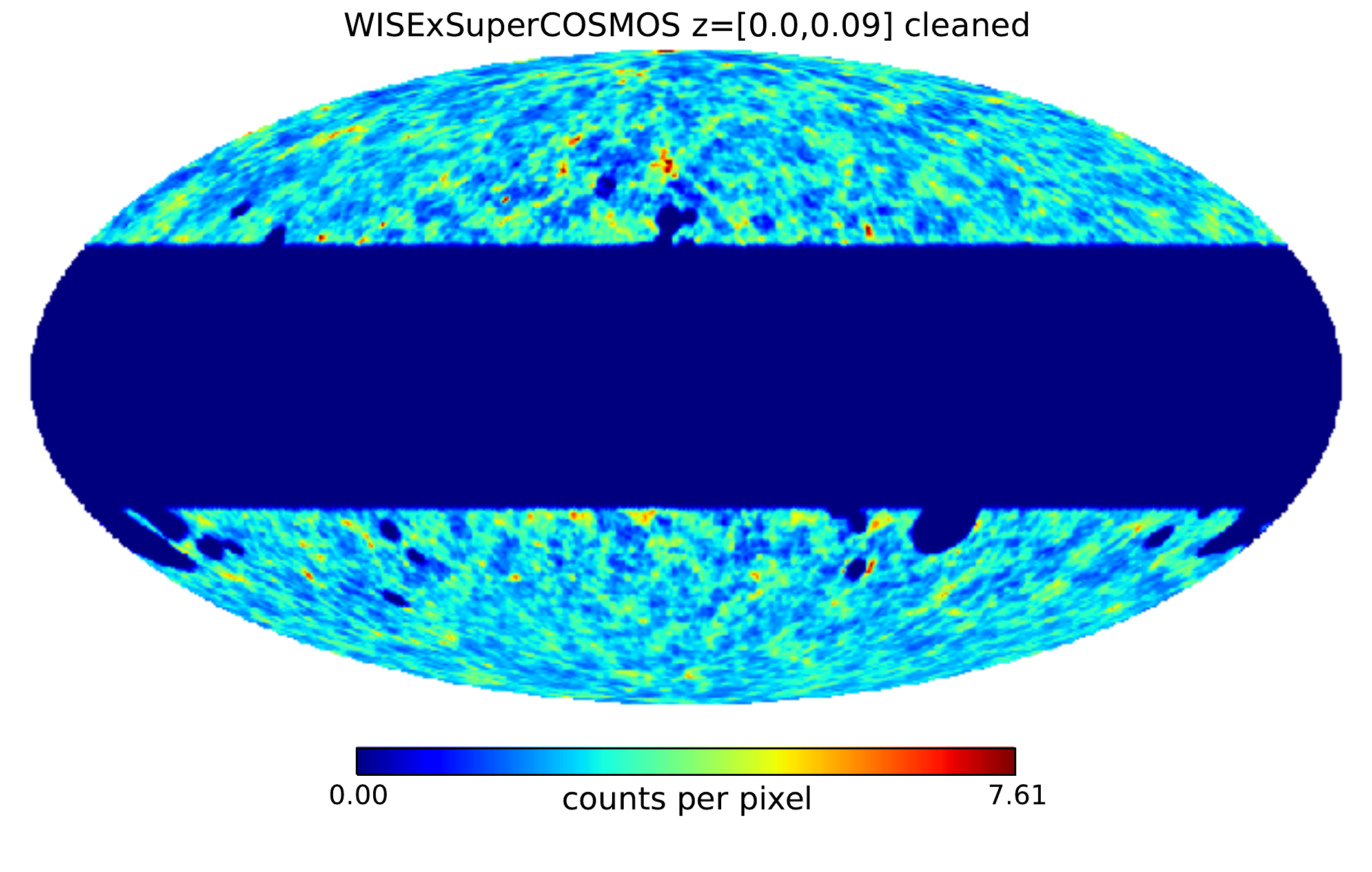}
\includegraphics[width=0.33\textwidth]{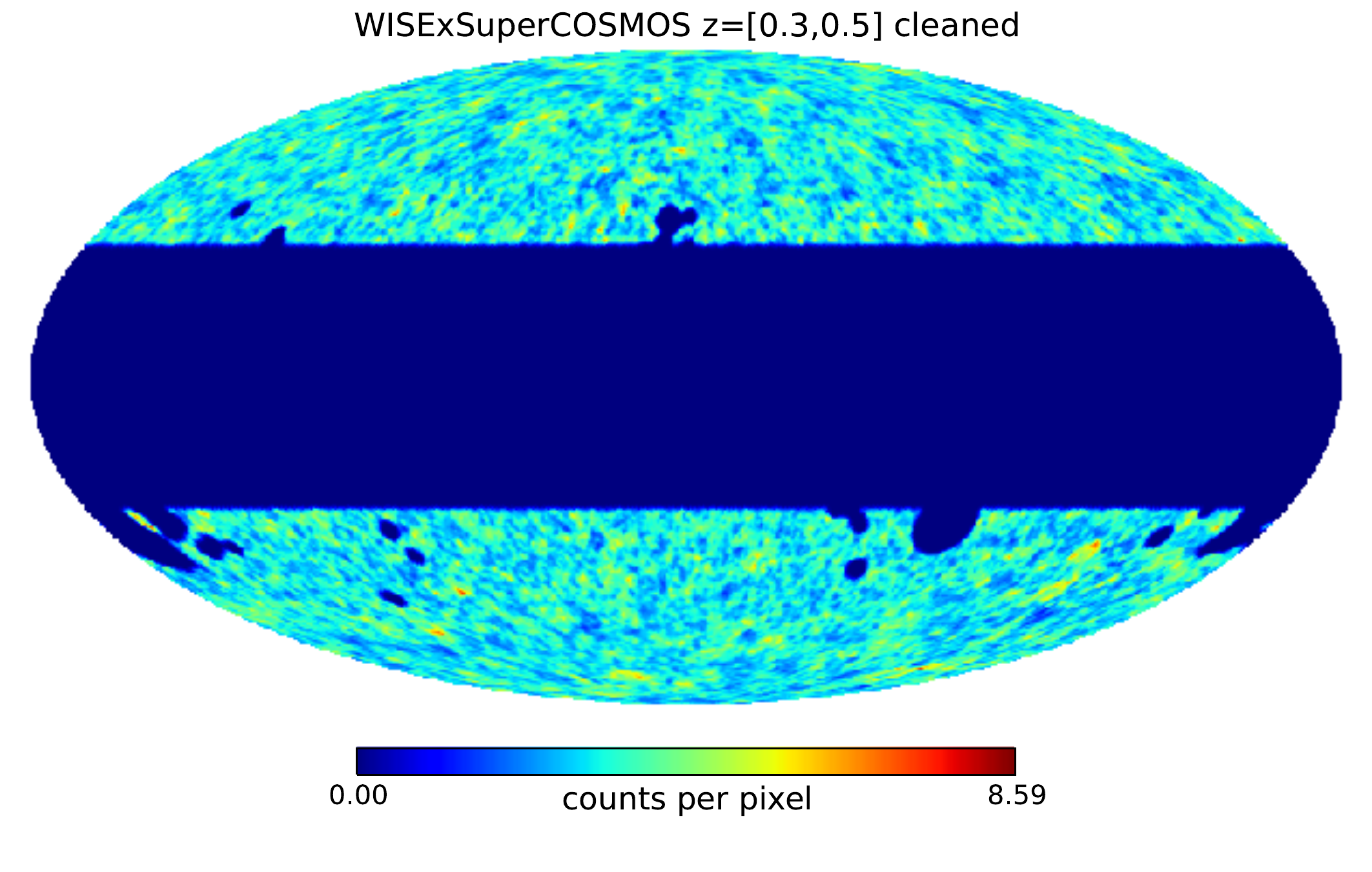}
\caption{All-sky projections of the WISE~$\times$~SuperCOSMOS galaxy distribution in the four redshift shells adopted in the analysis. The maps have HEALPix resolution $N_\mathrm{side}=256$ and include additional Gaussian smoothing of $\mathrm{FWHM}=1^\circ$ for better visualization.
The two bottom-right panels show the first and the last redshift bins after the $\ell10$ cleaning procedure, and applying the catalog mask and the fiducial Galactic plane mask of $|b|<30^\circ$.}
\label{fig:maps-WIxSC}
\end{figure*}

The WISE~$\times$~SuperCOSMOS photometric redshift catalog \citep{bilicki16}, hereafter \WISC,
 is the result of cross-matching the
two widest galaxy photometric catalogs currently available: the mid-infrared WISE  and optical SuperCOSMOS datasets.
Information from GAMA-II \citep{liske15} and SDSS-DR12 \citep{dr12} was used to exclude stars and quasars,  to obtain a sample of $\sim 20$ million galaxies with  a mean surface density above 650 sources per square degree.
The resulting catalog is $\sim 95\%$ pure at high Galactic latitudes of $|b|>30^\circ$ and
highly complete over $\sim 70\%$ of the sky, outside the Zone of Avoidance ($|b|<10^\circ$ plus the area around the
Galactic bulge) and  other confusion regions.

Photometric redshifts for all  galaxies, calibrated on GAMA-II, were estimated with a systematic error $| \delta_z | \sim 10^{-3}$
and a random error $\sigma_z \sim 0.033$  with $\sim 3\%$ of outliers beyond $3\sigma_z$.
The redshift distribution of  \WISC\ galaxies has a mean  
 $\langle z \rangle=0.2$ and is characterized by a broad peak extending
from $z\sim0.1$ to $z\sim 0.3$ and a prominent high-$z$ tail reaching up to $z>0.4$, as shown in Fig.~\ref{fig:dNdzs} (blue curve).

After masking we are left with about 18.5 million galaxies that we
divided into four redshift bins:
$z \in [0.00,0.09]$, $[0.09,0.21]$, $[0.21, 0.30]$, and $[0.30, 0.50]$.
As in the 2MPZ case, the binning was chosen to guarantee a significant overlap with the other source catalogs used in our analysis.
The first bin, Z1, encompasses the first two redshift bins of the 2MPZ sample, as well as the first redshift bin of the SDSS-DR12 one.
Because of the bright cut used to build the catalog,  \WISC\ probes an intrinsically faint population and has very few sources in common with 2MPZ 
and SDSS-DR12 at $z\leq0.09$ \citep[for more details see][]{bilicki16}.
The two bins at $z>0.21$, which contain an approximately equal number of  \WISC\ galaxies, overlap with SDSS-DR12 bins Z3 (i.e. the third redshift bin) and Z4+Z5.

The sky maps of the \WISC\ sources in the four bins are shown in Fig.~\ref{fig:maps-WIxSC}. The problematic areas near the Galaxy and the Magellanic Clouds,
that feature prominently especially in the first bin
 have been masked out and excluded from the cross-correlation analysis.
A residual over-density of sources along the Galactic Plane, which is visible in the first and last redshift bin and is likely due to stellar contamination, survives the masking procedure. We decided to remove it by applying the same $\ell10$ cleaning procedure as adopted for the $\gamma$-ray map.
This conservative procedure has little impact on the cross-correlation analysis since these problematic areas are largely excluded by the \textit{Fermi} Galactic plane mask ($|b|<30^\circ$) which we apply in each cross-correlation
(see, e.g., the bottom-right panels of Fig. ~\ref{fig:maps-WIxSC}).

\subsection{SDSS DR12 photometric}
\label{sec:DR12}

\begin{figure*}
\includegraphics[width=0.33\textwidth]{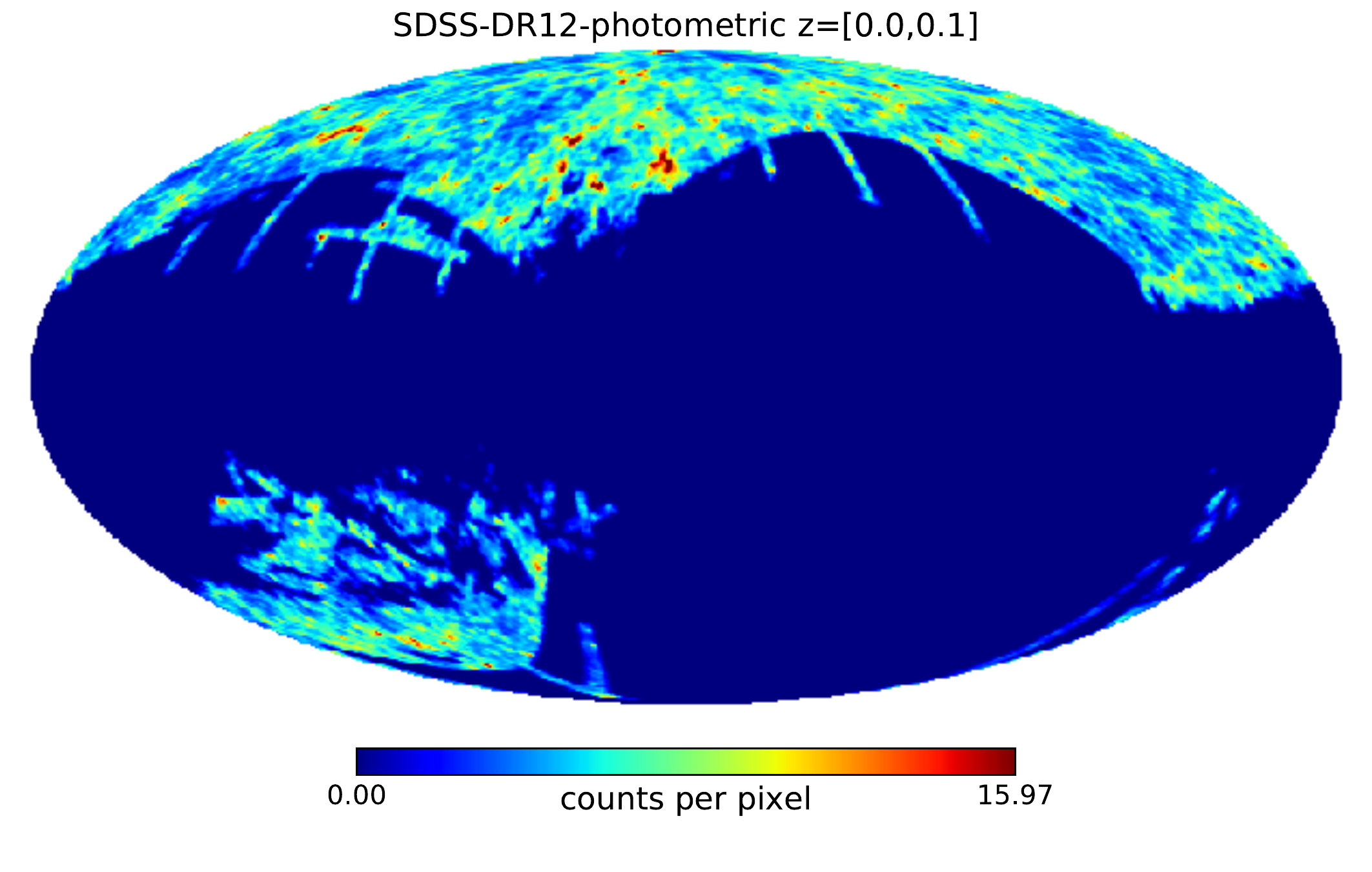}
\includegraphics[width=0.33\textwidth]{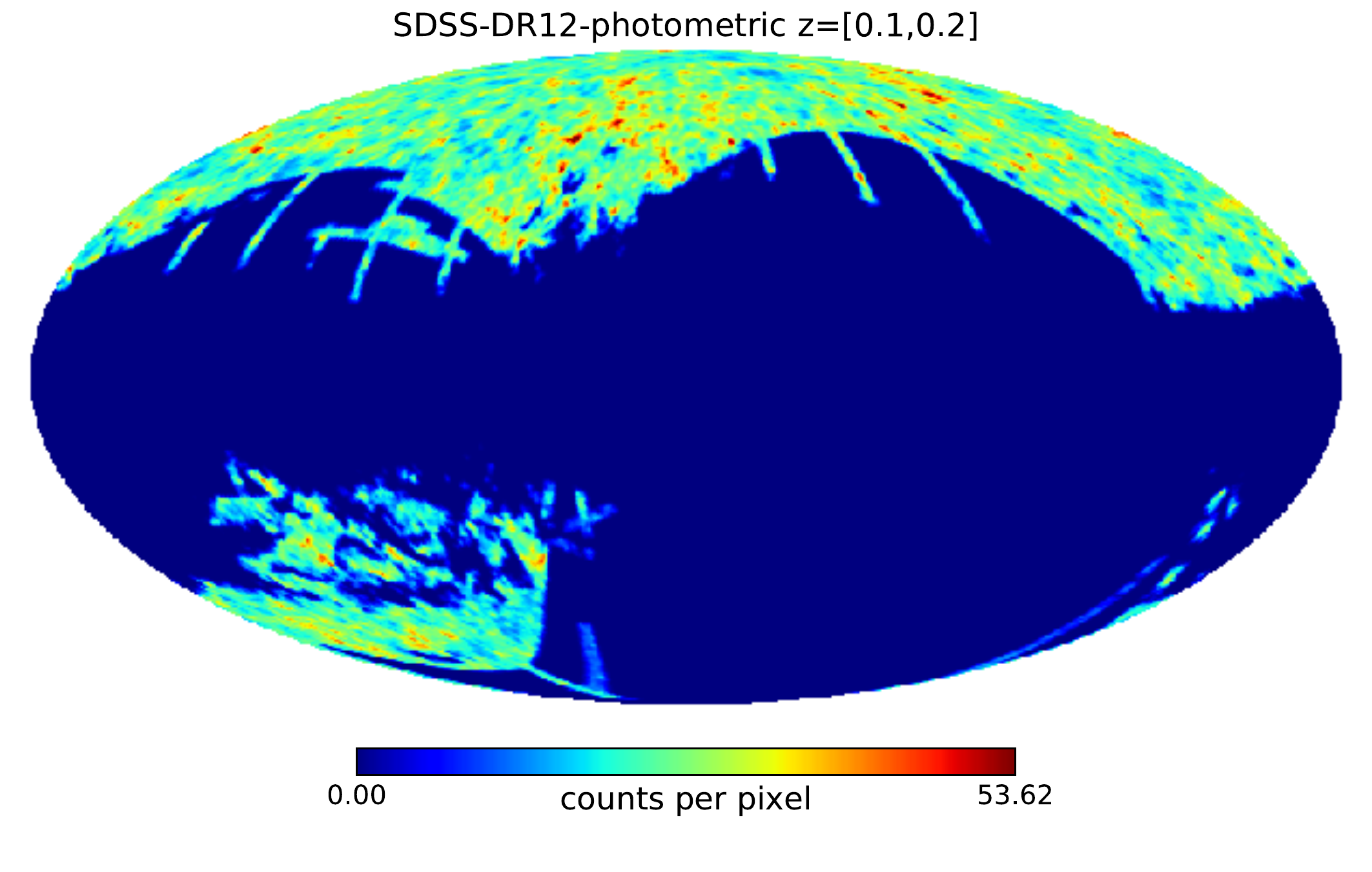}
\includegraphics[width=0.33\textwidth]{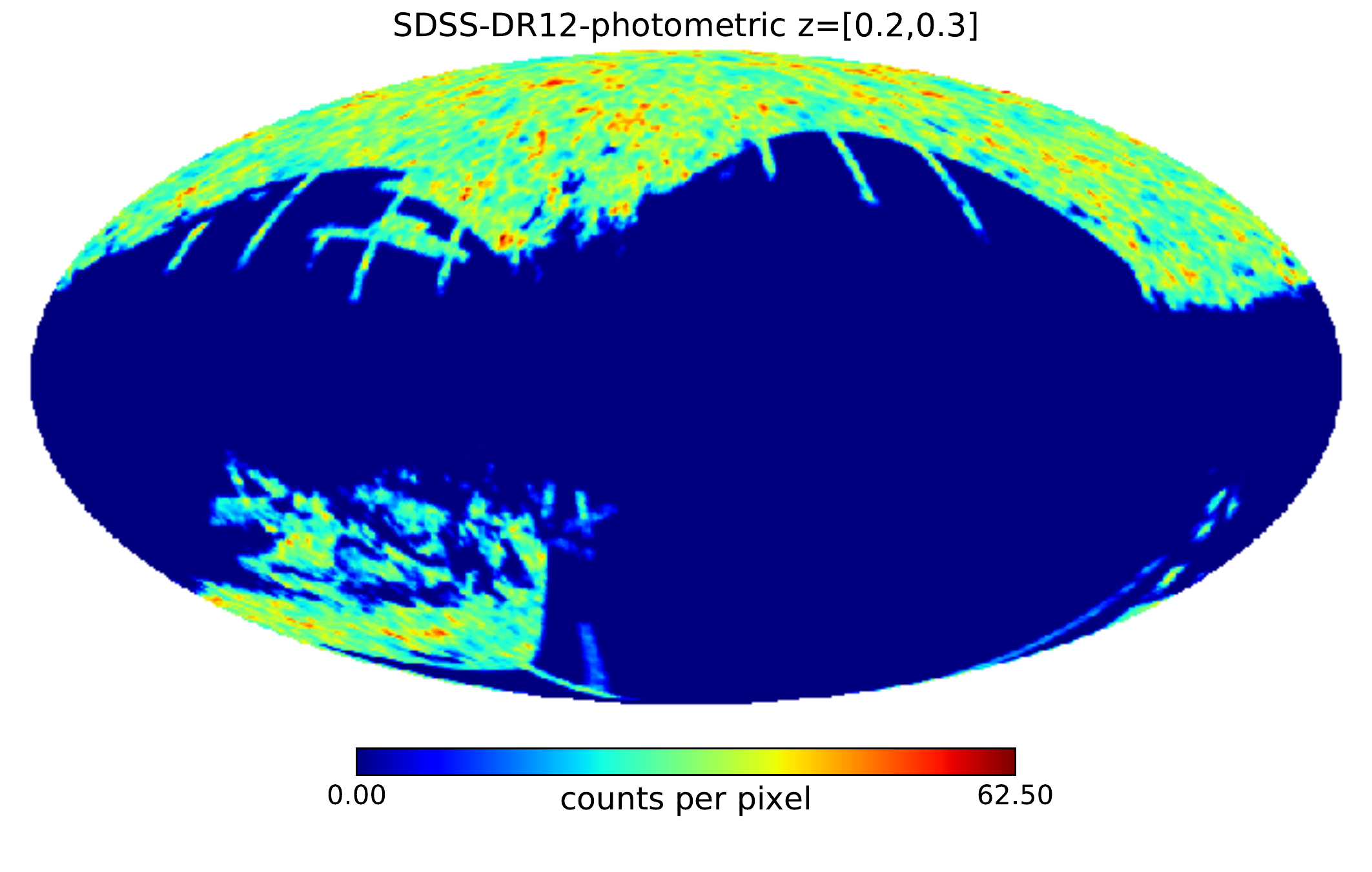}
\includegraphics[width=0.33\textwidth]{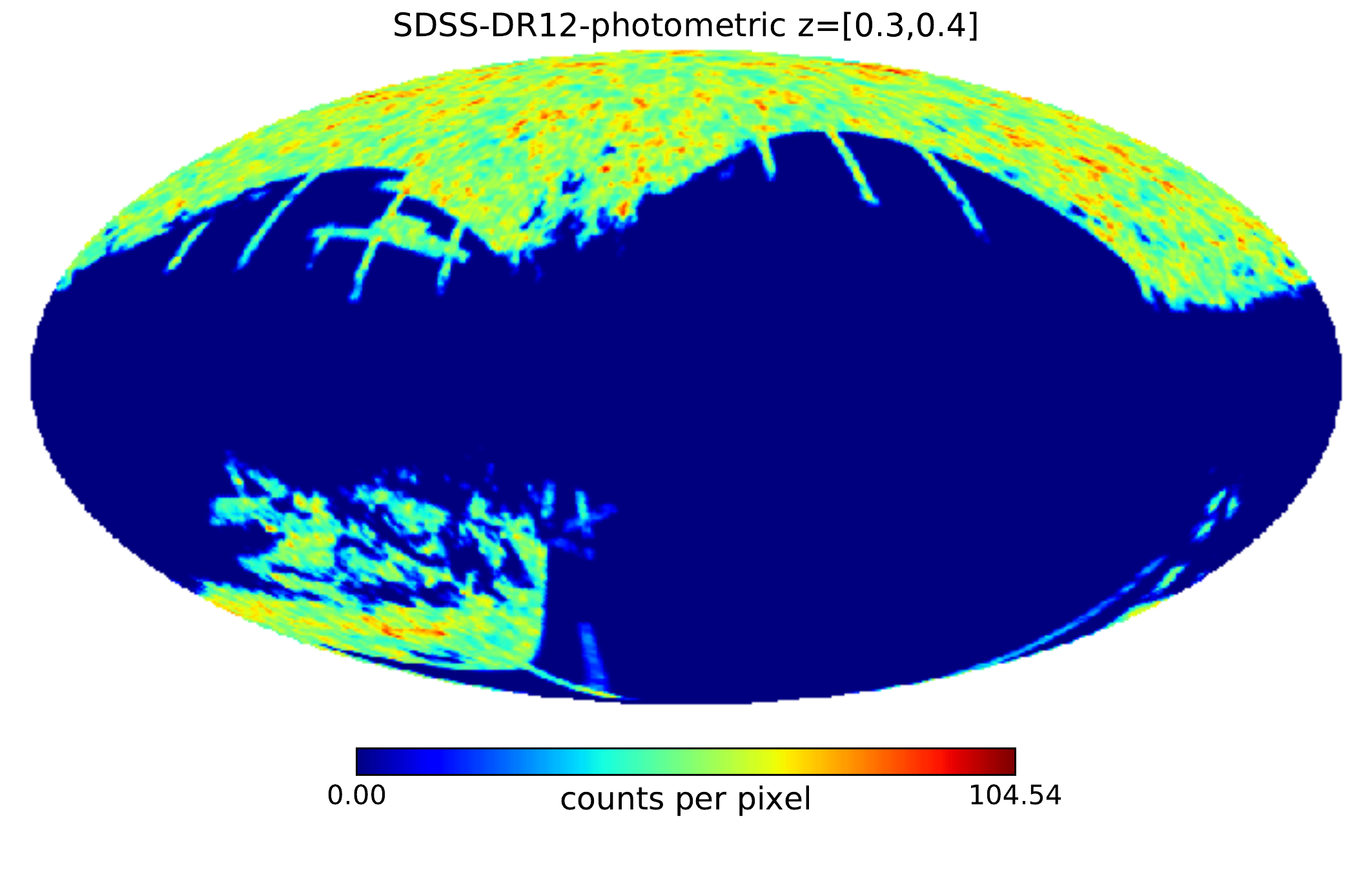}
\includegraphics[width=0.33\textwidth]{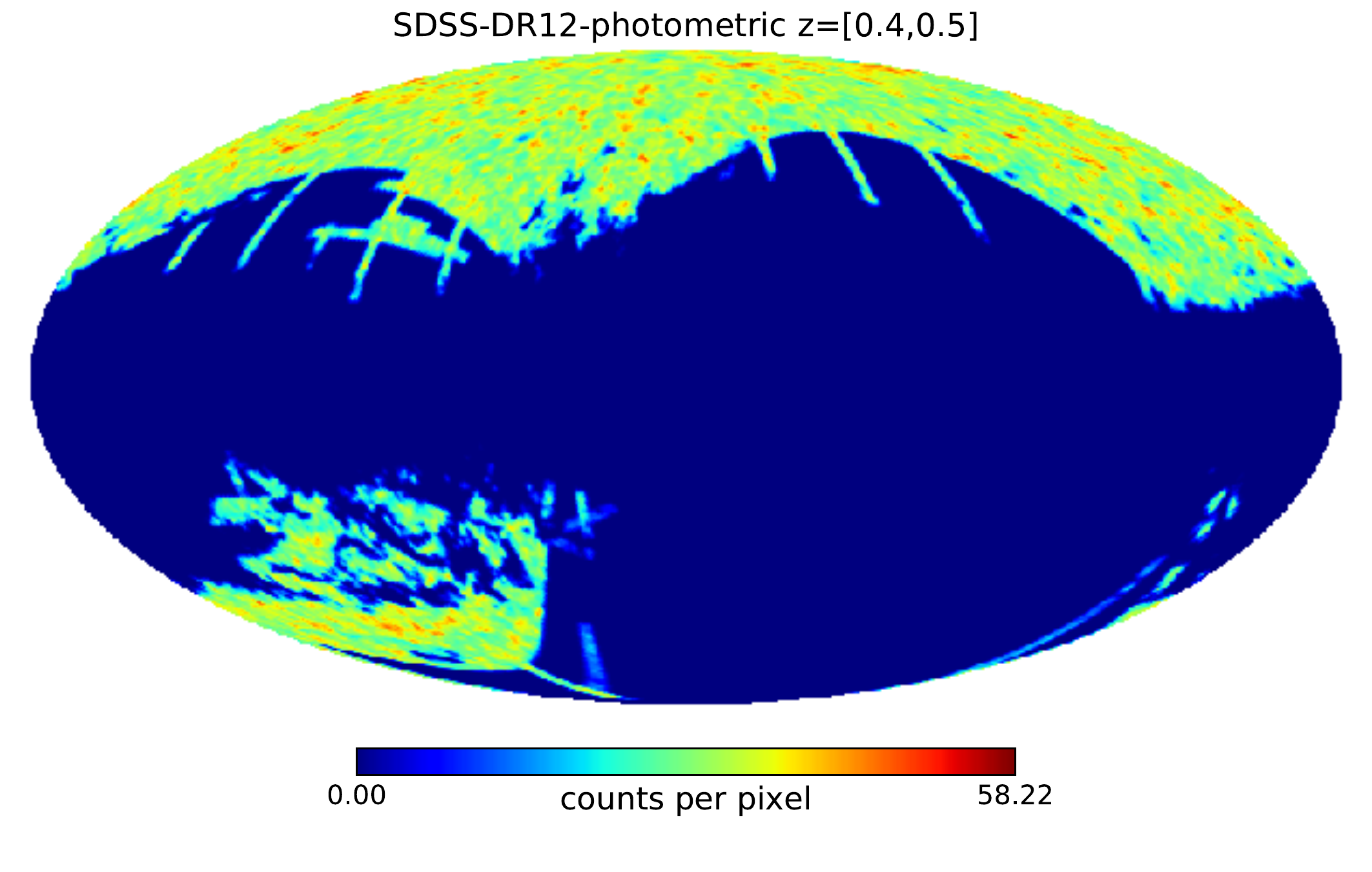}
\includegraphics[width=0.33\textwidth]{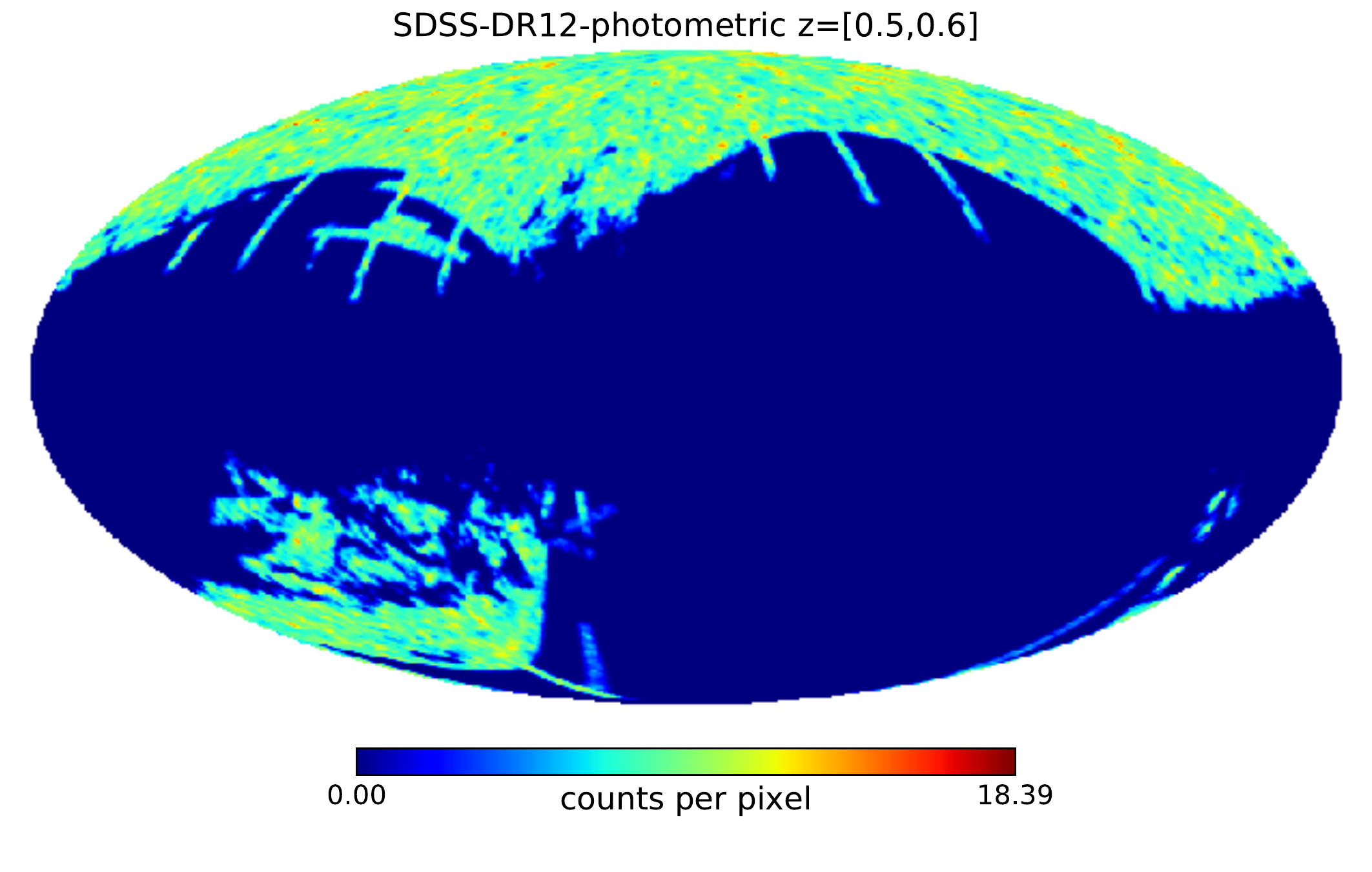}
\includegraphics[width=0.33\textwidth]{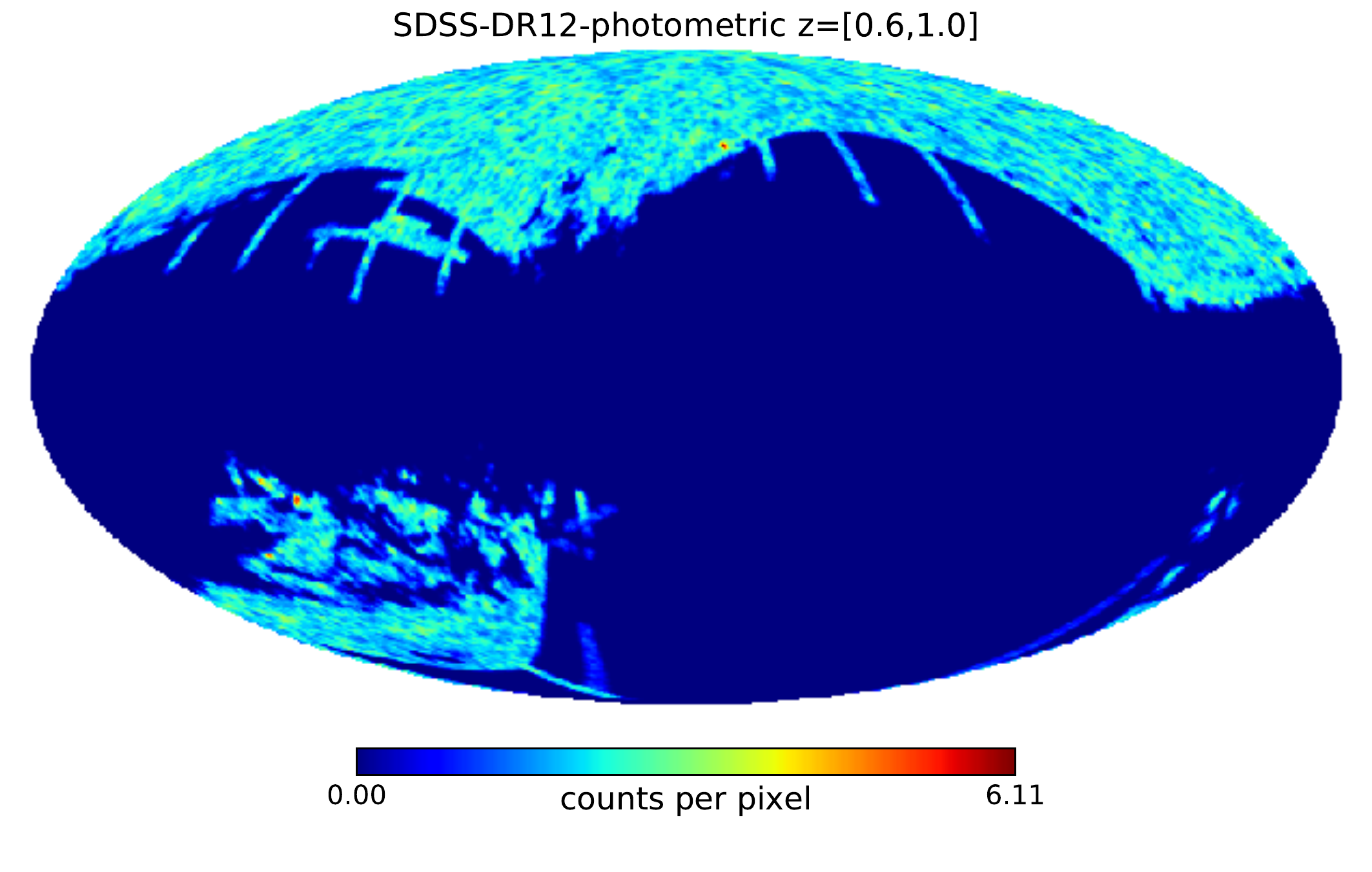}
\caption{All-sky projections of the SDSS DR12 photometric galaxy distribution in the seven redshift shells adopted in the analysis. The maps have HEALPix resolution $N_\mathrm{side}=256$ and include additional Gaussian smoothing of $\mathrm{FWHM}=1^\circ$  for better visualization.}
\label{fig:maps-SDSS-DR12}
\end{figure*}

This catalog is a revised version of the one used in \cite{xia11,xia15}. It has been derived from the SDSS photometric redshift  catalog
of  \cite{beck16}, which includes over 200 million sources classified by SDSS as galaxies, and  provides photometric redshifts for a large part of them.
Following authors' recommendations\footnote{See also \url{http://www.sdss.org/dr12/algorithms/photo-z/}.}, we
considered only the sources with a photometric error class $-1$, 1, 2, or 3, whose {\it rms} redshift error is $\sigma_z\leq0.074$  \citep{beck16}.
This dataset was further purified to obtain uniform depth over the observed area. For that reason we considered only galaxies with extinction-corrected
$r$ band magnitudes in the range $13<r<21$ outside the Zone of Avoidance $-10^\circ<b<15^\circ$,  as well as areas of  $r$-band extinction $A_r < 0.18$.
After applying these  selection criteria, we are left with 32 million sources with a mean redshift  $\langle z \rangle = 0.34$ and mostly
within $z < 0.6$. The sky coverage is $\sim 27\%$  and the mean surface density is $\sim 2900$ deg$^{-2}$.
As shown in Fig.~\ref{fig:dNdzs} (green line), their redshift distribution features a main peak at $z\sim0.38$ and a secondary one at $z\sim0.19$, possibly
indicating some issue with the photo-z assignment.
 Given the very large density of objects, we can split the sample into several redshift bins, keeping low shot-noise in each shell.
In our analysis we  divided the dataset  into seven redshift bins: six of width $\Delta z=0.1$, starting from $z=0$ to $z=0.6$, and the seventh encompassing the
wider range $0.6<z<1$
 (to compensate the source fall-off in the redshift distribution). These shells are illustrated in all-sky maps of Fig.~\ref{fig:maps-SDSS-DR12}.
 It can be seen that the SDSS galaxies are distributed into two disconnected regions in the Galactic  south and north, with most of the area in the north part.
Furthermore, as shown in the Figure, the southern region has quite uneven sampling. For this reason we have excluded this region from the analysis and use only the northern part. 

\section{Cross-Correlation Analysis}
\label{sec:corranalysis}

In the previous section we have presented the catalogs of extragalactic objects that
we use in the analysis. Their format is that of a 2D pixelized map of object counts
$n(\hat{\Omega}_i)$, where $\hat{\Omega}_i$ specifies the angular coordinate of the $i$-th pixel.
For the cross-correlation analysis we consider  maps of normalized counts $n(\hat{\Omega}_i)/\bar{n}$, where $\bar{n}$
is the mean object density in the unmasked area, and the  {\it Fermi}-LAT residual flux sky-maps, also pixelized with a matching angular resolution.

In our  analysis we compute both the angular 2-point cross-correlation function, CCF,
$w^{(\gamma c)}(\theta)$, and its harmonic transform, the
angular power spectrum $\bar C_\ell^{(\gamma c)}$, CAPS.
In particular, we shall use the CCF for visualization purposes but we restrict the quantitative analysis to the CAPS only.
The reason for this is choice is that the CAPS has the advantage that the different multipoles are almost uncorrelated, especially after binning.
Their covariance matrix is therefore close to diagonal, which simplifies the comparison  between models and data.
Furthermore, it is easier to subtract off instrumental effects like the point-spread function (PSF) smearing, since
a convolution in configuration space is just a multiplicative factor in harmonic space.
On the other hand, its interpretation is not so intuitive since the CAPS signal typically extends over a broad range of multipoles.
The CCF offers the advantage of a signal concentrated within a few degrees that can be intuitively associated to the angular size of the
$\gamma$-ray emitting region.
The quantitative analysis of the CCF is, instead, more challenging because the cross-correlation signals in different angular bins are highly correlated
and the PSF convolution effect is more difficult to account for.

Following \Xia, we use the  {\it PolSpice}\footnote{See \url{http://www2.iap.fr/users/hivon/software/PolSpice/}.}
statistical toolkit \citep{szapudi01,chon04,efstathiou04,challinor05} to estimate both CCF and CAPS.
{\it PolSpice} automatically corrects for the effect of the mask.
In this respect, we point out that the effective geometry of the mask used for the correlation analysis is
obtained by combining that of the LAT maps with those of each catalog of astrophysical objects.
The accuracy of the {\it PolSpice} estimator has been assessed in \Xia\ by comparing the measured CCF
with the one computed using the popular Landy-Szalay  method \citep{ls93}. The two were found to be in very good agreement.
 {\it PolSpice} also provides the  covariance matrix for the angular power spectrum,
 $\bar V_{\ell\ell'}$ \citep{2004MNRAS.349..603E}.

The instrument  PSF and the map pixelization affect the
estimate of the CAPS.
To remove these effects we proceed as in \Xia:
we first derive the beam window function $W_\ell^{B}$ associated to the LAT PSF, and
the pixel window function $W_\ell^{\rm pixel}$ associated to the map pixelization.
{Since the LAT PSF varies significantly with energy, we derive $W_\ell^{B}$
on a grid of 100 energy values from 100 MeV to 1 TeV. This is then used
to derive the average $W_\ell^{B}$ in the specific energy bin analyzed.
The procedure is described in detail in \Xia.}
Then we exploit the fact that convolution in configuration space is a multiplication in harmonic space and
estimate the deconvolved CAPS $C_\ell^{(\gamma c)}$ from the measured one $\bar C_\ell^{(\gamma c)}$ as
$C_\ell^{(\gamma c)}=(W_\ell)^{-1}\,\bar C_\ell^{(\gamma c)}$, where $W_\ell = W_\ell^{B} \left( W_\ell^{\rm pixel} \right)^2 $
is the global window function.
The window function $W_\ell$ has two contributions, from the LAT and cross-correlating catalog: 
it is the double product of the beam window $W_\ell^{B}$ and the pixelization window function $W_\ell^{\rm pixel}$ from each catalog. 
However, we neglect a factor of $W_\ell^{B}$ related to the catalog maps
since the typical angular resolution of the catalogs ($<10"$) is much smaller than the pixel size,
 so that the associated $W_\ell^{B} \simeq 1$.
The square in the $W_\ell^{\rm pixel}$ term takes into account the pixel window functions of both maps. Its effect is minor since, 
as shown in Fig.3 in \citealt{FermiAPS_16}, its value is 
close to unity up to $\ell=2000$, which is the maximum multipole considered in our analysis. 
The covariance matrix for the deconvolved  $C_\ell^{(\gamma c)}$ is then expressed as
$V_{\ell\ell'}  = \bar V_{\ell\ell'} W_\ell^{-2}W_{\ell'}^{-2}$.
Finally, to reduce the correlation in nearby multipoles induced by the angular mask, we bin the measured CAPS
into 12 equally spaced logarithmic intervals in the range $\ell\in[10,2000]$.
We choose logarithmic bins to account for the rapid loss of power at high $\ell$ induced by the PSF.
We indicate the  binned CAPS with the same symbol as the unbinned one $C_\ell^{(\gamma c)}$.
It should be clear from the context  which one is used.
The $C_\ell^{(\gamma c)}$ in each bin is given by the simple unweighted average
of the $C_\ell^{(\gamma c)}$ within the bin.
For the binned $C_\ell^{(\gamma c)}$ we build
the corresponding covariance matrix as a block average of the unbinned covariance matrix, i.e., $\sum_{\ell\ell'} V_{\ell\ell'}/\Delta\ell/\Delta\ell'$,
where $\Delta\ell, \Delta\ell'$ are the widths of the two multipole bins, and $\ell, \ell'$ run over
the multipoles of the first and the second bin.
The binning procedure is very efficient in removing correlation among nearby multipoles,
resulting in a block covariance matrix that is, to a good approximation, 
diagonal\footnote{{Note that, in the case of non-Gaussian fluctuations, like the one considered here, 
the non vanishing trispectrum could induce, possibly, extra correlation among the multipoles
\citep{Komatsu:2002wc,Ando:2017wff}. These terms are not considered in the covariance matrix computed by {\it PolSpice},
that we use in our analysis. The importance of these terms for our analysis is uncertain, although we have  found that errors computed using the 
 {\it PolSpice} covariance matrix are compatible to those computed using Jackknife resampling techniques (\Xia). 
 A dedicated analysis would be required to  properly quantify the impact of these terms, which is beyond the scope of this work.}}
For this reason
we will neglect the off-diagonal terms in our analysis and only use the diagonal ones
$\left(\Delta C_{\ell} \right)^2=\sum_{\ell\ell'} V_{\ell\ell'}/\Delta\ell^2$.

The CCF covariance matrix can be computed from the CAPS covariance as \citep{2013arXiv1303.5075P}
\begin{equation}
C_{\theta\theta'}^{\rm PS}=\sum_\ell\sum_{\ell'}\frac{2\ell+1}{4\pi}\frac{2\ell'+1}{4\pi}P_\ell(\cos \theta)P_{\ell'}(\cos \theta')
\bar{V}_{\ell\ell'}~ \; .
\end{equation}
An average over the angular separations $\theta$ and $\theta'$ within each bin can be performed to obtain a binned covariance matrix.
In the following, we will compute the $w^{(\gamma c)}(\theta)$ in the range $\theta\in[0.1^{\circ},100^{\circ}]$ binned into 10 logarithmically spaced bins.
Since, as already mentioned, we limit our quantitative analysis to CAPS, we shall not use the CCF covariance matrix nor we make any attempt to
deconvolve the measured  $w^{(\gamma c)}(\theta)$ to account for the effects of the PSF and pixelization.
We do, however, show the measured CCF and its errors in our plots. The errorbars there correspond to the diagonal element of the binned CCF covariance matrix.
Error covariance is therefore not represented in the plots.

\section{CAPS models and $\chi^2$ analysis}
\label{sec:chi2}

In this section we illustrate our models for the CAPS and the CCF
 and how we compare them with the
measurements.

We consider a simple, phenomenological CAPS model, 
inspired by the {\it halo model},
\citep{Cooray:2002dia} in which the angular spectrum in each energy bin
is a sum of two terms
\begin{equation}
  \hat{C}^{\rm \gamma c}_{\ell} = C^{\rm 1h} + A^{\rm 2h}C^{\rm 2h}_{\ell},
 \label{eq:1h2h}
\end{equation}
named 1-halo and 2-halo terms, respectively.
The halo model assumes that all matter in the universe is contained in DM halos populated by
 baryonic objects, like galaxies, AGNs, and, in particular, $\gamma$-ray sources.
In this framework, $C^{\rm 1h}$  quantifies the spatial correlation  within a single halo
i.e.,  $\gamma$-ray sources and extragalactic objects that reside in the same DM halo. 
{The special case in which  the $\gamma$-ray and the extragalactic sources are the same object detected at 
different wavelengths is, sometimes, treated differently, since it formally corresponds to a Dirac-delta
correlation in real space.
Nonetheless, since halos are typically smaller than the available angular resolution of Fermi-LAT,
it is, in practice, hard to distinguish the degenerate case from the case of two separate objects
Thus, for simplicity, we include both into a single term which contribute to the  the 1-halo correlation.}
The $C^{\rm 2h}$ term describes the halo-halo clustering. If non zero it indicates that both 
$\gamma$-ray sources and extragalactic objects trace the same large scale structure.

{The Fourier transform of the 1-halo term, $C^{\rm 1h}$ is therefore made of two components.
The first one, which comes from the Dirac-delta, is a constant term in the
 $\ell$-space. The second one, which is the   Fourier transform of the halo profile, does depend on the
multipole $\ell$.
In practice, however, its $\ell$ dependence is very weak because DM halos are almost point-like
at the resolution set by the LAT PSF.
Therefore we model the total $C^{\rm 1h}$ as a constant and ignore any multipole dependence.
We believe that this is a fair hypothesis for all analyses performed in this study except, perhaps, the
cross correlation with the  2MPZ catalogue since some of the halo hosts are close enough to us to appear wider than
the LAT PSF.
In this case the modeling of $C^{\rm 1h}$ is probably inaccurate at the highest $\ell$.
Nonetheless, this inconsistency should have a negligible impact on our analysis 
because of the large errors on the $C_\ell$ measured at large multipoles which reduce  substantially the sensitivity to 
the shape of the  $C^{\rm 1h}$ at high $\ell$ values.}
$A^{\rm 2h}$ is the second free parameter of the model which sets the amplitude of the 2-halo term, $C^{\rm 2h}_{\ell}$,
that accounts for the  correlation among halos. Its $\ell$ dependence reflects the angular correlation properties of
the DM halo distribution.
To first approximation it can be expressed as
\begin{equation}\label{xcross}
  C^{\rm 2h}_{\ell}=\frac{2}{\pi}\int{k^2P(k)[G^{\rm \gamma}_{\ell}(k)][G^{\rm c}_{\ell}(k)]dk}~,
\end{equation}
where $P(k)$ is the power spectrum of  matter density fluctuations.
We take the linear prediction of  $P(k)$ from the {\it camb} code \citep{camb} for
the \cite{Ade:2015xua} cosmological parameters specified in Sec.~\ref{sec:intro}, and apply
a non-linear correction using {\it halofit} \citep{halofit,Takahashi:2012em}.
The functions $G(k)$ specify the contribution of each field to the cross-correlation signal.
{More specifically, the contribution from the field of number
density fluctuations in a population of discrete objects is given by
\begin{equation}
  G^{\rm c}_{\ell}(k) = \int{\frac{dN}{dz}b_{\rm c}(z)D(z)j_\ell[k\chi(z)]dz}~,
\end{equation}
where $dN/dz$ is the redshift distribution of the objects, 
$j_\ell$ are spherical Bessel functions, $D(z)$ is the linear growth factor of density fluctuations,
$b_{\rm c}$ is the linear bias parameter of the objects, and $\chi(z)$ is the comoving distance to redshift $z$.
The analogous quantity for the diffuse UGRB field is
\begin{equation}
  G^{\rm \gamma}_{\ell}(k) = \int{\bar{\rho}_\gamma(z)b_\gamma(z)D(z)j_\ell[k\chi(z)]dz}~,
\end{equation}
where $b_\gamma(z)$ is the linear bias of the $\gamma$-ray emitters,
and $\bar{\rho}_\gamma(z)$ is their average flux density.}

When the cross-correlation is computed for the whole  catalog of sources, we
consider the full $dN/dz$ shown in Fig.~\ref{fig:dNdzs}. When, instead,  the cross-correlation is computed in
a specific redshift bin, then we set the $dN/dz$ equal to zero outside
the redshift bin and equal to the original $dN/dz$ inside the bin. The amplitude of the corresponding $dN/dz$ is normalized to unity.
{For the distribution of the $\gamma$-ray emitters,  $\bar{\rho}_\gamma(z)$,
the situation is more complicated, since we do not observe  $\bar{\rho}_\gamma(z)$ directly.} 
{In principle, the aim of the cross-correlation analysis  is, indeed, to constrain this quantity, i.e., to assume a model 
 $\bar{\rho}_\gamma(z)$, predict the expected cross correlation and compare it with the observed one.
 This will be pursued in a follow-up analysis in which we shall consider physically motivated $\bar{\rho}_\gamma(z)$ models.}
 {Instead, here, where we aim at an illustrative, model-independent approach,
we choose to have the average  $\bar{\rho}_\gamma(z)$ 
in a given redshift bin as a free parameter.
In this way, the absolute normalization of $\bar{\rho}_\gamma(z)$ is absorbed in  the parameter  $A^{\rm 2h}$.
More precisely, when cross-correlating the UGRB with a catalog in a given redshift bin,
the measured  $A^{\rm 2h}$ will be the product of three quantities.
The first two are the average bias factors $b_{\rm c}(z)$ and $b_\gamma(z)$ in the redshift range of the bin, and
the third will be the average $\bar{\rho}_\gamma(z)$ in that bin.}

We stress that this simple model tries to capture the angular correlation features of the
expected cross-correlation signal without assuming any specific model for the sources of
the UGRB. Its main goal is to separate the signal
into 1-halo and 2-halo components, and study their energy dependence.
In a follow up paper, we shall consider a physically motivated model, similar to that of  \cite{Cuoco:2015rfa},
including the contribution from all potential unresolved $\gamma$-ray sources
(blazars, misaligned AGNs, star forming galaxies, decaying or annihilating non-baryonic matter).
Within this framework, it will be possible to explicitly specify the bias of the sources, their number density as a function of redshift, $\rho_\gamma(z)$,
as well as their clustering.

Eq.~\eqref{eq:1h2h} models the CAPS for a single energy bin.
However, since in this work we compute the cross-correlation signal in several energy bins,
we can also use a CAPS model which includes an explicit energy dependence. For this purpose
we have considered three different models specified below:
\begin{itemize}
\item Single Power Law  [SPL]:
\begin{equation}
\hat{C}^{\rm \gamma c}_{\ell} (E) = \Delta E \left( C^{\rm 1h} + A^{\rm 2h}C^{\rm 2h}_{\ell} \right) \cdot (E/E_0)^{-\alpha} \, ,
\label{eq:spl}
\end{equation}
where, $\Delta E$ is the width of the energy bin considered in the cross-correlation analysis, $\alpha$ is the slope and
$E_0=1$ GeV is a normalization energy scale.
\item Double Power Law [DPL]:
\begin{equation}
 \hat{C}^{\rm \gamma c}_{\ell} (E) = \Delta E\  C^{\rm 1h} (E/E_0)^{-\alpha_{\rm 1h}}  + \Delta E\ A^{\rm 2h}C^{\rm 2h}_{\ell} (E/E_0)^{-\beta_{\rm 2h}} \, ,
\label{eq:dpl}
\end{equation}
where the 1-halo and 2-halo terms are allowed to have two different  power laws with slopes $\alpha$  and $\beta$.
\item Broken Power Law  [BPL]:
\begin{equation}
\hat{C}^{\rm \gamma c}_{\ell} (E) = \Delta E \left( C^{\rm 1h} + A^{\rm 2h}C^{\rm 2h}_{\ell} \right) \cdot   \left\{  \!\!\!\!
\begin{tabular}{cc}
$(E/E_b)^{-\alpha}$, &  $E >E_{b}$\\
$(E/E_b)^{-\beta}$,  &   $E< E_{b}$
\end{tabular}
\right.
\, ,
\label{eq:bpl}
\end{equation}
characterized by a broken power law with slopes $\alpha$  and $\beta$ respectively above and below the break energy $E_{b}$.
\end{itemize}

To compare the data and models we use standard $\chi^2$ statistics for which we consider two implementations.
When we focus on a single energy range and thus we ignore energy dependence, then we use
\begin{equation}
  \chi^2 \equiv \chi^2(E,z,c)=\sum_{\rm \ell~bins} {\frac{(\hat{C}^{\rm \gamma c}_{\ell} - C^{\rm \gamma c}_{\ell})^2}{(\Delta C^{\rm \gamma c}_{\ell})^2} } \, ,
\label{eq:chi21}
\end{equation}
where $\hat{C}^{\rm \gamma c}_{\ell}$ and $C^{\rm \gamma c}_{\ell}$  represent the model and the measured CAPS, the sum is over all $\ell$ bins
and the triplet $(E,z,c)$ identifies the energy range, redshift bin and object catalog considered in the analysis.
The best-fitting $C^{\rm 1h}$ and $A^{\rm 2h}$ parameters are found by the minimization  of the  $\chi^2$ function.
Note that in the following, together with $C^{\rm 1h}$  we shall list the normalized value $A^{\rm 2h}C^{\rm 2h}_{\ell=80}$  that has the same dimension as  $C^{\rm 1h}$. 
This choice is motivated by the fact that the fit constrains the product $A^{\rm 2h}C^{\rm 2h}_{\ell}$, rather than the single terms separately.
The rationale for setting $\ell=80$ is twofold. First of all, random errors are small at $\ell=80$. 
Second,  $C^{\rm 2h}_{\ell}$ peaks at $\ell \lesssim 100$ and then steadily declines and becomes sub-dominant with respect to the 1-halo term
 (see the relevant plots in \Xia\ and \citealt{Branchini:2016glc}). Considering $\ell \sim 80$ thus allows us to reasonably compare both the 1-halo and 2-halo terms.
{Note also that in the product $A^{\rm 2h}C^{\rm 2h}_{\ell=80}$ the second term is the {\it model} $C^{\rm 2h}_{\ell=80}$.
As a result, the errors in $A^{\rm 2h}C^{\rm 2h}_{\ell=80}$ are propagated from the $A^{\rm 2h}$ term only.}

When we consider different energy bins and explicitly account for the CAPS energy dependence, then we use
\begin{equation}
  \chi^2_{\rm e} \equiv  \chi^2_{\rm e}(z,c)=\sum_{\ell~\rm bins}\, \sum_{E~\rm bins} {\frac{(\hat{C}^{\rm \gamma c}_{\ell} (E_i) - C^{\rm \gamma c}_{\ell}(E_i))^2}{(\Delta C^{\rm \gamma c}_{\ell}(E_i))^2} } \, ,
\label{eq:chi22}
\end{equation}
where the sum is over both $\ell$ and energy bins, while
the pair $(z,c)$ identifies the redshift bin and object catalog considered in the analysis
and the label $e$ characterizes the model energy dependence of the CAPS,
i.e., $e$ = SPL, DPL or BPL.
In this case the number of fitting parameters varies depending on $e$, i.e.,
3 parameters for SPL, 4 for DPL and 5 for BPL.

To quantify the significance of a measurement we use as test-statistic
the quantity
\begin{equation}
\rm{TS} = \chi^2(0)-\chi_{min}^2 \; ,
\end{equation}
 where $\chi_{min}^2$ is the minimum $\chi^2$,
and $\chi^2(0)$  is the $\chi^2$  of the null hypothesis, i.e.\ of the case $C^{\rm 1h} = A^{\rm 2h} = 0$.
TS is expected to behave asymptotically as a $\chi^2$ distribution with a number of degrees of freedom
equal to the number of fitted parameters, allowing us to derive the significance level
of a measurement based on the measured TS.

Note that in Eq.~\eqref{eq:1h2h}  $\hat{C}^{\rm \gamma c}_{\ell}$, $C^{\rm 1h}$, and $A^{\rm 2h} C^{\rm 2h}_{\ell}$ all have units of (cm$^{-2}$s$^{-1}$sr$^{-1}$)sr,
since they refer to CAPS of $\gamma$-ray flux  maps integrated over the given energy bin.
Instead, in Eqs.~(\ref{eq:spl}-\ref{eq:bpl}) $C^{\rm 1h}$ and $A^{\rm 2h} C^{\rm 2h}_{\ell}$ have units of (cm$^{-2}$s$^{-1}$sr$^{-1}$GeV$^{-1}$)sr
so that  $\hat{C}^{\rm \gamma c}_{\ell}(E)$ still has units of (cm$^{-2}$s$^{-1}$sr$^{-1}$)sr.
The results obtained in the two implementations described above, i.e. for the single and combined energy bins are shown in 
Table~\ref{tab:T01} and in Table~\ref{tab:T02}, respectively.
Each sample in the Tables is identified by the following label: CCCC ZX EY, where CCCC indicates the catalogs of extragalactic objects used in the
cross-correlation (e.g. NVSS, 2MPZ etc.), ZX identifies the redshift bin (e.g. Z1 for the first $z$ bin, Z2 for the second .... and ZA for the full redshift range)
and EY identifies the energy bin (e.g. E1 for the first $E$ bin, E2 for the second etc.).
In Table~\ref{tab:T01} we list the best-fit values of the parameters and their 1 $\sigma$ errors, whereas only the best fit values are 
shown in Table~\ref{tab:T02}.  
To perform the fit we have assumed a frequentist approach. 
To derive the errors we build for each parameter its 1-d profile $\chi^2$ minimizing the $\chi^2$ with respect to the other parameters,
and calculate the 1 $\sigma$ errors  from the condition $\Delta \chi^2$=1.
In our analysis we assume that CAPS is a positive quantity. Therefore, in the fit we impose that both the 1-halo and 2-halo terms are non-negative. 
For this reason, when 1 $\sigma$ is not limited from below we just quote the 1 $\sigma$ upper limit.

In principle, cross-correlations can be negative.
However, in our model, the cross-correlation
between $\gamma$-ray sources and extragalactic objects  is induced by the fact that both trace the same large scale structure in 
some relatively compact redshift range. In this case the cross-correlation function is not expected to be negative,
motivating our constrain.
Nonetheless, for the sake of completeness, we did perform the same fit after relaxing this constraint.
We found that the 2-halo component can be negative when cross-correlating some catalogs with 
low ($<$ 1 GeV) energy URGB maps. However, the preference for this fit over the non-negative one is typically below 1 $\sigma$ and just in very few cases
slightly above 1 $\sigma$.

\renewcommand\tabcolsep{5.0pt}
\begin{table*}
\
\caption{Best fit to CAPS. Col. 1: subsample name.
Col. 2:  minimum $\chi^2$ value.
Cols. 3 and 4: values of the test statistic
$\rm{TS} = \chi^2(0)-\chi_{min}^2$ and corresponding statistical significance.
Cols. 5 and 6:  68\% C.L. constraints on the one-halo term $C_{\rm 1h}$ and on the
two-halo term  $A_{\rm 2h} \times C_{\rm \ell=80}$
both expressed in units of $10^{13} ($cm$^{-2}$s$^{-1}$sr$^{-1})\,$sr. 
{The fit in each row is performed using 12 data points and 2 fit parameters,
for a total of 10 degrees of freedom.}}
\label{tab:T01}
\begin{center}
\begin{tabular}{c|c|c|c|c|c}
\hline \hline
Sample & $\chi^2_{\rm min}$&TS& $\sigma$ & $C_{\rm 1h}$ & $A_{\rm 2h}C_{80}$  \\
\hline
\hline
NVSS ZA E1 &$20.3$ & $15.0$  & 3.5 & $47^{+12}_{-13}$ & $<15.7$  \\
NVSS ZA E2 & $32.7$ & $110$  & 10.3 &  $26.1^{+2.6}_{-2.7}$ & $<4.95$  \\
NVSS ZA E3 & $5.49$ & $64.4$ & 7.7 &  $0.94^{+0.12}_{-0.11}$&$<0.372$  \\
\hline
QSO ZA E1 & $5.53$ & $7.25$  & 2.2 &  $<29.9$ & $<23.8$ \\
QSO ZA E2 & $11.3$ & $12.0$  & 3.0 &  $5.7^{+1.7}_{-2.1}$ & $<5.18$ \\
QSO ZA E3 & $11.4$ & $12.3$ & 3.1 &   $<0.22$&$0.71^{+0.26}_{-0.288}$  \\
\hline
2MPZ Z1 E1 & $4.40$&$0.20$& 0.1 & $<90.7$ & $<59.8$  \\
2MPZ Z1 E2 & $7.97$&$4.27$& 1.6 &  $<24.9$ & $<24.8$  \\
2MPZ Z1 E3 & $15.5$&$4.04$& 1.5 &  $<0.780$ & $<1.79$  \\
\hline
2MPZ Z2 E1 & $ 8.64$ & $0.168$ & 0.1 &   $<62.7$ & $<49.7$   \\
2MPZ Z2 E2 & $6.35$ & $1.11$ & 0.6 &  $<12.5$ & $<21.7$   \\
2MPZ Z2 E3 & $9.33$ & $3.25$& 1.3 &  $<0.448$ & $<2.15$   \\
\hline
2MPZ Z3 E1 &$6.89$&$1.88$& 0.9 &  $<94.9$& $<47.5$   \\
2MPZ Z3 E2  & $2.44$&$15.4$& 3.5 &  $19.8^{+5.1}_{-7.0}$ & $<20.7$   \\
2MPZ Z3 E3  & $8.26$ & $17.1$ & 3.7 &  $0.71^{+0.21}_{-0.23}$ & $<2.15$   \\
\hline
2MPZ ZA E1  & $7.85$&$0.911$ & 0.5 &  $<59.8$ & $<37.7$   \\
2MPZ ZA E2  &$8.18$&$8.18$ & 2.4 &  $8.6^{+3.5}_{-4.3}$ & $<18.7$   \\
2MPZ ZA E3  & $12.3$&$13.3$ & 3.2 &  $0.31^{+0.11}_{-0.13}$ & $<1.63$   \\
\hline
WIxSC ZA E1 & $16.2$ & $22.0$ & 4.3 &  $32.8^{+7.3}_{-7.0}$ & $<5.95$   \\
WIxSC ZA E2 & $9.32$ & $26.5$ & 4.8 &  $4.1^{+1.5}_{-1.7}$ & $<11.4$   \\
WIxSC ZA E3 & $1.99$ & $35.3$ & 5.6 &  $0.098^{+0.040}_{-0.040}$ & $0.56^{+0.26}_{-0.27}$   \\
\hline
MG12 ZA E1  &$6.90$&$11.5$& 2.9 &  $21.7^{+9.8}_{-10.8}$ & $<31.4$   \\
MG12 ZA E2   &$7.69$&$26.9$& 4.8 &   $ 3.0^{+1.6}_{-1.5}$ & $6.8^{+3.4}_{-3.3}$    \\
MG12 ZA E3  &$8.73$&$23.5$& 4.5 &  $ 0.098^{+0.032}_{-0.034}$ & $<0.780$   \\
\hline
\hline
\end{tabular}
\end{center}
\end{table*}

\begin{figure*}
\centering \epsfig{file=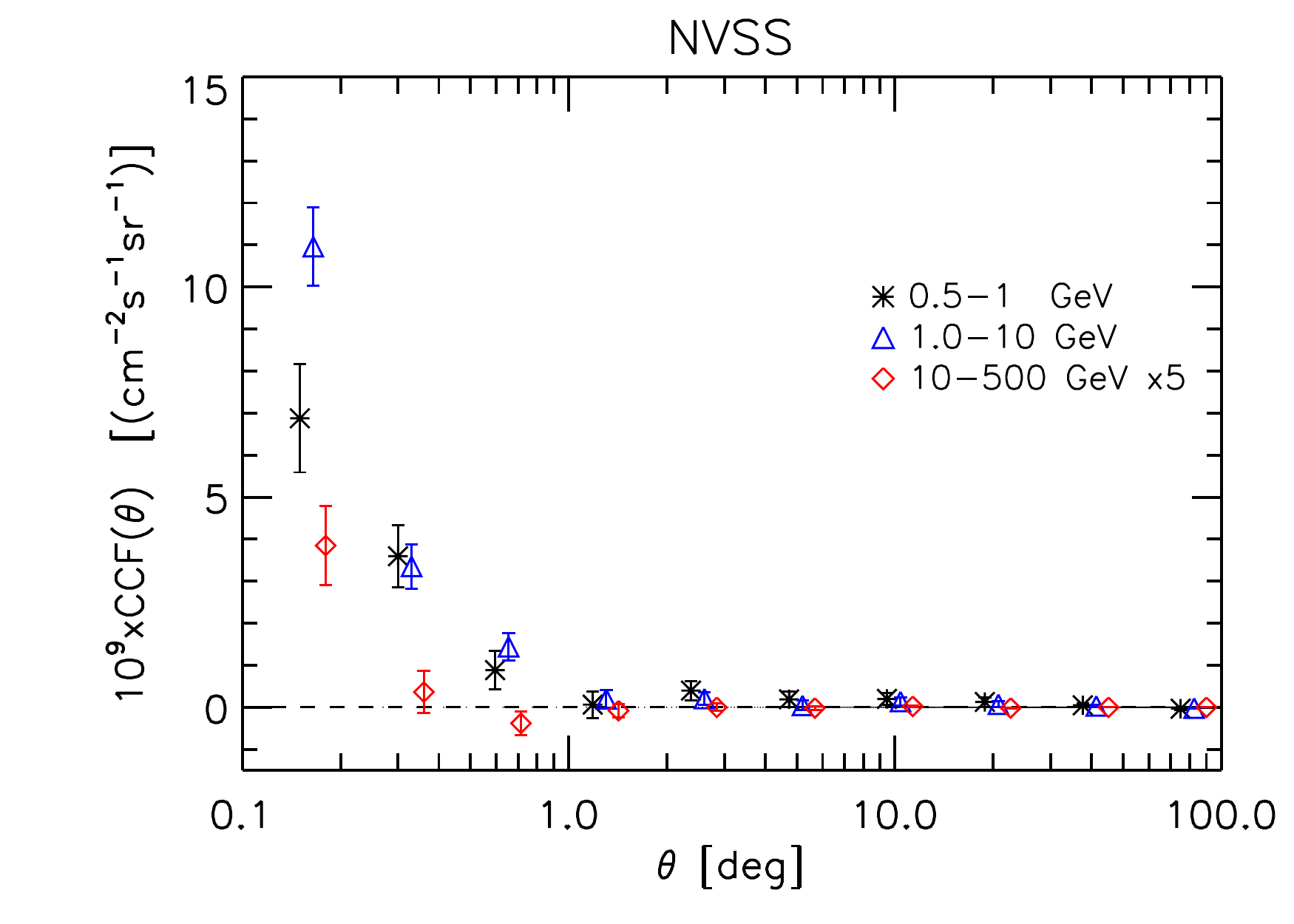, angle=0, width=0.45 \textwidth}
\centering \epsfig{file=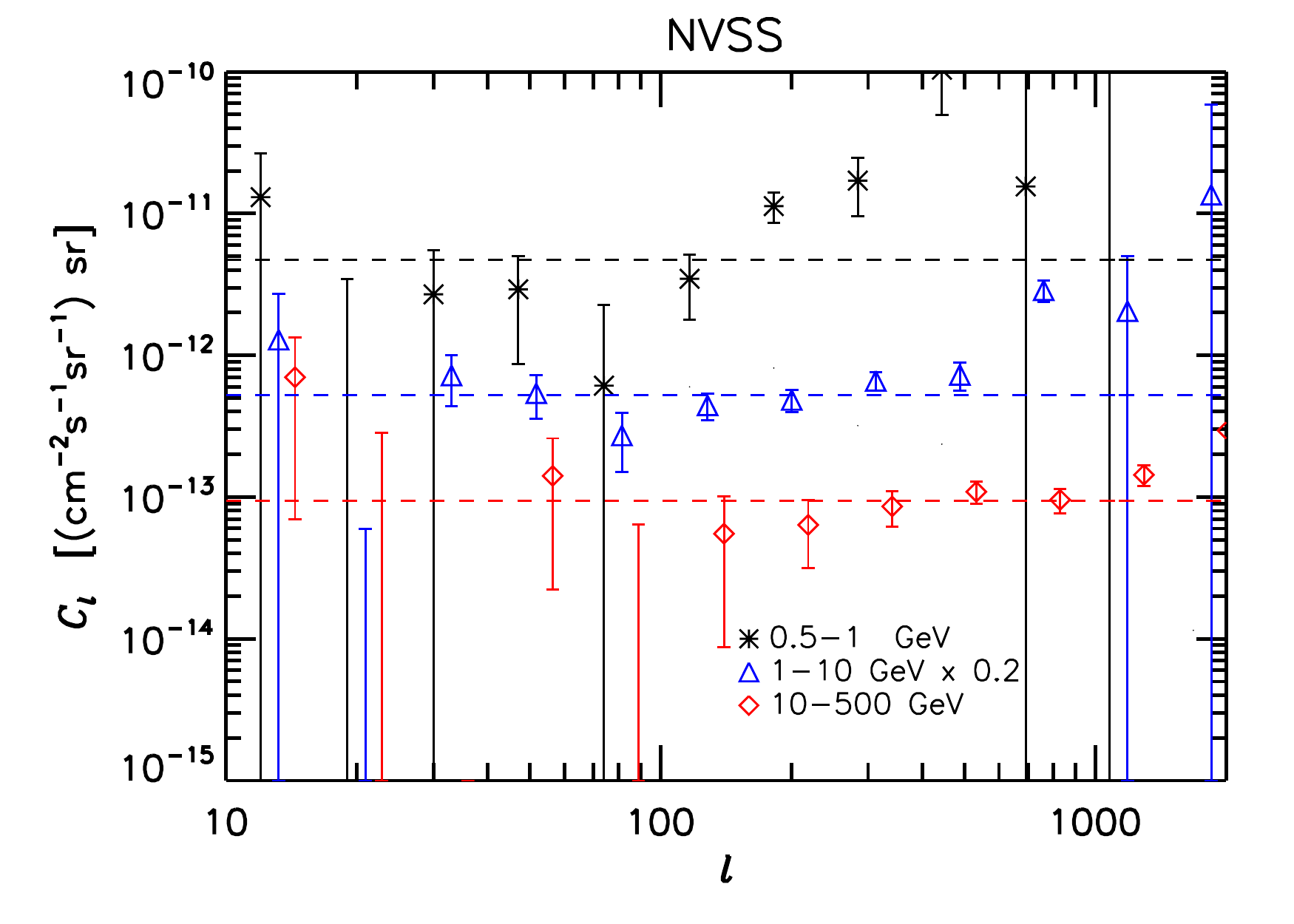, angle=0, width=0.45 \textwidth}
\caption{Angular CCF (left panel) and CAPS (right panel) for NVSS galaxies. Different symbols indicate the three
energy bins: $[0.5,1]$, $[1,10]$ and $[10,500]$ GeV. Error bars represent the square root of the diagonal elements
of the covariance matrix (which, for the CAPS, is to a good approximation diagonal). Furthermore, the CAPS
have been deconvolved by the
PSF and pixel effects.
Dashed lines in the right plots show the best-fit values of the 1-halo term $C_{\rm 1h}$ from Table~\ref{tab:T01}. }
\label{fig:nvss_ccf_fermi}
\end{figure*}

\begin{figure}
\includegraphics[angle=0,width=0.45\textwidth]{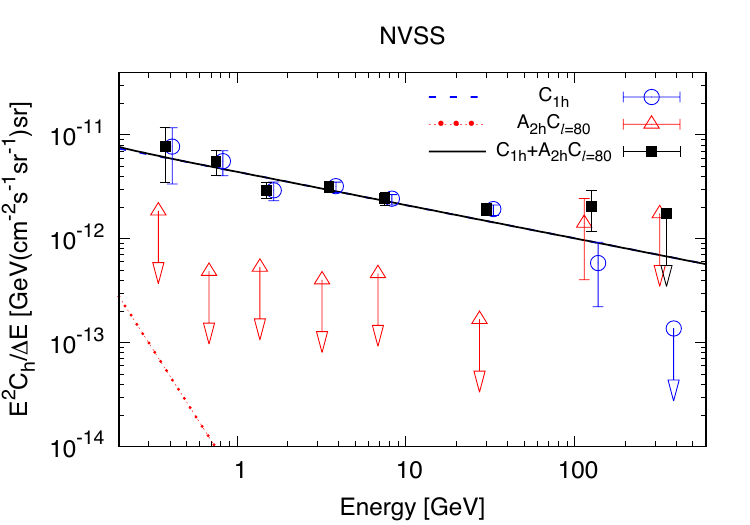}
\caption{Energy dependence of the $C_{\rm 1h}$ and $A_{\rm 2h} C_{80}$ terms and
of their sum.
The symbols represent the best-fit values in each energy bin. Bars represent 1-$\sigma$ errors.
In the case of upper limits, a downward arrow is shown.
The plot also shows the best-fit DPL model (black solid) as well as the 1-halo  (blue dashed)
and 2-halo (red dotted) components. Their numerical values are listed in
  Table~\ref{tab:T02}. 
 {Note that in this case the blue line is not visible, since it overlaps completely with the black one.} }
\label{fig:nvss_Espec}
\end{figure}

\renewcommand\tabcolsep{3.8pt}
\renewcommand{\arraystretch}{1.1}

\begin{table}
\caption{CAPS energy dependence.
Results of the best fit when the Double Power Law model is assumed. Col. 1: sub-sample considered.
Col 2:  minimum $\chi^2$ value (the $\chi^2$ is calculated as a sum over 8 energy bins and 12 multipole bins, i.e., 96 bins in total.
{The number of fitted parameters is 4, for a total of 92 degrees of freedom}).
Cols 3 and 4: values of the test statistic
$\rm{TS} = \chi^2(0)-\chi_{min}^2$ and corresponding statistical significance.
Col 5 and 6: best-fit slopes of the 1-halo and 2-halo power-law energy dependence.
Col 7 and 8: best-fit values of the one-halo term $C_{\rm 1h}$ and
 two-halo term  $A_{\rm 2h} \times C_{\rm \ell=80}$
 both expressed in units of of $10^{13} \times$  (cm$^{-2}$s$^{-1}$sr$^{-1}$GeV$^{-1}$)sr.
}
\label{tab:T02}
\begin{tabular}{c|c|c|c|c|c|c|c}
\hline \hline
Sample&$\chi^2_{\rm min}$&TS& $\sigma$ &$\alpha_{\rm 1h}$&$\beta_{\rm 2h}$ & $C_{\rm 1h}$ & $A_{\rm 2h} C_{80}$ \\
\hline
NVSS ZA &  $126.$ & $274.0$ &  16.1 &   $2.32$ & $4.54$ & $44.1$ & $0.0466$ \\
\hline
QSO6 Z1 & $108.$ & $18.0$ &  3.2 &    $3.59$ & $1.50$ & $17.0$ & $0.769$\\
QSO6 Z2 & $96.0$ & $5.14$ &  1.1 &    $3.30$ & $2.17$ & $5.72$ & $2.62$ \\
QSO6 Z3 & $94.3$ & $20.5$ &  3.5 &    $3.19$ & $2.05$ & $8.31$ & $11.8$ \\
\hline
WIxSC Z1& $93.7$ & $19.8$ &  3.5 &    $2.48$ & $1.58$ & $11.3$ & $0.0357$  \\
WIxSC Z2& $96.3$ & $25.7$ &  4.1 &    $2.39$ & $1.80$ & $5.39$ & $ 1.85$ \\
WIxSC Z3&  $71.4$ & $62.9$ &  7.2 &    $2.30$ & $1.87$ & $7.00$ & $ 3.60$ \\
WIxSC Z4& $82.2$ & $34.1$ &  5.0 &    $1.90$ & $2.67$ & $1.29$ & $ 23.6$ \\
\hline
2MPZ Z1 & $83.7$ & $8.1$ &  1.7 &    $2.53$ & $3.53$ & $21.7$ & $0.0328$ \\
2MPZ Z2 &  $61.7$ & $6.58$ &  1.4 &    $1.89$ & $2.51$ & $1.76$ & $ 8.45$ \\
2MPZ Z3 &  $69.6$ & $38.3$ &  5.3 &    $2.22$ & $1.77$ & $22.8$ & $1.35$ \\
\hline
MG12 Z1 & $56.4$ & $13.9$ &  2.7 &    $2.03$ & $1.91$ & $3.11$ & $ 2.05$\\
MG12 Z2 &  $82.1$ & $19.2$ & 3.4 &    $4.47$ & $2.02$ & $2.97$ & $ 12.3$ \\
MG12 Z3 &  $86.4$ & $46.7$ &  6.0 &    $2.23$ & $2.00$ & $3.86$ & $ 10.3$ \\
MG12 Z4 &  $69.4$ & $42.4$ &  5.7 &    $3.59$ & $1.95$ & $16.2$ & $ 8.23$ \\
MG12 Z5 &  $80.5$ & $41.4$ &  5.6 &    $3.79$ & $2.07$ & $14.1$ & $ 9.99$ \\
MG12 Z6 & $61.4$ & $27.1$ & 4.3 &    $2.36$ & $2.38$ & $6.22$ & $ 11.2$ \\
MG12 Z7 & $69.7$ & $12.4$ &  2.5 &    $2.28$ & $2.34$ & $6.64$ & $ 2.44$ \\
\hline\hline
\end{tabular}
\end{table}


\section{Results}
\label{sec:results}

In this section we show the results of our cross-correlation analysis of the cleaned {\it Fermi}-LAT UGRB maps with
the angular distributions of objects in the various catalogs presented in Section~\ref{sec:maps}.
As already mentioned, we shall plot the CCFs, whose visual
interpretation in the framework of the halo model is more transparent.
However, the statistical analyses and the results listed in the Tables are obtained from the measured CAPS, after deconvolution from
pixel and PSF effects.

For each catalog we show three  sets of results.
The first one includes the results of the CAPS $\chi^2$ analysis (Eq.~\ref{eq:chi21})
restricted to well defined, relatively wide
energy bins $E=[0.5,1]$ GeV, $E=[1,10]$ GeV and $E=[10,500]$ GeV.
The results of this analysis are listed in Tables~\ref{tab:T01} and ~\ref{tab:T1apdx}.
The first one contains  the results
of the plots that are shown in the main text.
This subset includes all analyses of the full sample  catalogs (ZA case) and, for the 2MPZ case only, also the
analyses of the individual redshift bins. The latter serves to illustrate the advantage of performing a tomographic approach
with respect to that of considering the full redshift range, as \Xia\ did using the whole 2MASS sample.
The second table, located in Appendix~\ref{apdx:moreplots},
contains all results from the subsamples considered in the analysis. The corresponding plots are also shown in the same Appendix.

The second set of results is similar to the first one but we consider eight narrow energy bins, instead of the three wide ones.
In this case, we do not quote results of the fit in a table, but display them in plots in which we show the best fit
1-halo and 2-halo terms as well as their sum, as a function of energy.
As a general remark we note that errors on the 1-halo and 2-halo terms measured in the narrow energy bins
are large, often resulting only in upper limits. This is due the fact that the two terms are typically not clearly separable given the large CAPS error bars.
For this reason, in the plots we shall show also the sum of the two,
which  is  more tightly constrained and thus has smaller errors.

The results of the CAPS energy-dependent fit are part of the third set of results. 
In this case we considered three models: SPL, DPL and
BPL (Eqs. \ref{eq:spl}--\ref{eq:bpl}). The statistical significance of the results is similar in the three cases,
thus, the SPL model  is satisfactory.
Nonetheless, since in a few cases the DPL  gives a slightly better fit  
{(in particular, for the MG12-Z4, Z5 cases, which both have  
$\Delta \chi^2 = \chi^2_{\rm SPL}- \chi^2_{\rm DPL}\approx 6$  corresponding to $\sim 2.4\, \sigma$
improvement)}, we decided to focus
mainly on this latter model, whose results are reported in Table~\ref{tab:T02},
while results for all the three models are listed in Appendix~\ref{apdx:moreplots}.
We will show in each plot the best-fit DPL model, together with the 1-halo and 2-halo terms and their errors
derived from the fit  in each narrow energy bin separately.
Note that, for better clarity of the plots,  we will show only the best fit model and we will omit 
the associated error band, which is typically quite large, 
especially for the 1-halo and 2-halo component singularly.
We will, in the following, use this best-fit model  to make some qualitative  comment 
on the preferred energy spectrum of the correlation, and its eventual evolution in redshift, or
differences between  the catalogs.

\begin{figure}
\centering \epsfig{file=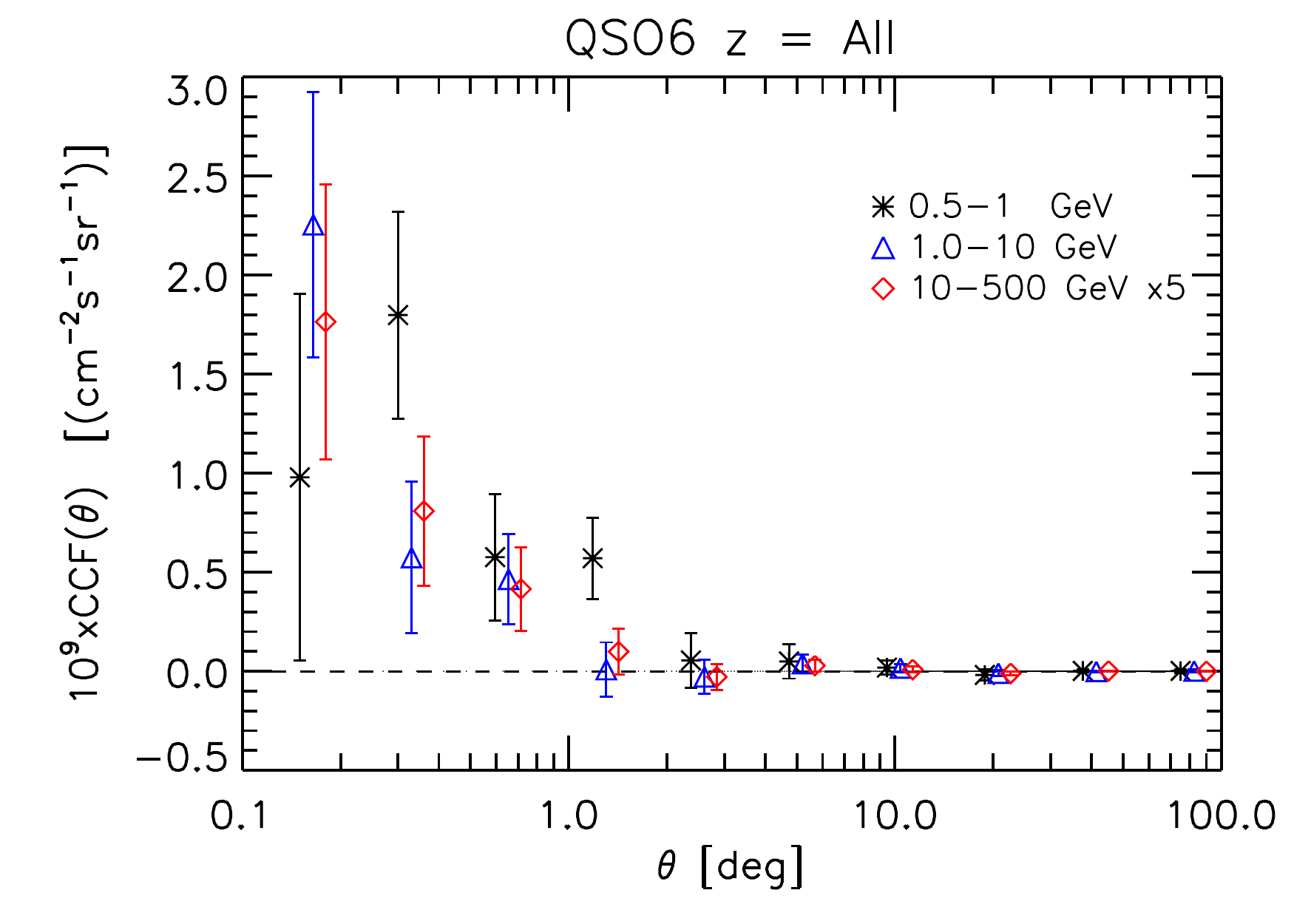, angle=0, width=0.45 \textwidth}
\caption{Same as the left panel of Fig.~\ref{fig:nvss_ccf_fermi} but for the cross-correlation
of the full SDSS DR6 QSO sample with \textit{Fermi}-LAT P8 data.}
\label{fig:qso_ccf_fermi}
\end{figure}

\begin{figure*}
\centering \epsfig{file=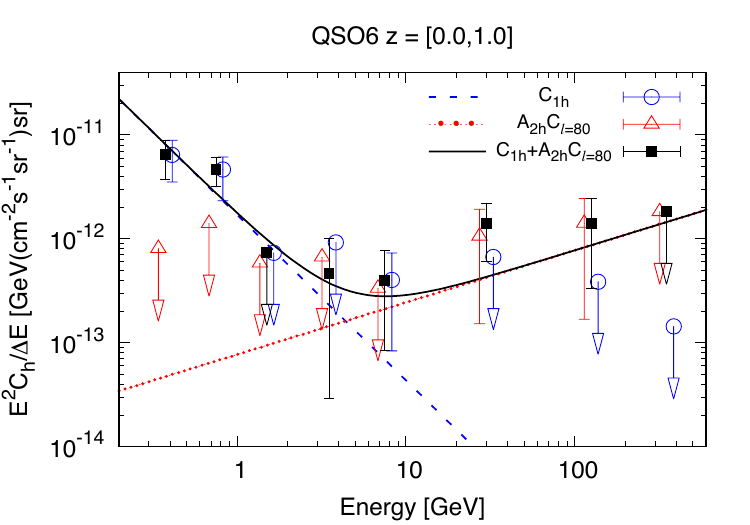, angle=0, width=0.32 \textwidth}
\centering \epsfig{file=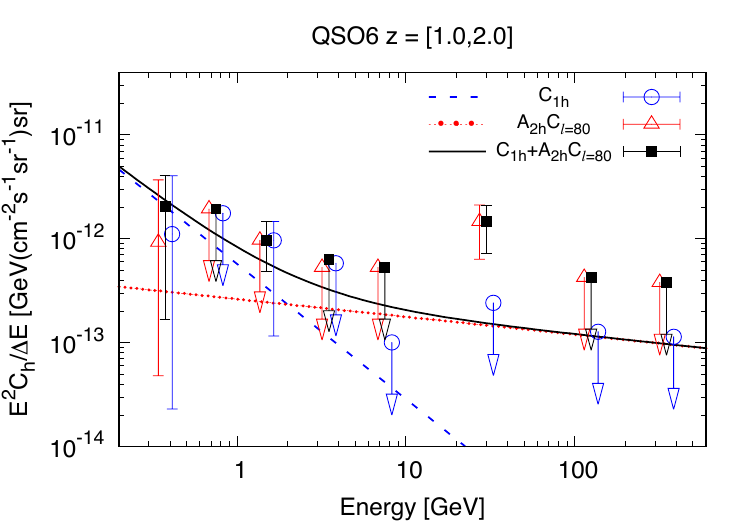, angle=0, width=0.32 \textwidth}
\centering \epsfig{file=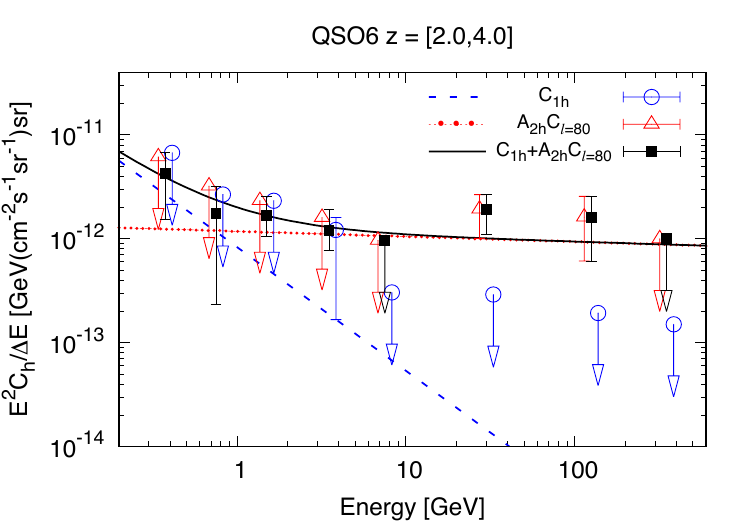, angle=0, width=0.32 \textwidth}
\caption{Same as  Fig.~\ref{fig:nvss_Espec} but for the
DR6-QSOs CAPS measured in three redshift  bins:  $z \in [0.0,1.0]$ (left),
$z \in [1.0,2.0]$ (middle), $z \in [2.0,4.0]$ (right). }
\label{fig:qso_Espec}
\end{figure*}

\begin{figure*}
\centering \epsfig{file=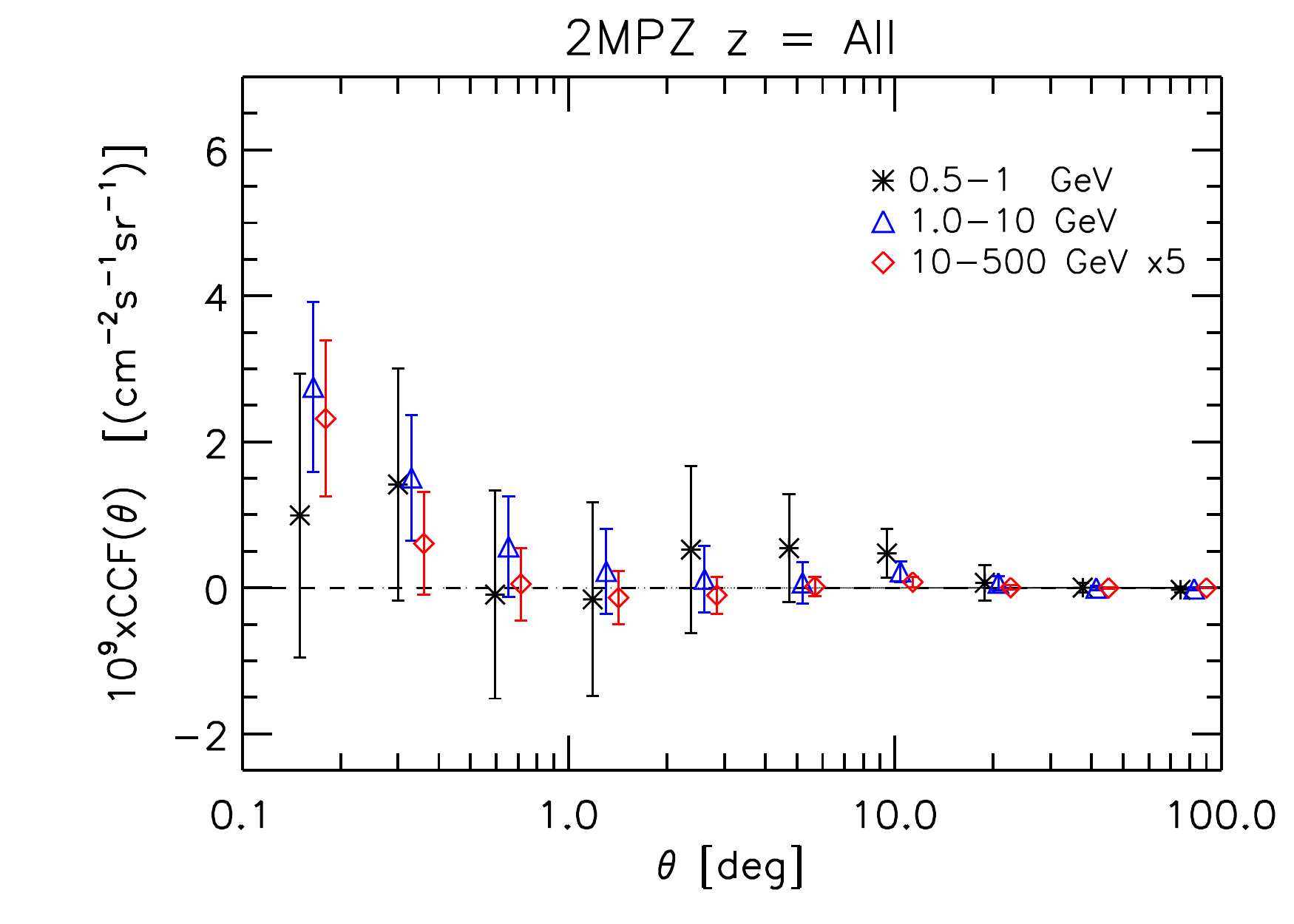, angle=0, width=0.45 \textwidth}
\centering \epsfig{file=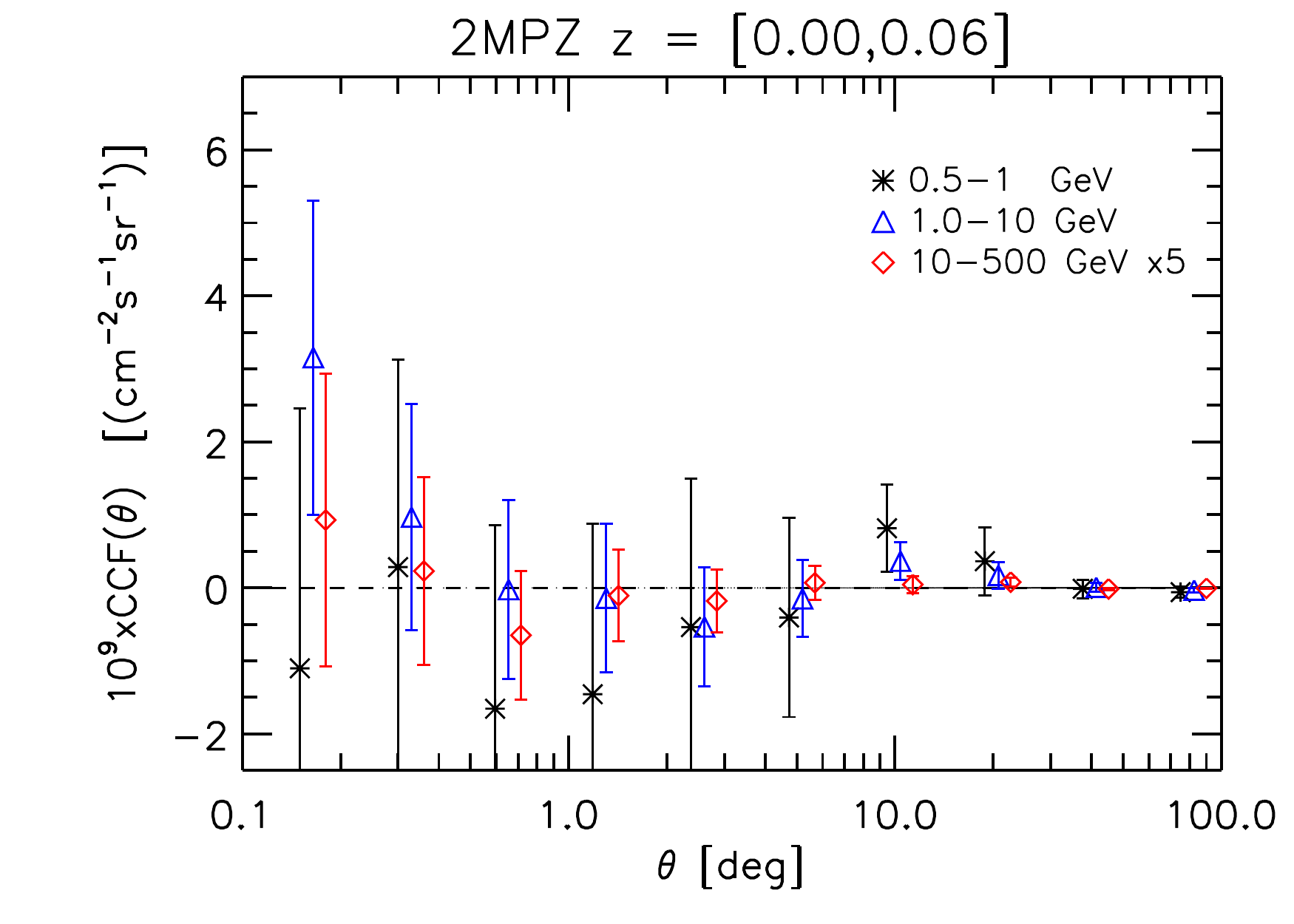, angle=0, width=0.45 \textwidth}
\centering \epsfig{file=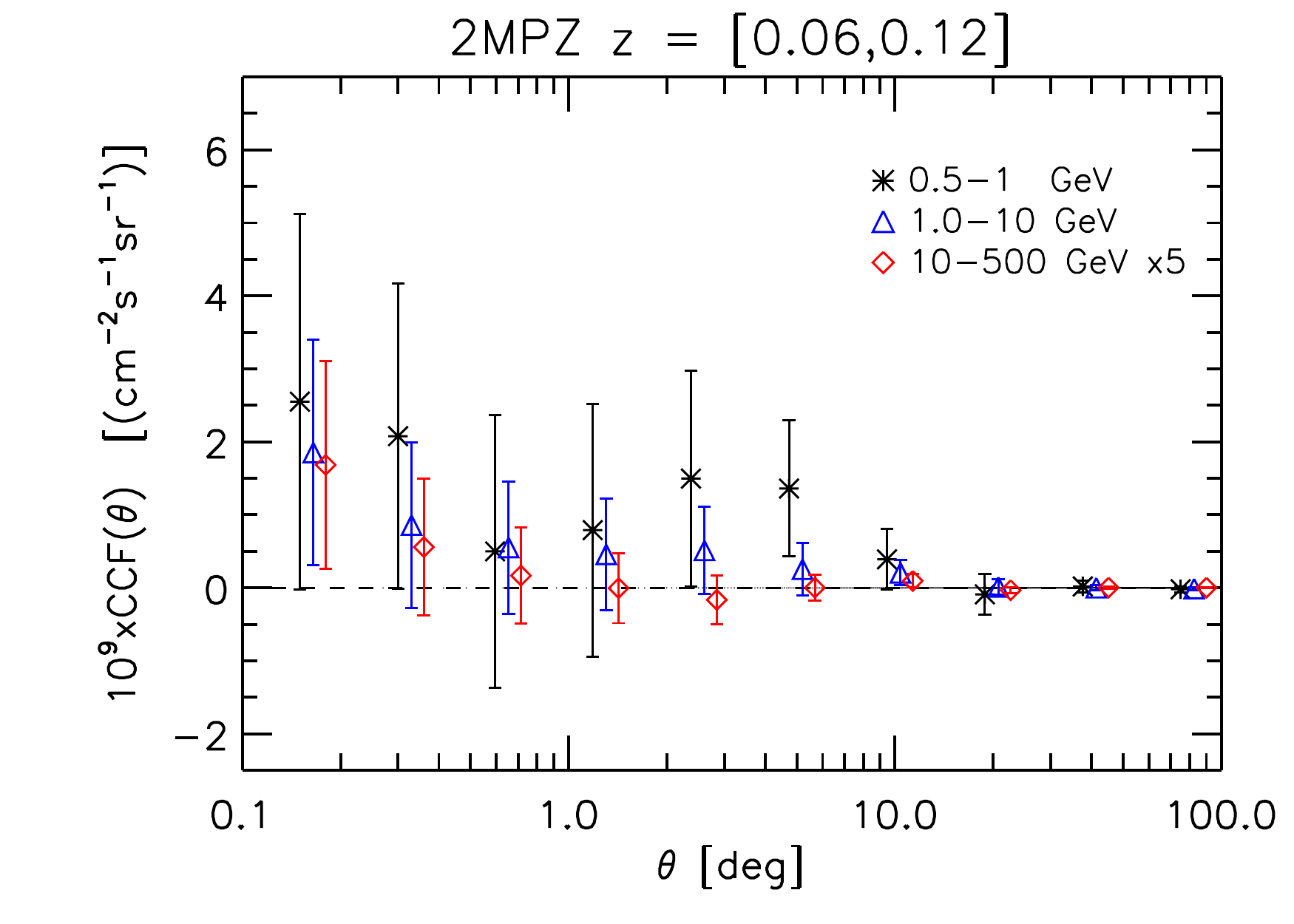, angle=0, width=0.45 \textwidth}
\centering \epsfig{file=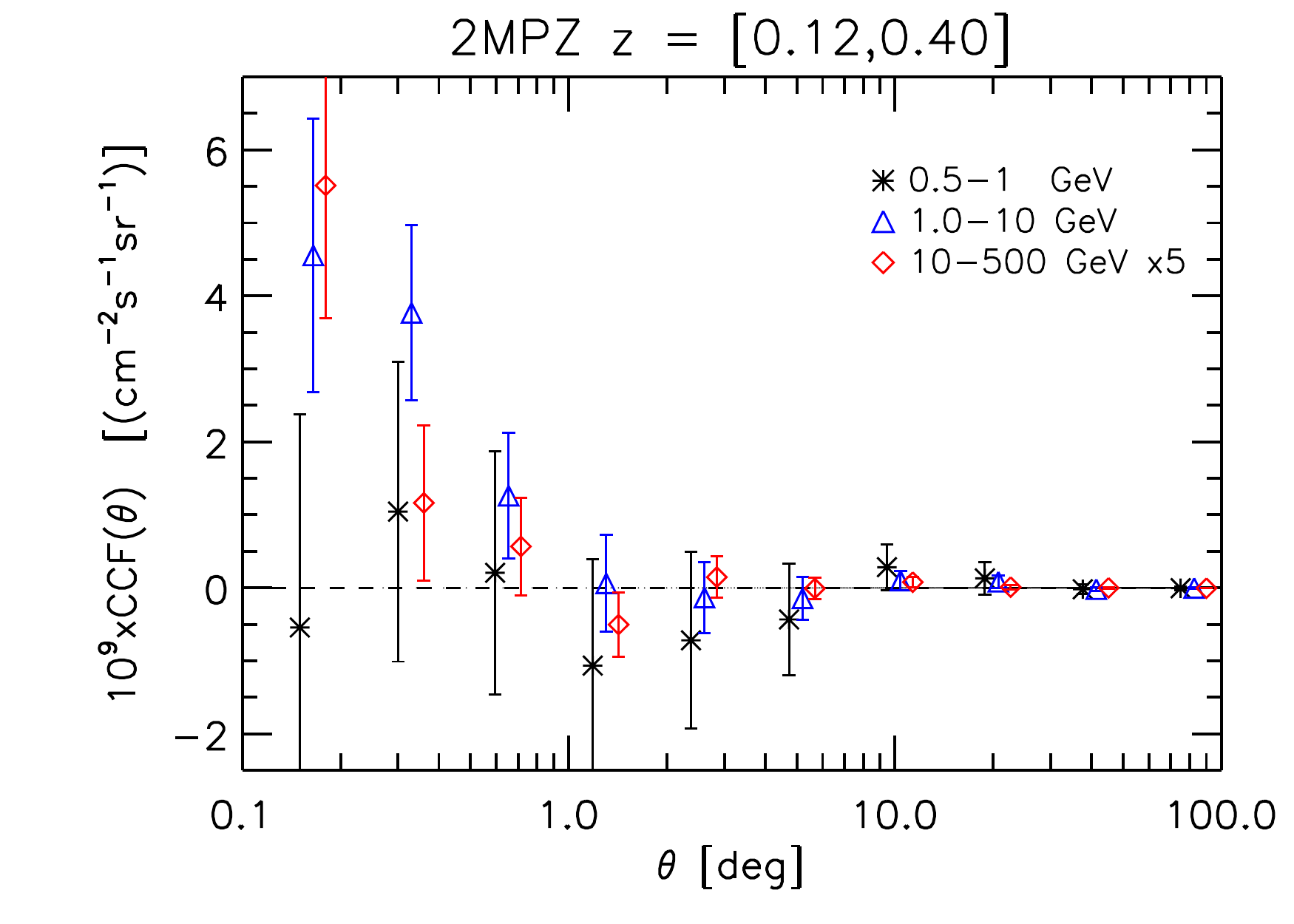, angle=0, width=0.45 \textwidth}
\caption{Same as the left panel of Fig.~\ref{fig:nvss_ccf_fermi} but for the cross-correlation
of the full 2MPZ sample with \textit{Fermi}-LAT P8 data, as well as for the three redshift slices adopted in the analysis.}
\label{fig:2mpz_ccf_fermi}
\end{figure*}

\subsection{Cross-correlation with NVSS galaxies}
\label{sec:ccfnvss}

The results of this analysis can be directly compared with those of \Xia\
to assess the improvement obtained by using the P8 LAT data.
In this case no tomographic analysis is performed here since
redshift measurements are not available for the majority of the NVSS objects.

The left panel of Fig.~\ref{fig:nvss_ccf_fermi}
shows the CCFs measured in three energy bins: $[0.5,1]$, $[1,10]$  and $[10, 500]$ GeV.
The corresponding CAPS are also shown in the right panel for reference.
A significant, positive correlation signal is detected  for $\theta<1^{\circ}$ at all  energies,
with a statistical significance of, respectively, 3.5, 10.3 and 7.7 $\sigma$ in the three energy bins.
The corresponding best-fitting 1- and 2-halo terms are  listed in Table~\ref{tab:T01}.
This result is similar to that of \Xia, indicating that, for the NVSS case, errors are
dominated by  systematic effects. In the lowest energy bin the significance has
decreased  (9.9 to 3.5 $\sigma$). This apparent inconsistency derives from the fact that
\Xia\ considered all photons with $E>0.5$ GeV, while we consider only those
with $0.5<E<1$ GeV.

As in \Xia, the CCF signal is quite localized. It is strongly dominated by the 1-halo term and the contribution of the 2-halo term is
negligible.
The $\chi^2$ analysis of the CAPS confirms this impression. Table~\ref{tab:T01} shows that
the cross-correlation signal is indeed dominated by the term $C_{\rm 1h}$, which is clearly detected in all energy bins, whereas for the
two-halo term, $A_{\rm 2h} C_{80}$, we  obtain only upper limits.
In the right-hand panel of Fig.~\ref{fig:nvss_ccf_fermi}, the best-fit values
of  $C_{\rm 1h}$ are shown together with the PSF-deconvolved CAPS.
The energy dependence of the best-fitting 1- and 2-halo terms in the eight narrow energy bins is presented in Fig.~\ref{fig:nvss_Espec}. The
1-halo term dominates over a large fraction of the energy range considered.
The contribution from the 2-halo term becomes  significant beyond 30 GeV and matches the 1-halo term
 at $\sim 100$ GeV.

Based on this evidence, we confirm the interpretation proposed  by \Xia: 
 the cross-correlation signal arises from NVSS objects also emitting in  $\gamma$-rays.
This is a sound argument  since radio galaxies  are often associated with $\gamma$-ray  emitters \citep{Acero:2015hja}.
However, this interpretation does not hold at very high energies.
At   $E\sim100$ GeV  the cross-correlation has a significant  2-halo component, and it is thus contributed by $\gamma$-ray sources residing in different
halos  than those of the nearest NVSS source.
From Tab.~\ref{tab:T02}, for  the DPL model the slope of the 1-halo term is  $\sim 2.3$,
while the 2-halo component is basically rejected by the fit, 
and in the plot is seen to give some contribution only at very low energies. 
{In particular, at $\sim$100 GeV the DPL fit predicts a
 2-halo term that is  several orders
of magnitude smaller than 
 the 2-halo datapoint inferred from a fit performed using
 eight narrow energy bins. 
This mismatch appears either because the DPL fit is dominated by the low energy data points,
where indeed the 1-halo term dominates, or
because a simple power law is not able to represent well the 2-halo component at $\sim 100$ GeV 
 without overpredicting the
amplitude of the 
the 2-halo term at lower energies.  }
The global significance of the NVSS signal in terms of the DPL (with 4 free  parameters) is $16.1\ \sigma$.
Adding more parameters using the BPL model does not improve the fit significantly (see Tab.~\ref{tab:T2apdx}).

\begin{figure*}
\centering \epsfig{file=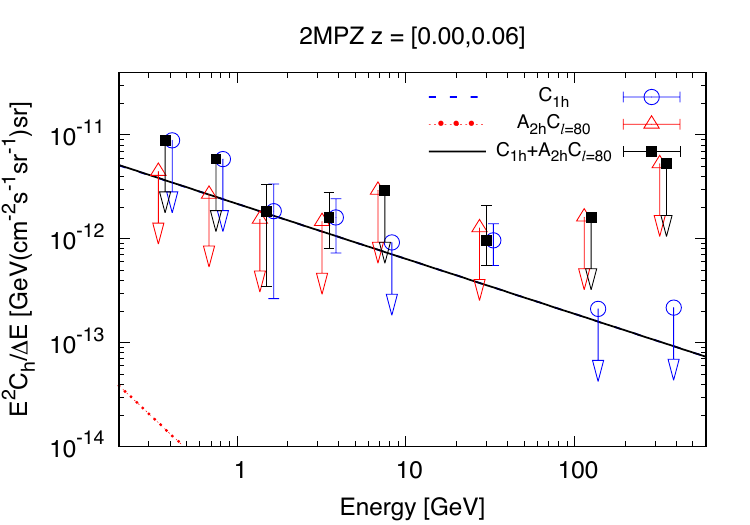, angle=0, width=0.32 \textwidth}
\centering \epsfig{file=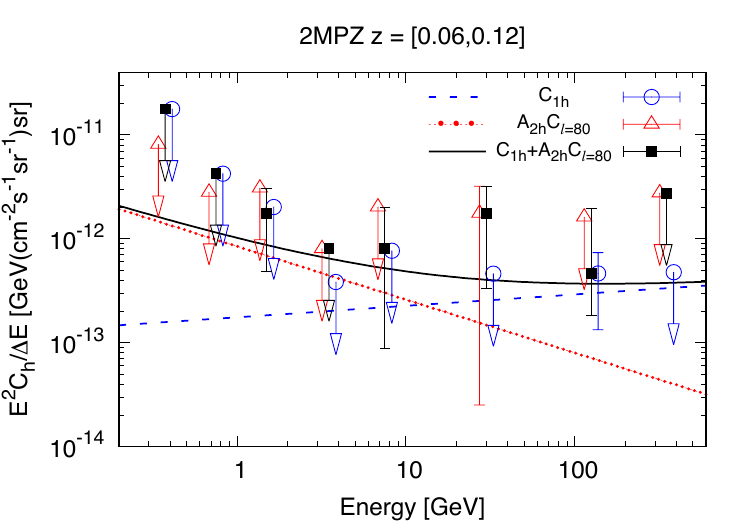, angle=0, width=0.32 \textwidth}
\centering \epsfig{file=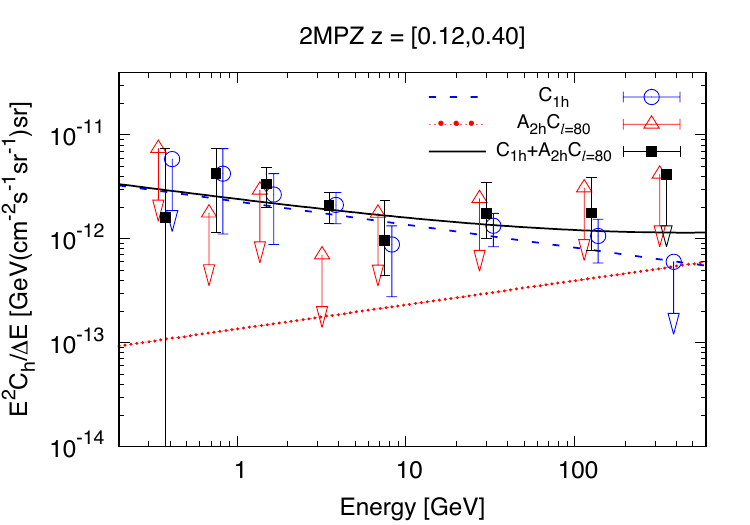, angle=0, width=0.32 \textwidth}
\caption{
Same as  Fig.~\ref{fig:nvss_Espec} but for the CAPS of 2MPZ galaxies measured in three redshift  bins:
$z \in [0,0.06]$ (left), $z \in [0.06,0.12]$ (middle), and $z \in [0.06,0.4]$ (right).}

\label{fig:2mpz_Espec}
\end{figure*}

\subsection{Cross-correlation with SDSS DR6 QSO}
\label{sec:ccfqso}

Fig.~\ref{fig:qso_ccf_fermi} is  analogous to Fig.~\ref{fig:nvss_ccf_fermi} and
shows the CCFs of P8 LAT data with the full SDSS DR6 QSOs sample, covering the whole redshift range $z\in[0,4]$,
in three energy bins.
The result is directly comparable with the one of \Xia\ where the same quasar sample was used.
A positive cross-correlation is detected out to  $\theta \sim 1^{\circ}$, with a significance of
$2.2\ \sigma$ in the low energy bin and $\sim 3\ \sigma$ in the two high energy ones
(see Table~\ref{tab:T01}).

The availability of photometric redshifts for this QSO sample allows us 
to decompose the signal tomographically which provides insight into the possible evolution of the $\gamma$-ray sources associated with the quasar distribution.
The results  are shown in Fig.~\ref{fig:qso_Espec}.
Contrary to the NVSS case, the 2-halo term is now prominent
except for, perhaps, at low energies and low redshifts.
The plots also show an evolution of the correlation signal as
function of redshift, suggesting that
the UGRB is contributed by different sources at different redshifts.
In particular, at $z < 1$  the CAPS energy spectrum has a two-component structure
with a steep 1-halo term below $E \simeq 10$ GeV and  a harder
 2-halo term above it.  Instead, at larger redshifts the 2-halo term is prominent at all energies with a flat spectrum with slope $\sim 2$.

\begin{figure}
\centering \epsfig{file=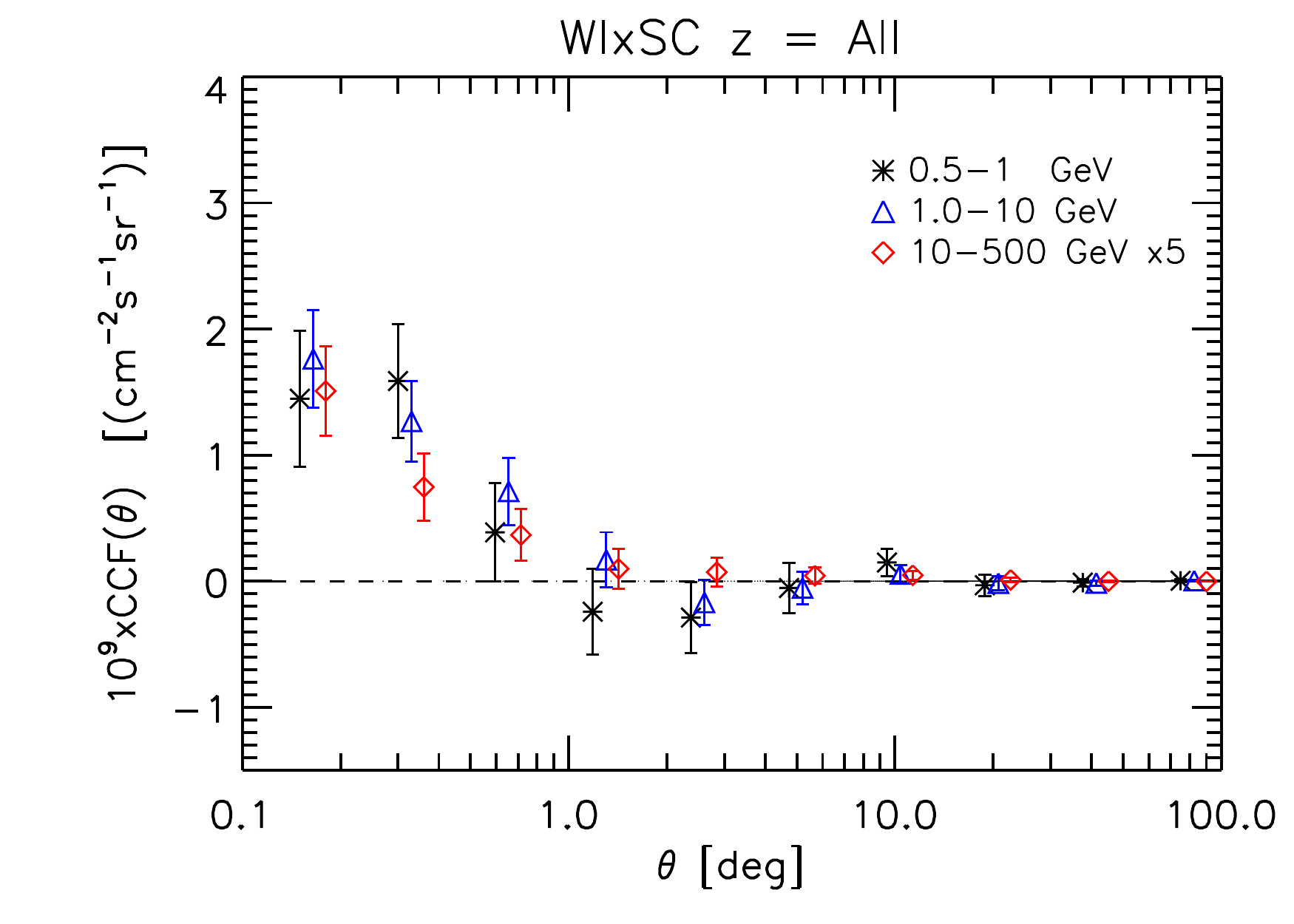, angle=0, width=0.5 \textwidth}
%
\caption{Same as the left panel of Fig.~\ref{fig:nvss_ccf_fermi} but for the cross-correlation
of the full \WISC\ sample with \textit{Fermi}-LAT P8 data.}
\label{fig:wixsc_ccf_fermi}
\end{figure}

\begin{figure*}
\centering \epsfig{file=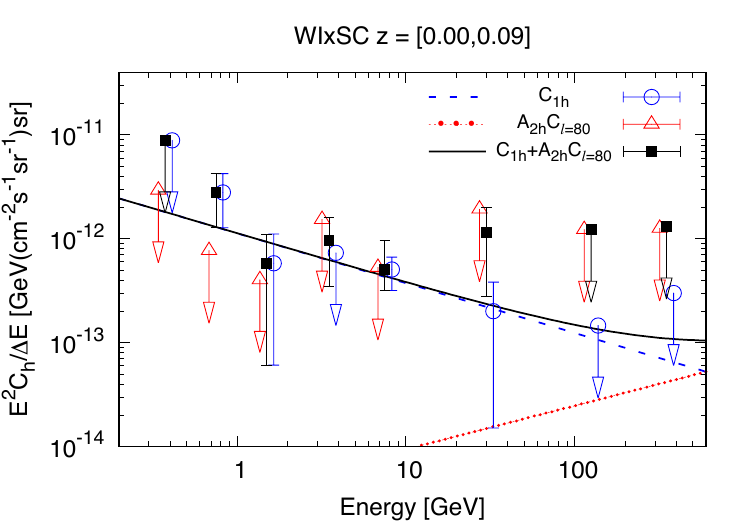, angle=0, width=0.45 \textwidth}
\centering \epsfig{file= 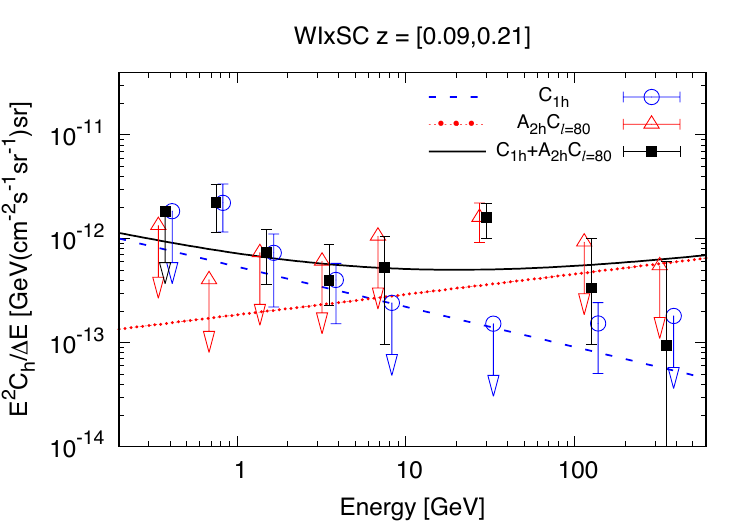, angle=0, width=0.45 \textwidth}
\centering \epsfig{file= 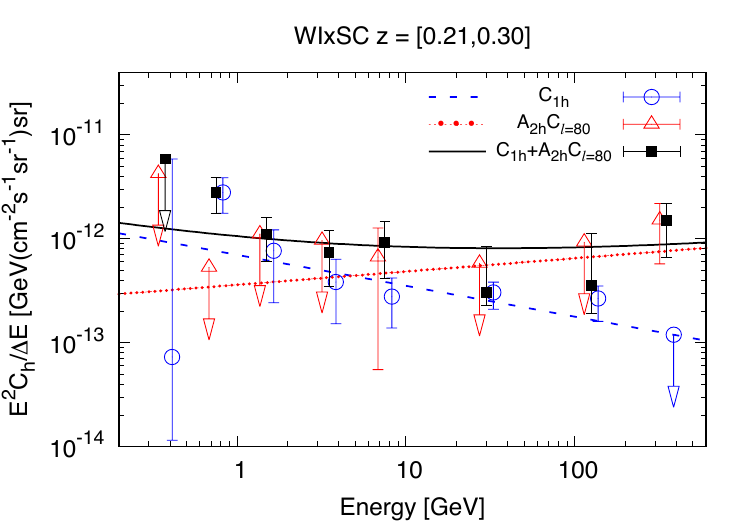, angle=0, width=0.45 \textwidth}
\centering \epsfig{file= 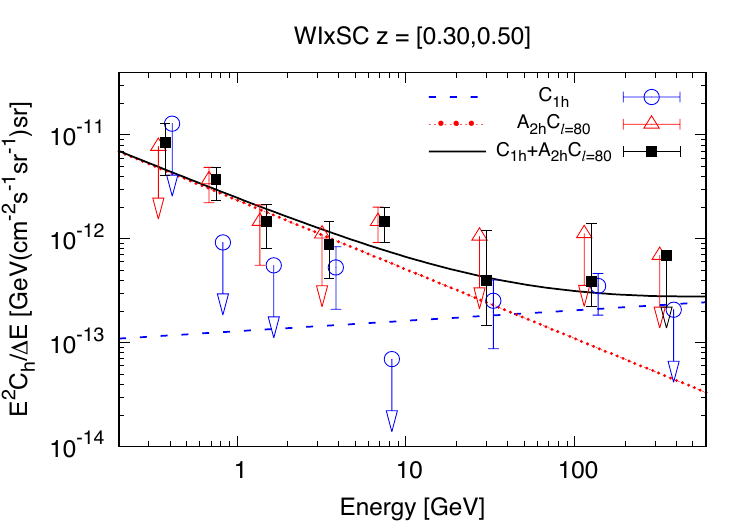, angle=0, width=0.45 \textwidth}
\caption{Same as  Fig.~\ref{fig:nvss_Espec} but for 
the CAPS of  \WISC\ galaxies measured in four redshift  bins:
$z \in [0.00,0.09]$ (top left),
$z \in [0.09,0.21]$ (top right), $z \in [0.21,0.30]$ (bottom left), $z \in [0.30,0.50]$ (bottom right).}
\label{fig:wixsc_Espec}
\end{figure*}

\subsection{Cross-correlation with 2MPZ galaxies}
\label{sec:ccf2mpz}

This catalog supersedes and largely overlaps with the 2MASS one used by \Xia.
The availability of photo-$z$'s for all
2MPZ objects allows us to slice up the sample  and carry out a
tomographic study in three independent redshift bins out to $z=0.4$
(although we note that there are practically no 2MPZ galaxies beyond $z \sim 0.3$, cf.\ Fig.~\ref{fig:dNdzs}).

The cross-correlation functions of 2MPZ galaxies and \textit{Fermi}-LAT P8 maps are shown in
Fig.~\ref{fig:2mpz_ccf_fermi}, for the full sample and for the three redshift shells
$z\in$ $[0,0.06], \, [0.06,0.12]$ and $[0.12,0.4]$.
Unlike the other catalogs, we show  in the main
text  the CCFs also for the redshifts bins, to discuss more in detail the comparison with the results of \Xia\ and to illustrate
the importance of performing tomographic studies.

For the full sample case (top left panel of Fig.~\ref{fig:2mpz_ccf_fermi}), the results are directly comparable with \Xia.
From Table~\ref{tab:T01} we see that  the statistical significance in the second energy bin is similar to the one found in   \Xia, while
for the third bin ($E>10$ GeV) the significance has increased noticeably thanks to the larger statistics.
Again, as for NVSS and SDSS QSOs, the significance in the first energy bin is smaller than
the one reported in \Xia, which is attributable to the different energy ranges of
the bins. This also means that the correlation seen in \Xia\ for the energy range $E>0.5$ GeV
had, apparently, a significant contribution from the $\gamma$-ray events with $E>1$ GeV.

Fig.~\ref{fig:2mpz_ccf_fermi}, Tab.~\ref{tab:T01}, and Tab.~\ref{tab:T02} all show
little or no correlation in the first two redshift bins of 2MPZ. The CCF signal
is instead largely generated in the third redshift bin, at $z>0.12$.
This is quite unexpected since in this redshift range we sample the tail of the 2MPZ distribution (see Fig.~\ref{fig:dNdzs}),
whereas  a large fraction of 2MPZ galaxies populate the second $z$-bin, where the peak is located.

This puzzling result suggests that the nature of 2MPZ objects changes  at these redshifts, which
is consistent with the fact that the bias of these sources also increases significantly
from $b\sim 1$ to $b\sim 2$ \citep{FP10,Steward14}.
This reflects, at least in part,
the flux-limited nature of the sample. 2MPZ galaxies at higher redshifts are intrinsically brighter and  trace the peaks of the
underlying density field which results in a larger auto-correlation signal and, thus, a larger $b$.

The result that $\gamma$-rays preferentially correlate with high-$z$ 2MPZ galaxies rather
than with the low-$z$ ones, illustrates explicitly  the added value of the tomographic approach.
It also shows that an analysis based on the
full sample, like in \Xia, can lead to partial, if not biased, conclusions.
The other advantage of the tomographic approach is that the above result can be cross-checked
using other catalogs and selecting objects in the same redshift interval.
We will, indeed, discuss this comparison in the next sections in relation to \WISC\ and SDSS DR12.

Comparing the CCF of the full 2MPZ  $z$-range (Fig.~\ref{fig:2mpz_ccf_fermi}) with the one of 2MASS from
 \Xia, a factor of $\sim 2$  mismatch in the normalization  is visible.
After cross-checks, we found the origin of this inconsistency. It was due to an error in the derivation of the
exposure map in each energy bin which led in \Xia\ to an  incorrect normalization of the
flux maps and thus of the derived CCF and CAPS.
The results of the present analysis thus supersede the ones in \Xia\ not only because of the better statistics and the tomographic approach,  but also due to the updated normalization.
We stress, nonetheless, that
the results obtained from the analysis of \Xia\ (e.g. \citealt{Cuoco:2015rfa} and \citealt{Regis:2015zka}) are generally valid except for the fact that the estimated quantities should be rescaled by a factor of $\sim 2$.

The plots in Fig.~\ref{fig:2mpz_Espec}  show the energy dependence of the correlation signal.
Again, it can be seen that  the signal is quite weak in the first two $z$ bins and stronger in the third one.
In this bin the signal is compatible with a flat energy  spectrum and shows
a preference for a 1-halo term, although a non-negligible 2-halo contribution is also present.

\begin{figure}
\centering \epsfig{file=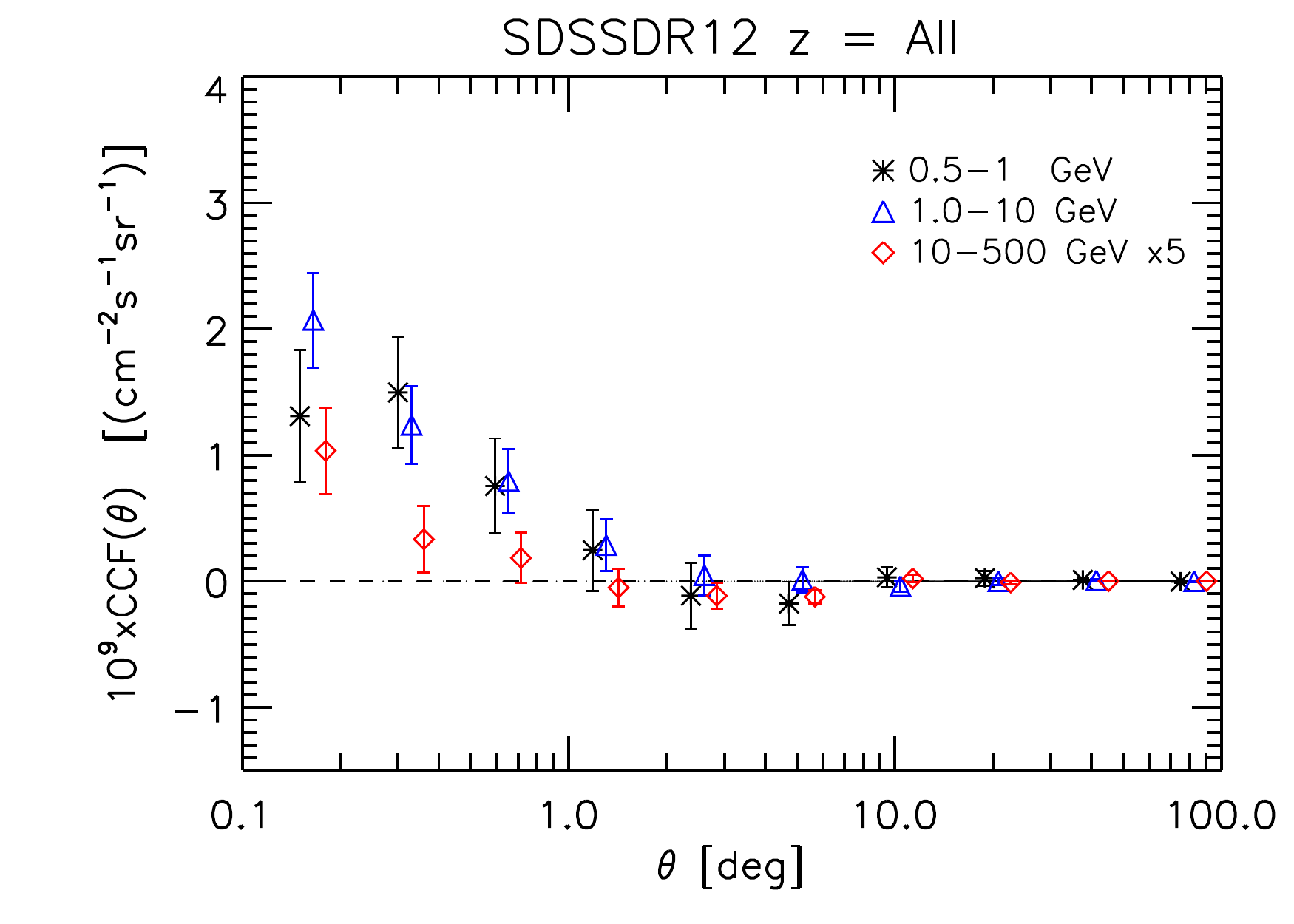, angle=0, width=0.45 \textwidth}
%
\caption{Same as the left panel of Fig.~\ref{fig:nvss_ccf_fermi} but for the cross-correlation
of the full  SDSS-DR12sample with \textit{Fermi}-LAT P8 data.}
\label{fig:sdss_ccf_fermi}
\end{figure}

\begin{figure*}
\centering \epsfig{file=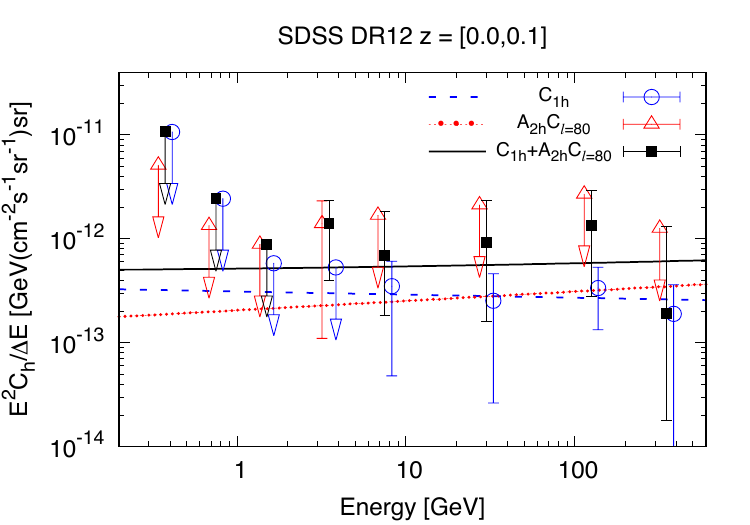, angle=0, width=0.32 \textwidth}
\centering \epsfig{file= 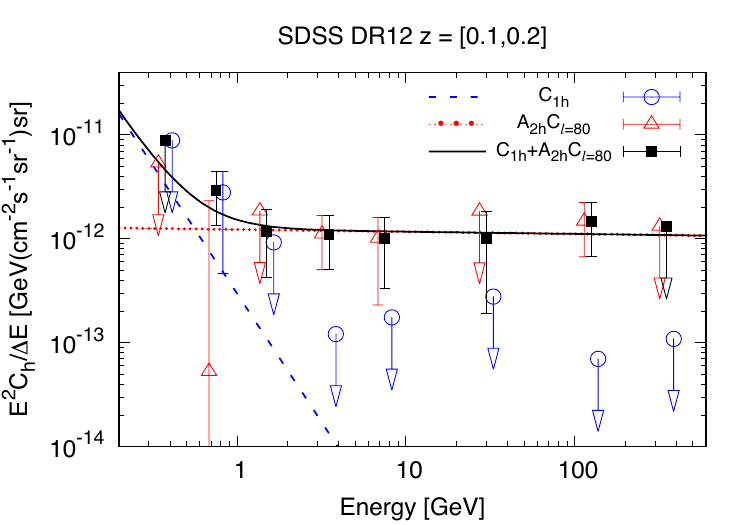, angle=0, width=0.32 \textwidth}
\centering \epsfig{file= 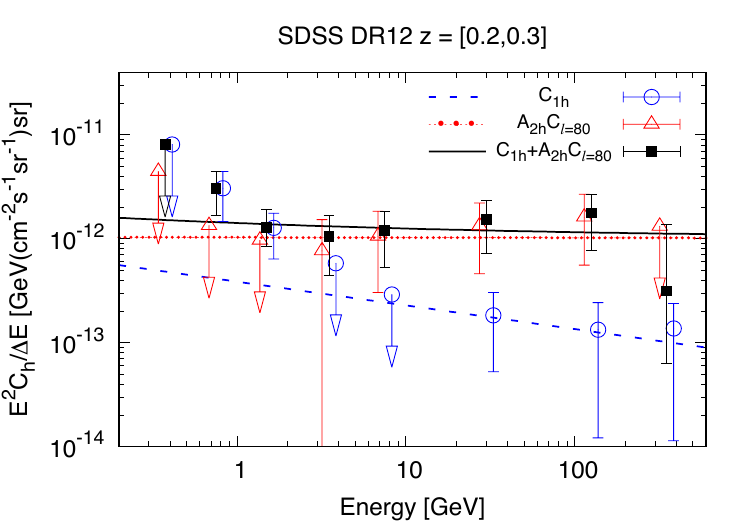, angle=0, width=0.32 \textwidth}
\centering \epsfig{file= 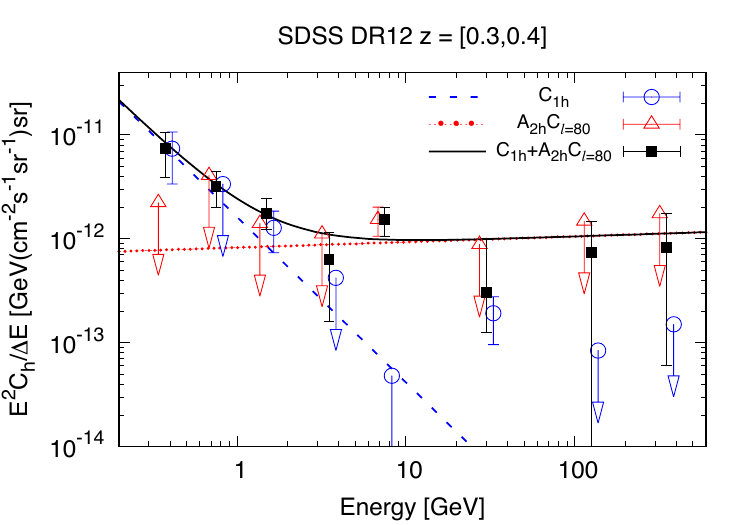, angle=0, width=0.32 \textwidth}
\centering \epsfig{file= 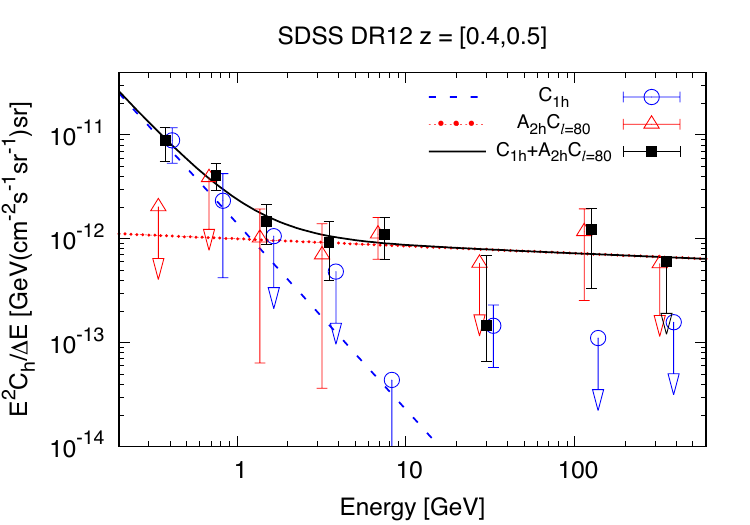, angle=0, width=0.32 \textwidth}
\centering \epsfig{file= 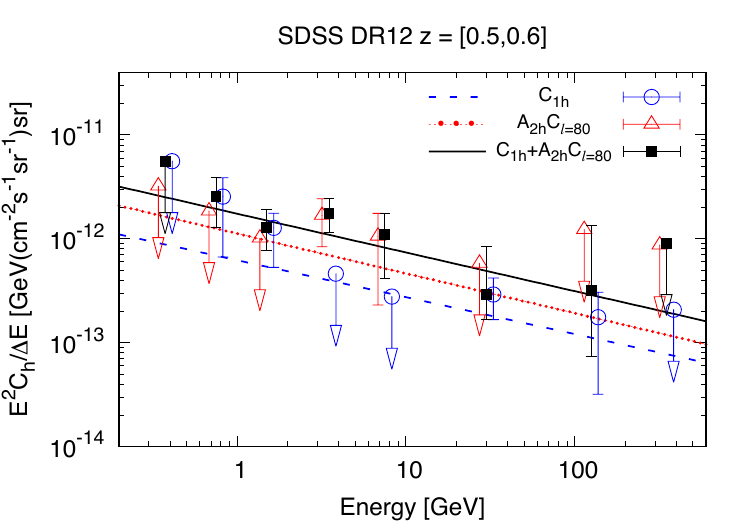, angle=0, width=0.32 \textwidth}
\centering \epsfig{file= 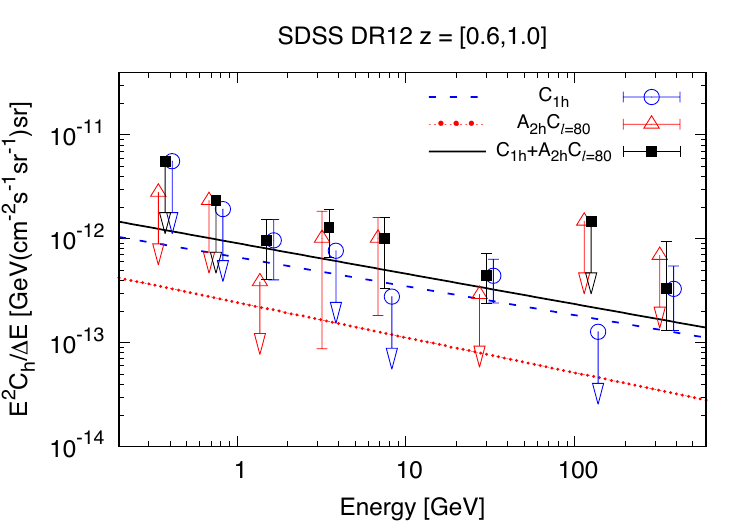, angle=0, width=0.32 \textwidth}
\caption{Same as  Fig.~\ref{fig:nvss_Espec} but for 
the CAPS of SDSS-DR12  galaxies measured in seven redshift  bins:
$z \in [0.0,0.1]$,
$z \in [0.1,0.2]$, $z \in [0.2,0.3]$, $z \in [0.3,0.4]$, $z \in [0.4,0.5]$, $z \in [0.5,0.6]$, $z \in [0.6,1.0]$.}

\label{fig:DR12_Espec}
\end{figure*}

\subsection{Cross-correlation with  WISE $\times$ SuperCOSMOS galaxies}
\label{sec:ccfwixsc}

The cross-correlation of the UGRB with  \WISC\ is performed here for the first time.
The \WISC\ catalog contains many more galaxies than the 2MPZ one, although its photometric redshifts are measured less precisely. However, thanks to the larger depth
of \WISC\, we are able to perform a similar, tomographic analysis using
four, thicker and not overlapping redshift slices.

As for all the other catalogs but the 2MPZ one, in the main text we only show the result  for the full $z$-range (Fig.~\ref{fig:wixsc_ccf_fermi}).
 The CCFs for the individual
redshift shells are shown in Appendix \ref{apdx:moreplots}.
The energy dependence of the correlation in the various redshift shells is shown
in  Fig.~\ref{fig:wixsc_Espec}.
A 1-halo component is favoured for $z<0.3$, although a 2-halo
contribution is allowed within the uncertainties,
except, perhaps at energies $<1$~GeV  and $z<0.09$.
In the range $z\in [0.3,0.5]$, instead, the 2-halo component is favored at all energies.
A redshift evolution of the energy spectrum is also evident.
The spectrum is close to flat for
$z\in [0.09,0.3]$ and much steeper with a prominent low energy tail for  $z\in [0.3,0.5]$.
This, again, confirms the importance of splitting the analysis into redshift shells.
The statistical  significance of the signal is above $3.7\ \sigma$ in all $z$ bins reaching
$7.2\ \sigma$ for $z\in [0.21,0.3]$ (Table~\ref{tab:T02}).
Combining the significances from all the $z$ bins gives a global significance for
the \WISC\ signal of $\sqrt{ \sum_i \sigma_i^2 } \approx 10.4\ \sigma$

\subsection{Cross-correlation with SDSS DR12 photometric galaxies}

\Xia\ cross-correlated  the SDSS DR8 datasets with 60-month \textit{Fermi}-LAT data.
Here we update that analysis using the \textit{Fermi}-LAT P8 maps and the SDSS DR12 photometric catalog sliced up into
seven redshift bins.

The CCF obtained by considering the catalog of all objects (reaching out to $z=1.0$)  is shown in Fig.~\ref{fig:sdss_ccf_fermi}.
A cross-correlation signal is detected within $1^{\circ}$ in all energy bands,
with a significance of  about 3.0, 4.7, $4.5\ \sigma$ respectively (see Table~\ref{tab:T01}),
which corresponds to a global significance of about $\sqrt{ \sum_i \sigma_i^2 } \approx 7\ \sigma$.
Much more information can be, however, extracted from the tomographic analysis.

The CCFs measured in the seven $z$ bins are shown
in Appendix \ref{apdx:moreplots} while their corresponding energy spectra are shown in Fig.~\ref{fig:DR12_Espec}.
The amplitude and the nature of the cross-correlation  signal varies significantly with redshift.
One remarkable feature is that at high energy ($E>10$ GeV)
the signal is quite local, with an amplitude that is the  largest at  $z\sim 0.3$ and negligible at higher redshifts.
A  second characteristic is the bimodal nature of the signal. The 2-halo component typically dominates above $\sim 5$ GeV at all redshifts
whereas the 1-halo term, characterized by a steeper spectrum, is more important  below $\sim 5$ GeV.
This suggests that SDSS galaxies trace  two different populations of $\gamma$-ray emitters.
The first one is made of  relatively low energy $\gamma$-ray sources, with steep spectrum (slope of $\sim 2.3$ or larger, from Table~\ref{tab:T02}),
that typically reside in the same DM halos as the SDSS galaxies.
The second population is composed of
high energy sources typically located in a different halo and with a flat (slope $\sim 2$, see again Table~\ref{tab:T02}) energy spectrum.

The energy spectrum also shows an interesting feature in the form of a bump at about $\sim 10$ GeV
in the redshift range $z \in [0.3, 0.4]$. Such feature is also seen in the
\WISC\ correlation at $z \in [0.3, 0.5]$.
The bump is seen in the 2-halo term only.
Moreover, the bump seems to be present, although less prominently,  also at $z \in [0.4, 0.5]$ and at
$z \in [0.5, 0.6]$, but at energies slightly below 10 GeV,
as could be expected from a cosmologically redshifted signal,
further suggesting that the bump may be a real feature instead of a statistical fluctuation.
If this is indeed the case, then it would be difficult to justify the bump using 
conventional astrophysical processes. The tantalizing hypothesis of 
an exotic process, like that of DM annihilation, could be then advocated.
We do not attempt here to quantify the statistical significance of this feature. 
We postpone its quantitative analysis and  interpretation to a future work in which 
the exotic sources will be included among more conventional $\gamma$-ray source populations.

Table~\ref{tab:T02} shows that the significance of the cross correlation
signal ranges from   $2.5\ \sigma$, in the highest $z$ bin,
to $6\ \sigma$, in the third $z$-bin.
The difference with the unbinned case is striking: the statistical significance of the CCF 
signal measured in the full $z$-bin is $7\ \sigma$, while the one obtained from the 
tomographic analysis is $\sim \sqrt{\sum_i \sigma_i^2} \approx 12\ \sigma$.
This comparison demonstrates further the {\it huge} gain in signal and information obtained by adopting 
the tomographic approach.

\begin{figure}
\centering \epsfig{file=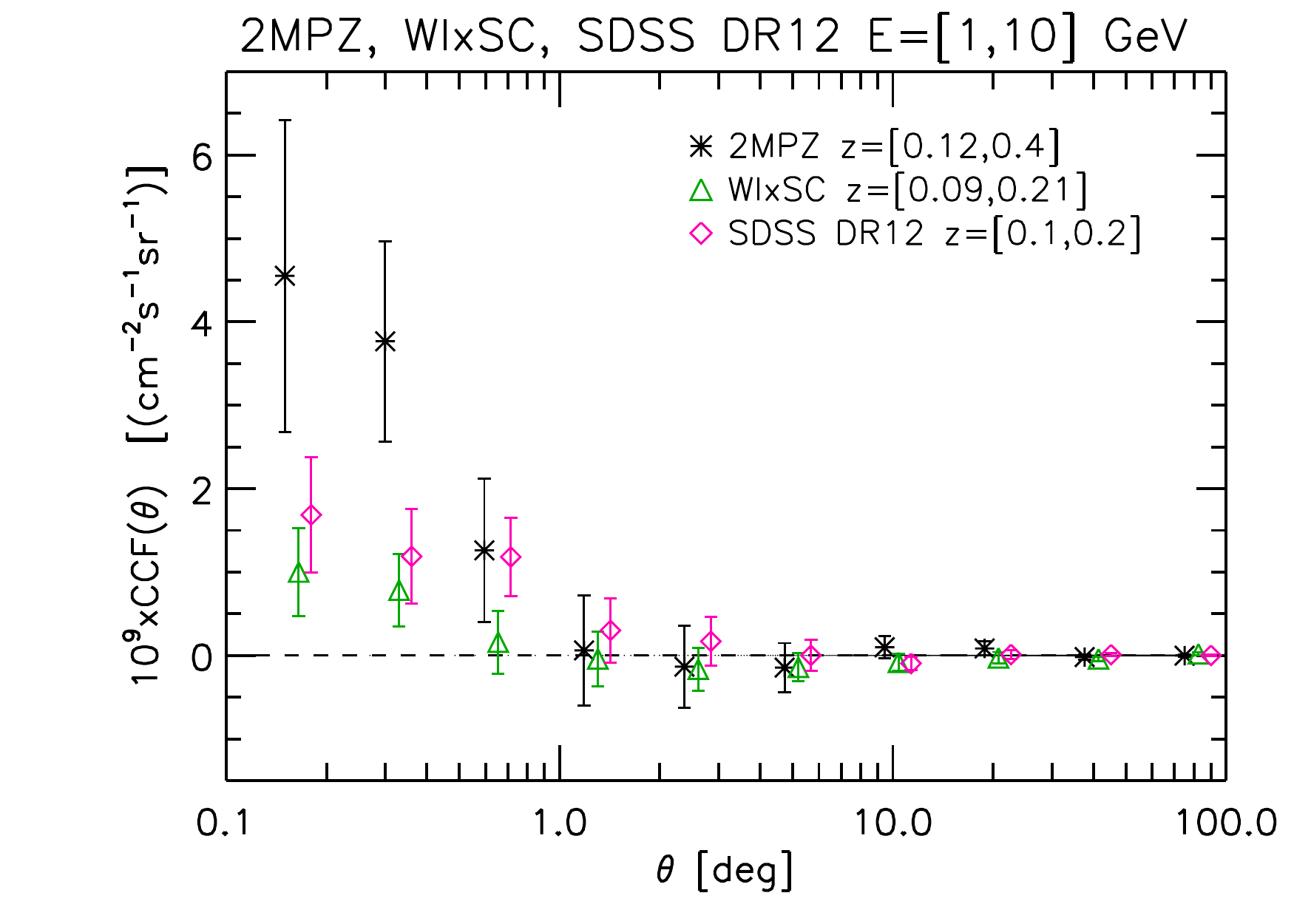, angle=0, width=0.45 \textwidth}
%
\caption{CCF of {\it Fermi}-LAT data in the energy range 1-10 GeV
with 2MPZ galaxies for $z\in [0.12,0.4]$, \WISC\  galaxies for $z\in [0.09,0.21]$
and SDSS DR12 galaxies for $z\in [0.1,0.2]$.
Note that due to the decreasing redshift tail of 2MPZ, the range $z\in [0.12,0.2]$
contains basically almost all the galaxies of $z\in [0.12,0.4]$.}
\label{fig:sdss_2mpz_wixsc}
\end{figure}

We conclude this section comparing the CCFs of 2MPZ \WISC\ and SDSS
in the range $z \in [0.12, 0.4]$.
In fact, given the fast decreasing number of 2MPZ galaxies for $z > 0.2$,
the vast majority of galaxies in this bin are  in the range $z \in [0.12, 0.2]$.
The most relevant comparison
is thus made with the \WISC\ and SDSS correlation in the range $z \in [0.1, 0.2]$.
This is shown in figure~\ref{fig:sdss_2mpz_wixsc}  for the energy bin $[1,10]$ GeV.
It can be seen that while the SDSS and \WISC\ cross-correlations are similar to each other, with
the SDSS one slightly larger, the 2MPZ one is quite different being higher by
a factor of $\sim 3$. This clearly suggests that  the population of 2MPZ galaxies in $z \in [0.12, 0.4]$
is quite different   from the one present in SDSS and \WISC\   in the same redshift range. The  high
normalization of the cross-correlation further suggests that   high-redshift 2MPZ sources have a very large
bias,  consistent with the one obtained from the
2MPZ auto-correlation analyses \citep{FP10,Steward14}.

\subsection{Redshift dependence of the cross-correlation signals}
\label{sec:UGRBz}

\begin{figure}
\centering \epsfig{file=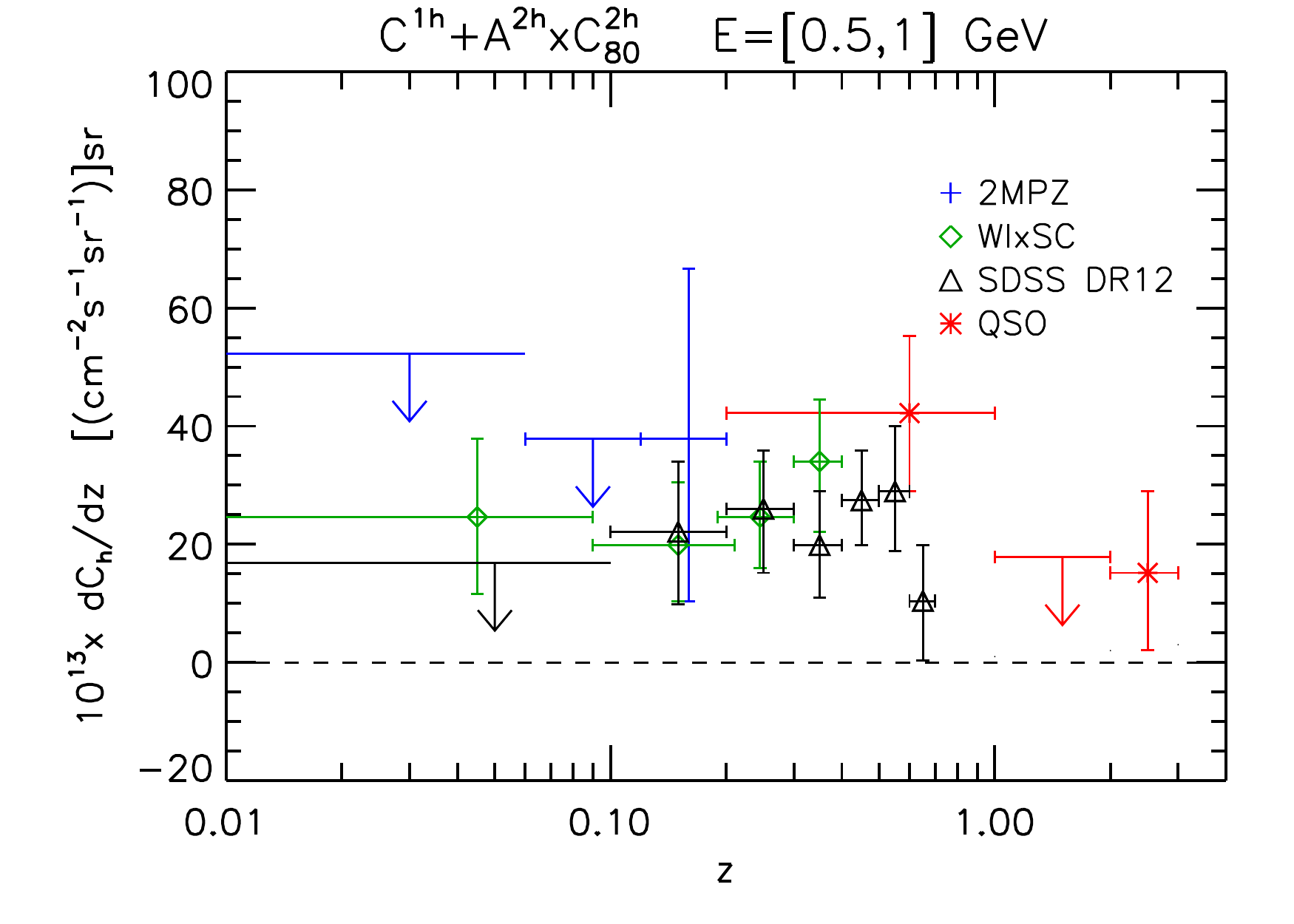, angle=0, width=0.45 \textwidth}
\centering \epsfig{file=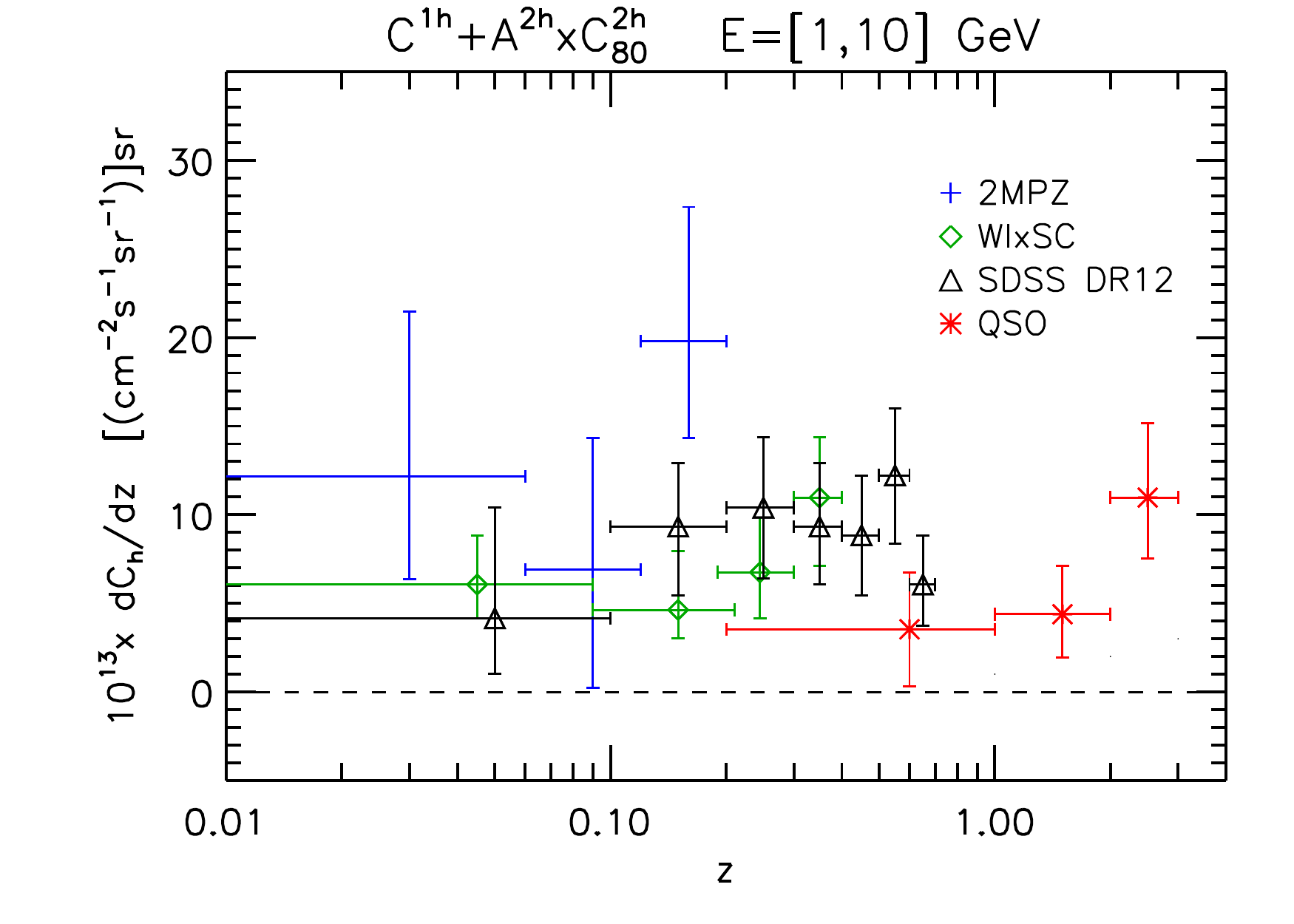, angle=0, width=0.45 \textwidth}
\centering \epsfig{file=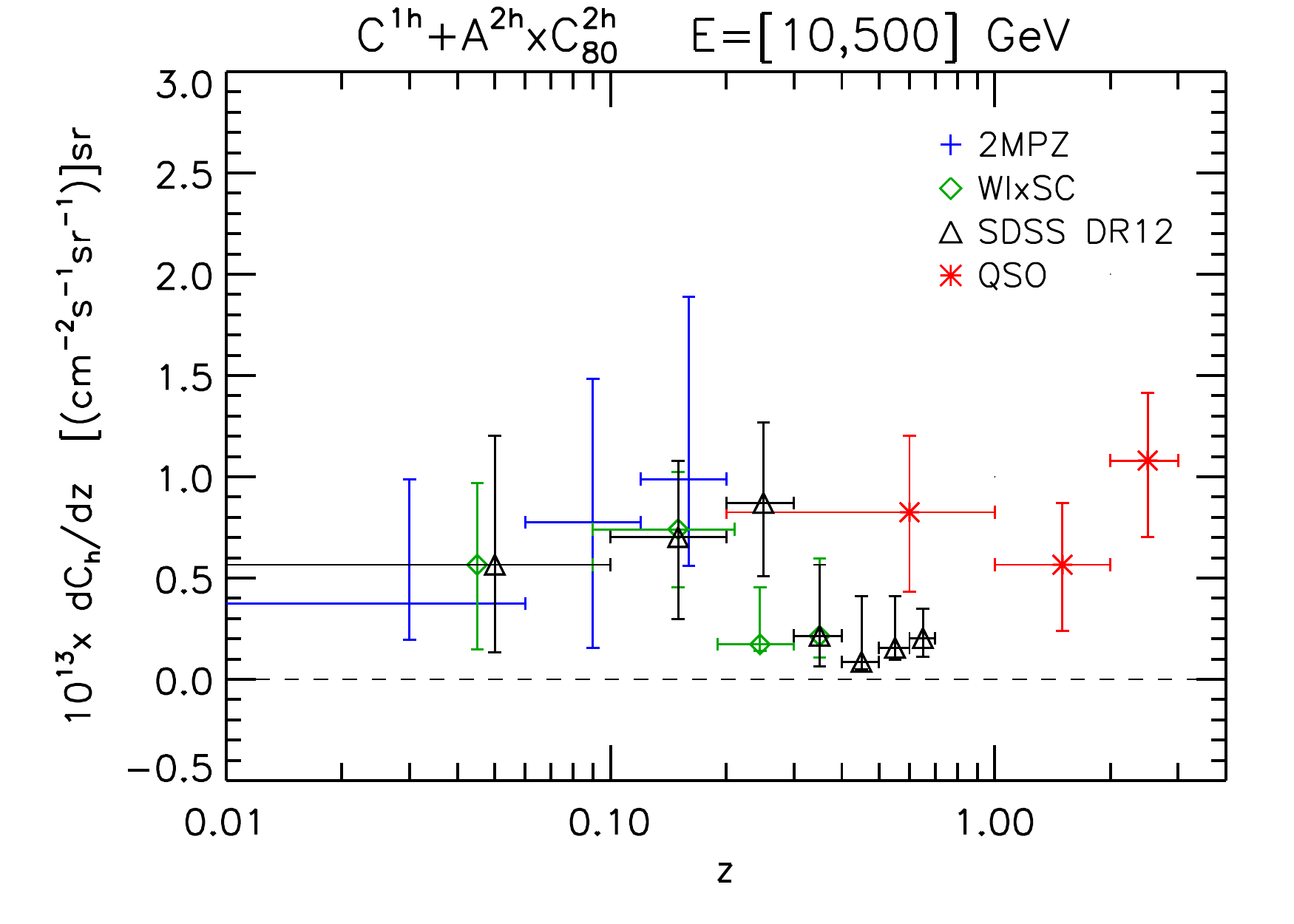, angle=0, width=0.45 \textwidth}

\caption{
Dependence of the fitted Fermi-LAT $\gamma$-ray data - catalogs cross-correlation signal,
$C^{\rm 1h} + A^{\rm 2h}C^{\rm 2h}_{\ell=80}$,  as a function of redshift, for 3 energy bins, as indicated in the plots.  }
\label{fig:UGRBz}
\end{figure}

Finally, we combine the information from all catalogs to investigate the redshift dependence of 
the cross-correlation signal. To this purpose,  we consider the sum $C_{\rm 1h}+A_{\rm 2h} \times C_{\rm \ell=80}$ 
measured in the three wide energy bins in all the catalogs and look for a dependence from $z$.
We did not consider the 1- and 2-halo terms individually since errors are too large for this analysis.
The results are summarized in the three panels of Fig.~\ref{fig:UGRBz}.
All types of sources have been considered here, except the NVSS ones for which we don't know the individual redshifts.
{ The data points represent effectively the correlation per unit redshift, and the plot can thus be seen
as the distribution in redshift of the correlation.}

The redshift distributions in the energy ranges 0.5-1 GeV and 1-10 GeV are quite similar. They both increase slowly from
$z=0$ to $z\sim 0.5$ and seem to drop at higher redshifts,
although the large errors in the QSO data points do not allow to draw a strong conclusion.
 At higher energy the behaviour  of the distribution is different:  the bulk 
of the correlation is generated at $z < 0.2$, while almost no correlation signal is detected 
at higher redshifts.
Again, the errors in the QSO data points are too large to derive firm conclusions, 
but, in this case, the above picture is supported
by the four high-$z$ SDSS data points, which have smaller errors.

These plots contain precious information on the sources that contribute to the UGRB. 
However, to infer the latter, one needs to make some hypothesis on the bias of the different objects.
We have assumed linear bias, and this allowed us to absorb it in  the normalization of the cross-correlation function.
However, different types of objects may have different bias factors. 
If all object considered had the same bias, then the plots would show the redshift distribution of the sources that generates the UGRB.
However, we do know that different types of sources are characterized by different bias factors.
For example the 2MPZ data point  at $z\sim 0.15$ in the energy range 1-10 GeV
\mbox{-- a clear outlier --} probably reflects the high bias of  bright 2MPZ galaxies at high redshifts.
{QSOs are also highly biased. Their large bias factor  ($b>2$ at high redshift), thus, significantly enhances 
the cross-correlation signal.}

A physically motivated cross-correlation model which includes hypothesis or independent constraints on the bias of the sources is therefore required to
interpret the intriguing results shown in Fig.~\ref{fig:UGRBz}. We postpone this task to a follow-up study.

\section{Discussion and Conclusions}
\label{sec:discussion}

In this work we have measured the angular cross-correlation between the cleaned P8 \textit{Fermi}-LAT maps of the
UGRB and different catalogs of extragalactic objects: NVSS, SDSS DR6 QSO, 2MPZ, WISE $\times$ SuperCOSMOS and SDSS-DR12 photometric.
These datasets have been selected using the following criteria: {\it i)}  large sky coverage to sample as many $\ell$-modes as possible;
{\it ii)} uniform preselection of objects across the relevant footprint; {\it iii)} wide span in redshift, from $z=0$ up to $z \sim 5$, with a significant spatial overlap between the samples.
The last requirement, also adopted by \Xia, has allowed those authors to perform a first, coarse-grained, tomographic analysis of the cross-correlation signal
which turned out to be a powerful tool to investigate the nature of the UGRB.
We took up from \Xia\ and improved the original analyses in several aspects:
\begin{itemize}

\item We used the Pass 8 \textit{Fermi}-LAT $\gamma$-ray data. Thanks to the improved photon statistics we were able to perform our cross-correlation study
in several (up to eight) non-overlapping energy bins.

\item Apart from NVSS, objects in the catalogs come with a redshift estimate, in the present analysis provided by \textit{photometric} redshifts.
Their error is much larger than of the spectroscopic ones but sufficiently small to enable us to slice up the catalogs in redshift bins, vastly improving
the tomographic aspect of the analysis.

\item We fixed a normalization issue that has affected the amplitude of the correlations measured  by \Xia.

\end{itemize}
Further, data files  of our cross correlation analysis  both in configuration and harmonic space are   publicly available
at \url{https://www-glast.stanford.edu/pub_data/}.

The combination of good energy resolution and the availability of photometric redshifts allowed us to explore the
energy and redshift dependence of the cross-correlation signal.
In our analysis we found that the UGRB is significantly correlated with the  spatial distribution of all types of mass tracers
that we have considered. The amplitude, angular scale and energy band in which the correlation is detected varies
with the type of objects  and their redshift.
A few general conclusions can be drawn:

\begin{itemize}

\item The CCF analysis of a   catalog  not divided in redshift bins, provides partial information on the
nature of the $\gamma$-ray sources. In fact, it may also lead to biased results in those cases in which the cross-correlation signal is generated
in different and   well-localized redshift bins.

\item The fact that in various cases a significant variation of the signal as a function of energy and redshift is observed,
strongly suggests that the UGRB is produced by different
types of sources, as  indicated also by  recent populations studies of resolved $\gamma$-ray sources \citep{Ajello:2015mfa,Fornasa:2015qua}.

\item When considering the same $z$-bin, different types of tracers produce different CCF signals. This is for example the case of the
CCFs of 2MPZ, \WISC\ and SDSS-DR12 in the range  $0.1 \lesssim z \lesssim 0.2$ and for $E \in [1,10]$ GeV.
These  dissimilarities reflect the differences  in   the relative bias between $\gamma$-ray sources and
galaxies   in the various catalogs, i.e., the fact that different types of galaxies are more or less effective tracers of the
unresolved  $\gamma$-ray sources.

\item The CCF signal is rather compact in size. It rarely extends beyond $\theta = 1^{\circ}$. In some cases it is even
more compact ($\theta \lesssim 0.4^{\circ}$ as in the NVSS case for $E>10$ GeV).  
To analyze quantitatively the information encoded in the CCF   as a function of
energy and redshift, we have compared our measurements with the predictions of a simple model, inspired by the halo model,
in which the cross-correlation signal is contributed by a compact 1-halo term and a more extended 2-halo term. Both the modeling
and the analysis were performed in harmonic rather than configuration space to minimize error covariance.
The use of this simple, yet physically motivated, general-purpose model, allows us to properly quantify the significance of the
CCF signal which, in several cases, can be quite large (i.e. $> 5 \sigma$, see Tables~\ref{tab:T01} and \ref{tab:T1apdx}).
The 1-halo term often dominates over the 2-halo one, hence justifying the compactness of the CCF. However, a   2-halo
term is clearly detected in several energy and redshift ranges and, in some cases, is more prominent than the 1-halo one.
This  diversity provides   further evidence in favor of the multi-source hypothesis for the UGRB.

\end{itemize}

We postpone a detailed study of these results to a follow-up analysis in which the wealth of information produced in this work
will be compared with more realistic UGRB models contributed by known (blazars, star forming galaxies, misaligned AGN)
as well as hypothetical (annihilating or decaying DM particles) $\gamma$-ray sources.
However, even  our simple model  can extract some additional information by exploring in more detail the energy
dependence of the cross-correlation signal.  Thanks to the exquisite photon statistics and energy resolution,
we  were able to compute  the cross-correlation in eight energy bins and  to compare
the results with our model in which
we allowed for an explicit energy dependence of the 1-halo and 2-halo terms. We modeled the energy dependence
in three different ways: A single, a double and a broken power law.   We found that 

\begin{itemize}

\item The SPL, DPL and BPL models typically provide similarly good fits.
Nonetheless, various cases show some hint of preference for the DPL model, i.e.,
a different slope for the 1-halo and 2-halo energy spectra.

\item More often than not the energy spectrum of the
2-halo term is harder than that of the 1-halo term. However, some counter examples are also seen.
This further suggests the presence of different populations of $\gamma$-ray sources characterized by
different spatial distributions and spectral properties.

\item An intriguing bump is seen at  $E\sim 10$ GeV and $z\in [0.3,0.5]$  in both
SDSS-DR12 and \WISC.  The bump is visible in the 2-halo term only.
Although we did not attempt to quantify the significance of this feature,
we note that an interpretation  in the framework of a UGRB generated by conventional astrophysical sources would be rather challenging,
while a bump in the energy spectrum in the 2-halo term  would have a natural  explanation in terms of DM annihilation.

\item Combining the information from all the catalogs, we have been able to
investigate the redshift distribution of the cross correlation signal as a function of the redshift.
We found that for energies below 10 GeV the signal increases with the redshift  up to $z\sim0.5$ and
then decreases. Above 10 GeV the correlation signal is  mostly confined to low redshift ($z\lesssim 0.3$) 
with some additional contribution above $z \sim 1$.
While these results support the hypothesis of multiple source populations contributing to the UGRB, drawing 
conclusions on the nature of these sources require a physically motivated model of the UGRB. We postpone such
analysis to a follow up study.

\end{itemize}

In conclusion, we present  a new way to characterize the UGRB by extracting accurate
spectral and   {\it redshift} information otherwise inaccessible when using $\gamma$-ray data alone.
In the present study we have only started to explore the   implications of the   wealth on new information
made available by the tomography technique.
In the near future, by exploiting these new data within the framework of well-motivated
$\gamma$-ray population models, we shall 
   set  tighter constraints on the nature of the
UGRB sources, whether of astrophysical origin or not.

\acknowledgments
MB is supported by the Netherlands Organization for Scientific Research, NWO, through grant number 614.001.451, and by the Polish National Science Center under contract \#UMO-2012/07/D/ST9/02785.

EB is supported by INFN-PD51 INDARK and Agenzia Spaziale Italiana agreement ASI/INAF/I/023/12/0.
EB and MB acknowledge support of the Ministry of Foreign Affairs and International
Cooperation, Directorate General for the Country Promotion (Bilateral
Grant Agreement ZA14GR02 - Mapping the Universe on the Pathway to SKA).

JX is supported by the National Youth Thousand Talents Program, the
National Science Foundation of China under grant Nos. 11422323,
11633001 and 11690023, and the Fundamental Research Funds for the
Central Universities, grant No. 2017EYT01.

The \textit{Fermi} LAT Collaboration acknowledges generous ongoing support
from a number of agencies and institutes that have supported both the
development and the operation of the LAT as well as scientific data analysis.
These include the National Aeronautics and Space Administration and the
Department of Energy in the United States, the Commissariat \`a l'Energie Atomique
and the Centre National de la Recherche Scientifique / Institut National de Physique
Nucl\'eaire et de Physique des Particules in France, the Agenzia Spaziale Italiana
and the Istituto Nazionale di Fisica Nucleare in Italy, the Ministry of Education,
Culture, Sports, Science and Technology (MEXT), High Energy Accelerator Research
Organization (KEK) and Japan Aerospace Exploration Agency (JAXA) in Japan, and
the K.~A.~Wallenberg Foundation, the Swedish Research Council and the
Swedish National Space Board in Sweden.
This work is performed in part under DOE Contract DE-AC02-76SF00515.

Additional support for science analysis during the operations phase is
gratefully acknowledged from the Istituto Nazionale di Astrofisica in
Italy and the Centre National d'\'Etudes Spatiales in France.

This publication makes use of data products from the Two Micron All Sky Survey, which is a joint project of the University of Massachusetts and the Infrared Processing and Analysis Center/California Institute of Technology, funded by the National Aeronautics and Space Administration and the National Science Foundation.

Funding for SDSS-III has been provided by the Alfred P. Sloan Foundation, the Participating Institutions, the National Science Foundation, and the U.S. Department of Energy Office of Science. The SDSS-III web site is \url{http://www.sdss3.org/}. SDSS-III is managed by the Astrophysical Research Consortium for the Participating Institutions of the SDSS-III Collaboration including the University of Arizona, the Brazilian Participation Group, Brookhaven National Laboratory, Carnegie Mellon University, University of Florida, the French Participation Group, the German Participation Group, Harvard University, the Instituto de Astrofisica de Canarias, the Michigan State/Notre Dame/JINA Participation Group, Johns Hopkins University, Lawrence Berkeley National Laboratory, Max Planck Institute for Astrophysics, Max Planck Institute for Extraterrestrial Physics, New Mexico State University, New York University, Ohio State University, Pennsylvania State University, University of Portsmouth, Princeton University, the Spanish Participation Group, University of Tokyo, University of Utah, Vanderbilt University, University of Virginia, University of Washington, and Yale University.

This research has made use of data obtained from the SuperCOSMOS Science Archive, prepared and hosted by the Wide Field Astronomy Unit, Institute for Astronomy, University of Edinburgh, which is funded by the UK Science and Technology Facilities Council.

We acknowledge using TOPCAT\footnote{\url{http://www.star.bristol.ac.uk/\~mbt/topcat/}} \citep{TOPCAT}
and STILTS\footnote{\url{http://www.star.bristol.ac.uk/\~mbt/stilts/}} \citep{STILTS}  software.
Some of the results in this paper have been derived using the HEALPix package\footnote{\url{http://healpix.sourceforge.net/}} \citep{2005ApJ...622..759G}.

We wish to thank Michela Negro and Luca Latronico for carefully reading the manuscript and providing comments,
and Nicolao Fornengo, Marco Regis and Matteo Viel for useful discussions on the subjects of this paper.


\bibliographystyle{aasjournal}
\bibliography{version_2}

\clearpage

\appendix

\section{Validation tests}
\label{apdx:validation}

\begin{figure*}
\centering \epsfig{file= 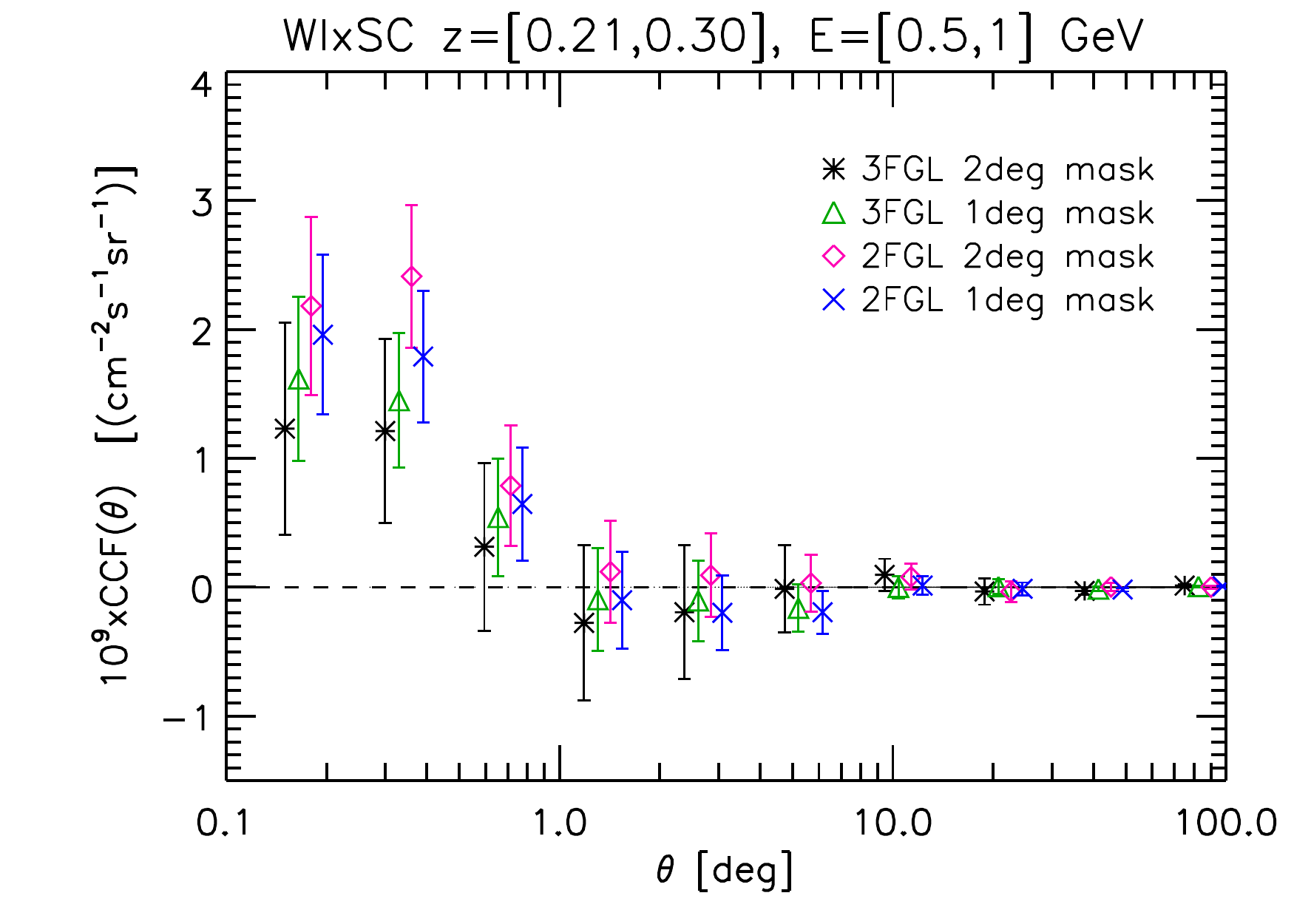, angle=0, width=0.45 \textwidth}
\centering \epsfig{file=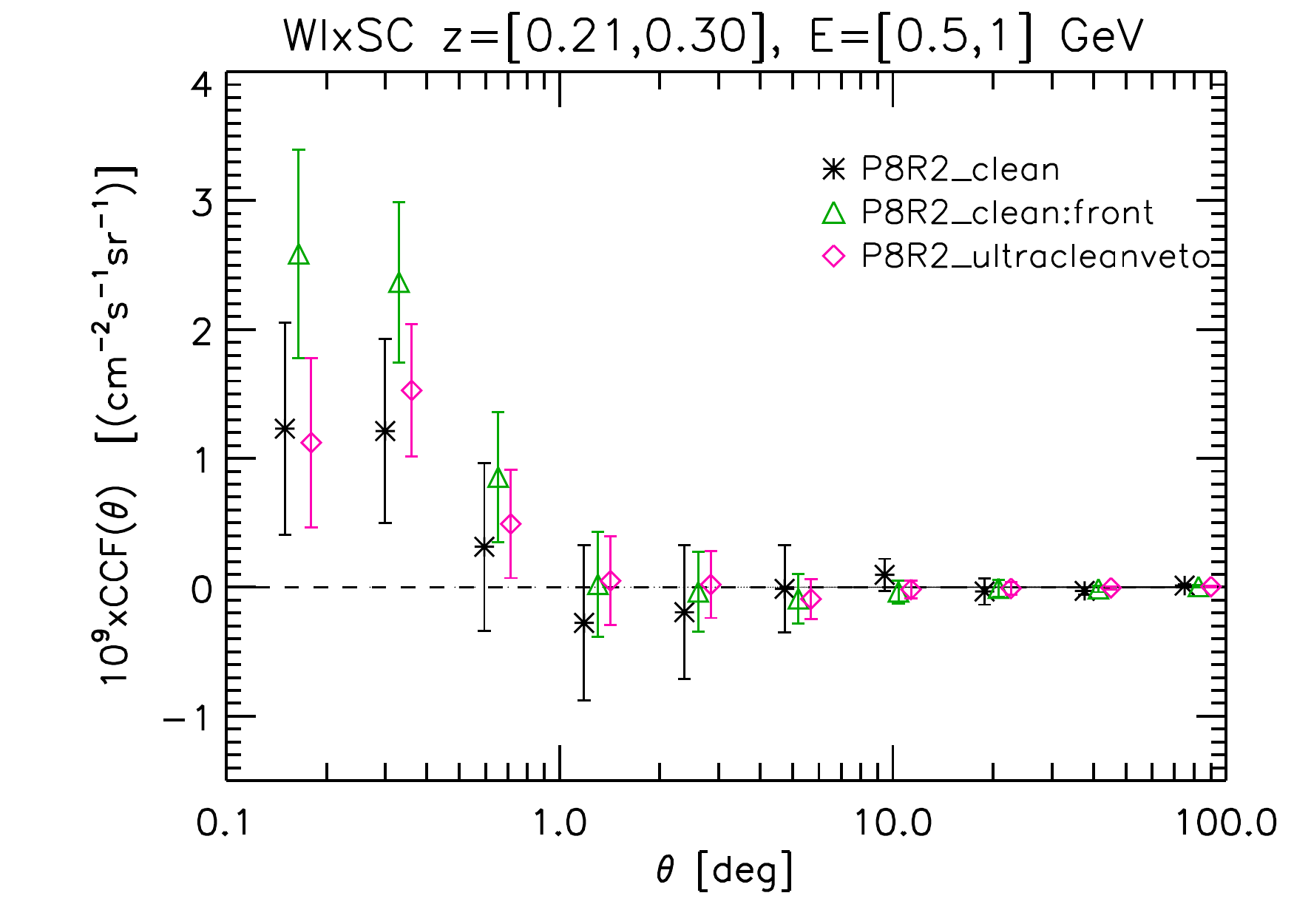, angle=0, width=0.45 \textwidth}
\centering \epsfig{file= 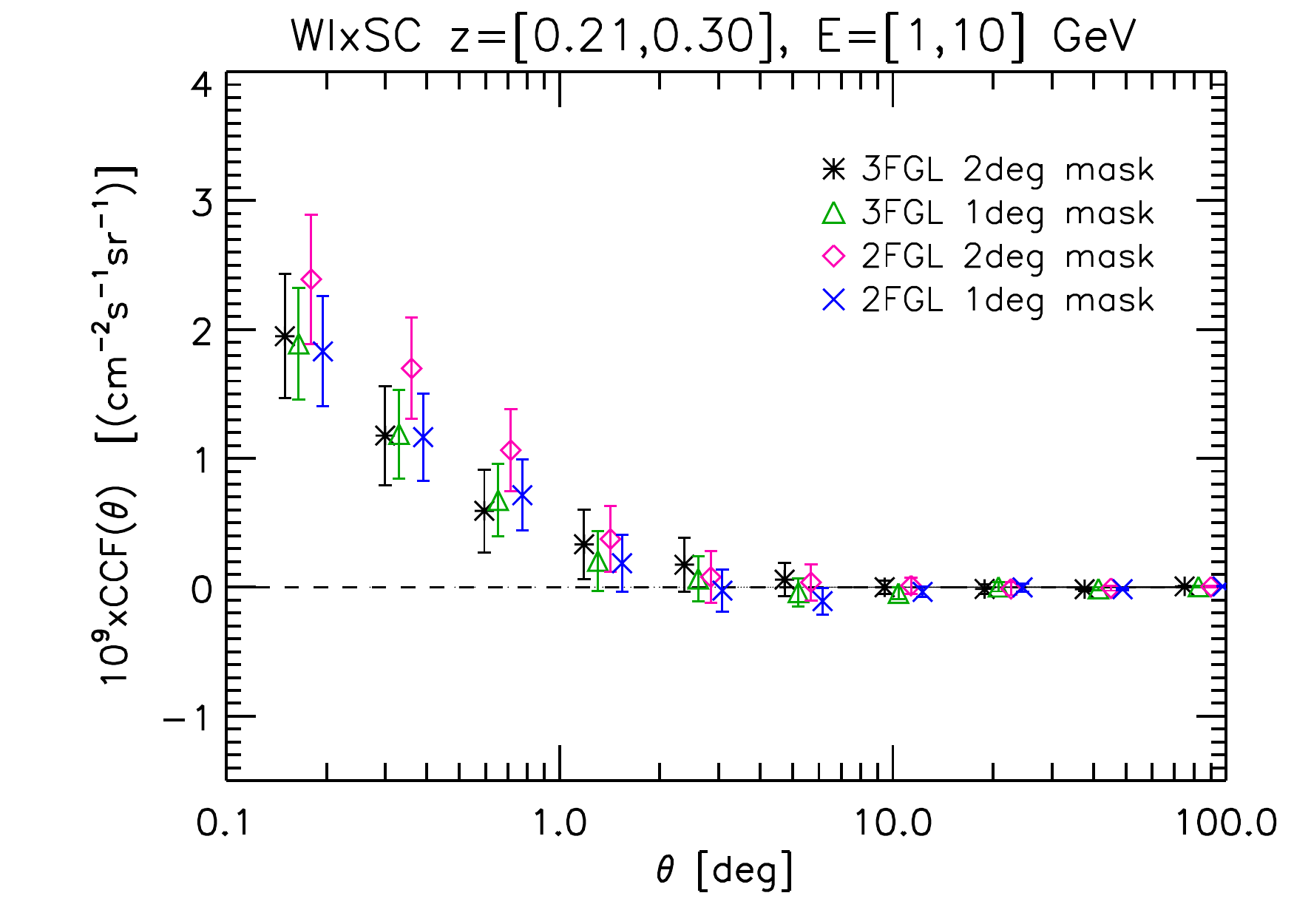, angle=0, width=0.45 \textwidth}
\centering \epsfig{file=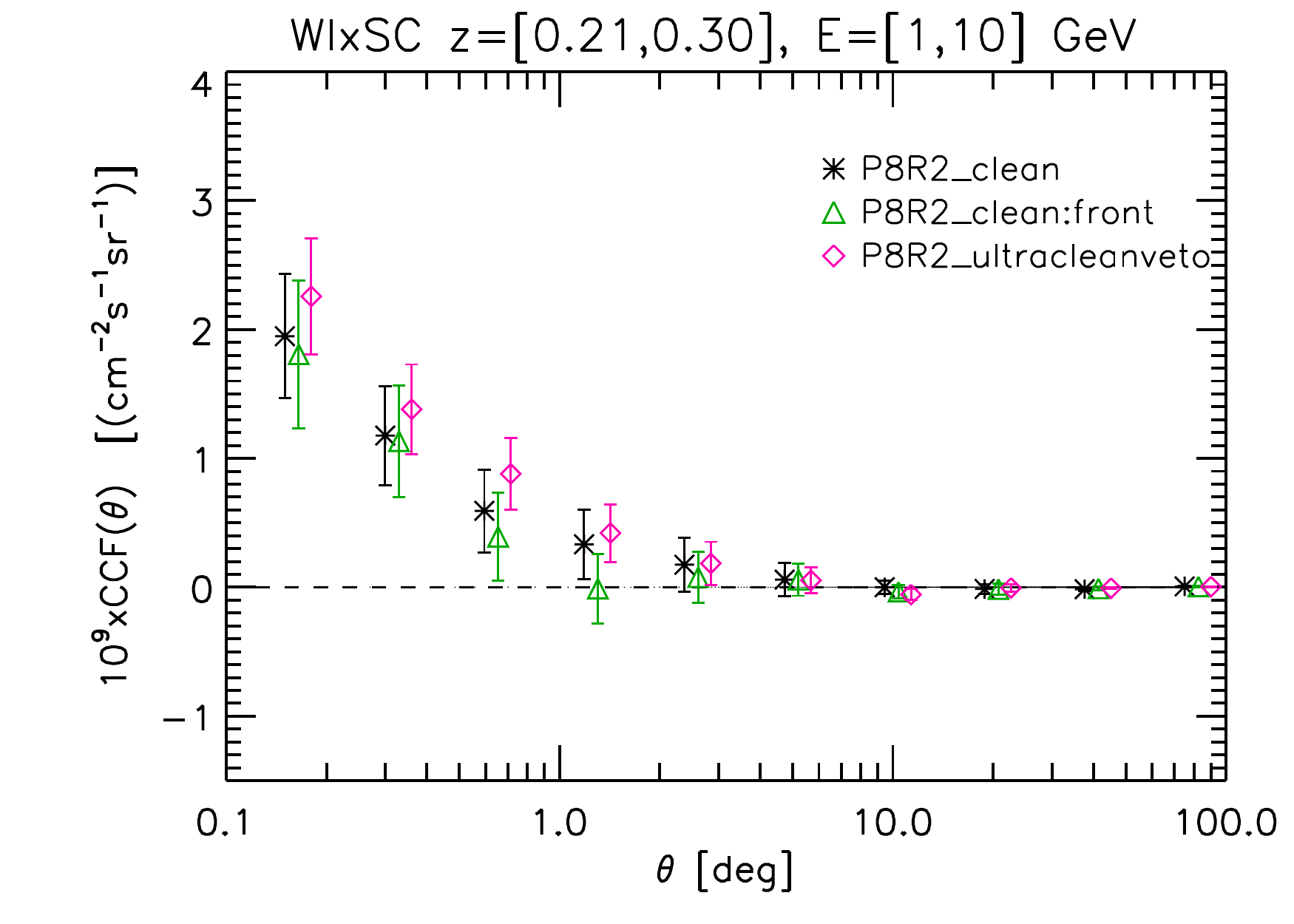, angle=0, width=0.45 \textwidth}
 \centering \epsfig{file= 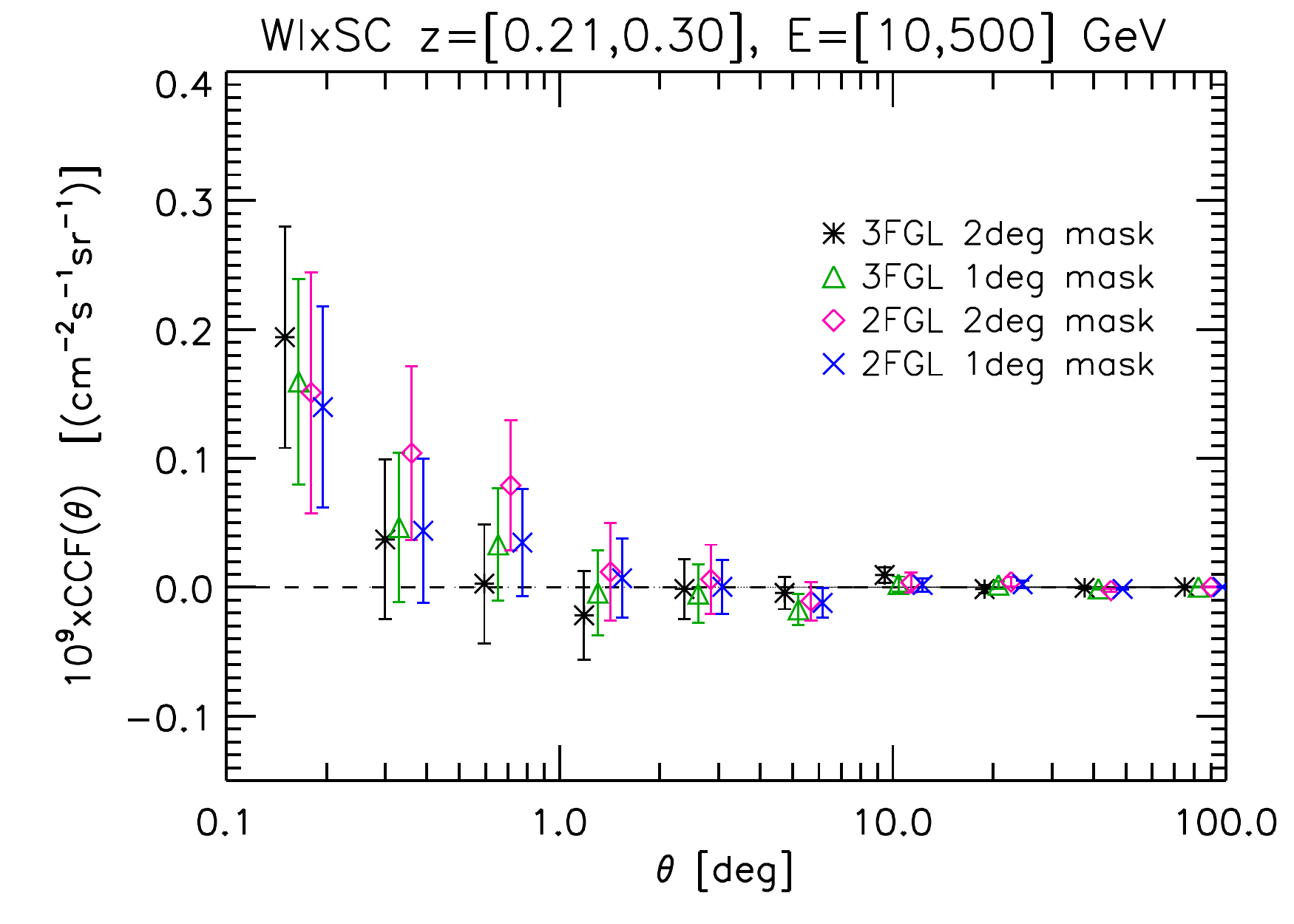, angle=0, width=0.45 \textwidth}
\centering \epsfig{file=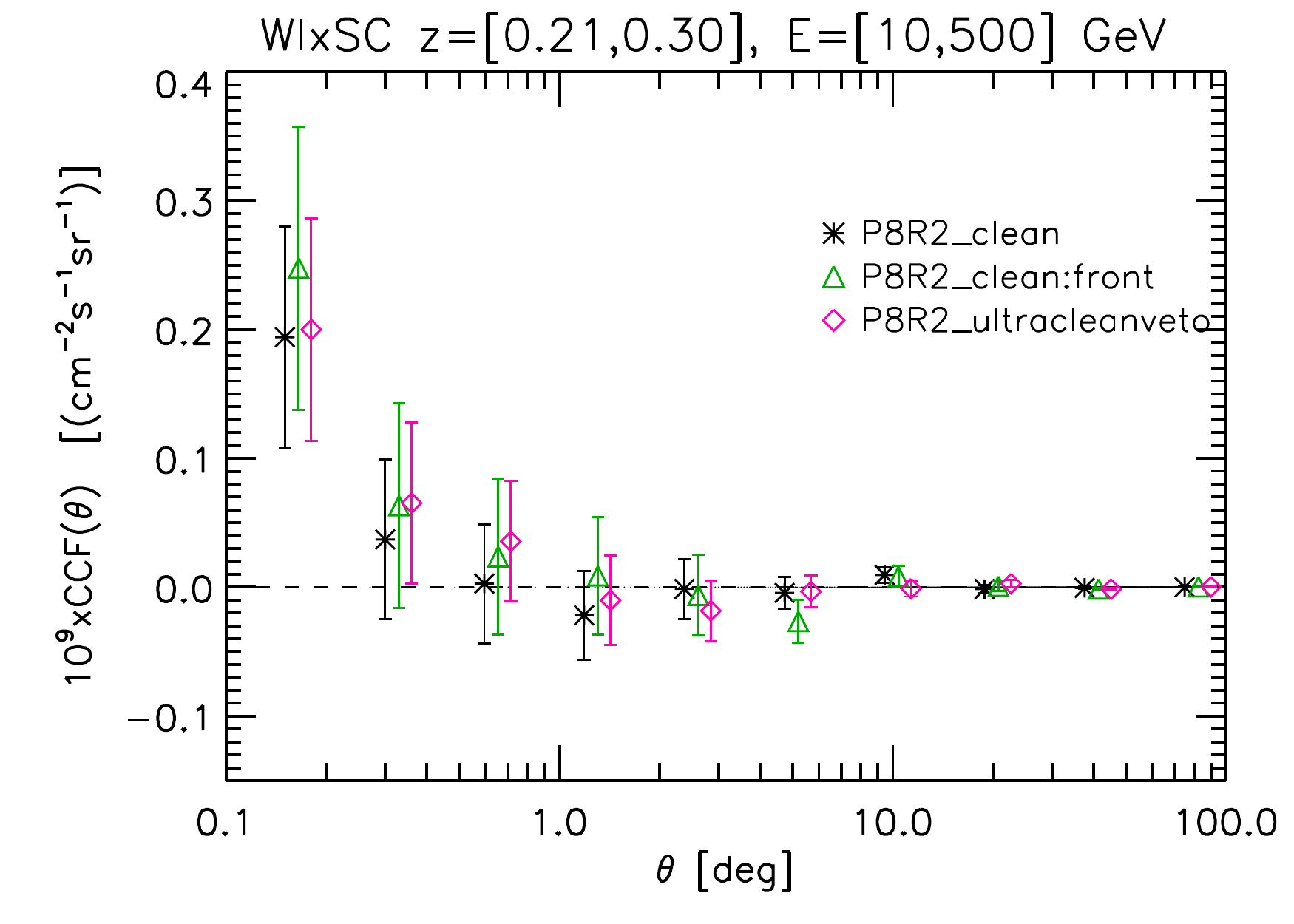, angle=0, width=0.45 \textwidth}
\vspace{-0.4cm}
\caption{{Angular CCF  for \WISC\ galaxies in the redshift bin $z \in [0.21,0.30]$ for different
$\gamma$-ray energy bins and different $\gamma$-ray point-source masks or $\gamma$-ray data selections,
as indicated in the plot labels and described in the text.
Error bars represent the square root of the diagonal elements of the covariance matrix. }}
\label{fig:checks-mask}
\end{figure*}

To validate the results presented in the main text, we have performed several tests   described  below.
These tests have been performed using all sub-catalogs considered in the cross-correlation analysis.
Here    we show the   representative case of the   \WISC\ galaxies  in the bin $z \in [0.21,0.30]$. 
Very similar results have been found   for all other subsamples analyzed.

{\bf Mask.}
{The left column of}  Fig.~\ref{fig:checks-mask} illustrates the impact of changing the mask used to remove
the $\gamma$-ray point sources.
Our baseline is that of masking the brightest (in terms of the integral photon flux in the energy range [0.1,100] GeV)
five hundred 3FGL point sources with a disk of $2^\circ$ radius and   the remaining ones with a $1^\circ$ disk.
We considered three more cases: {\it i)} all 3FGL sources are masked with $1^\circ$ disks;
  {\it ii)} all 2FGL  sources are masked with $2^\circ$ disks;   {\it iii)} all 2FGL sources are masked with $1^\circ$ disks.
It is clear from the plots that the impact of these different masks is negligible as all the CCFs are consistent with each other.
These results deserve some further considerations.
A dependence of the correlation on the mask is to be  expected.
For example, in \cite{FermiAPS_16} it was shown that the $\gamma$-ray auto-correlation depends significantly on the catalog used to mask the point sources, 
with the anisotropy for the case of the 2FGL mask being a factor of  $\sim 4$ larger than the case of the 3FGL mask.
This implies that sources that are in the 3FGL catalog but not in the 2FGL one are responsible for the bulk of the anisotropy
detected when using the 2FGL mask.
A similar effect would be expected also for the cross-correlation, although, evidently, at a much smaller level. 
The fact that, from Fig.~\ref{fig:checks-mask}, the cross-correlation for the 2FGL mask is consistent  with the 3FGL mask case
sets an upper limit on the magnitude of  this effect that cannot exceed the random error,  i.e. it has to be smaller than 20-30\%.
Indeed, we do detect this effect for the case of the NVSS catalog, for which the relative errors are the smallest, at the level of 5-10\%.
 In that case the cross correlation with 2FGL sources is 
$\sim$20\% larger than with the 3FGL ones.

\begin{figure}
\hspace{-2cm}
\centering \epsfig{file=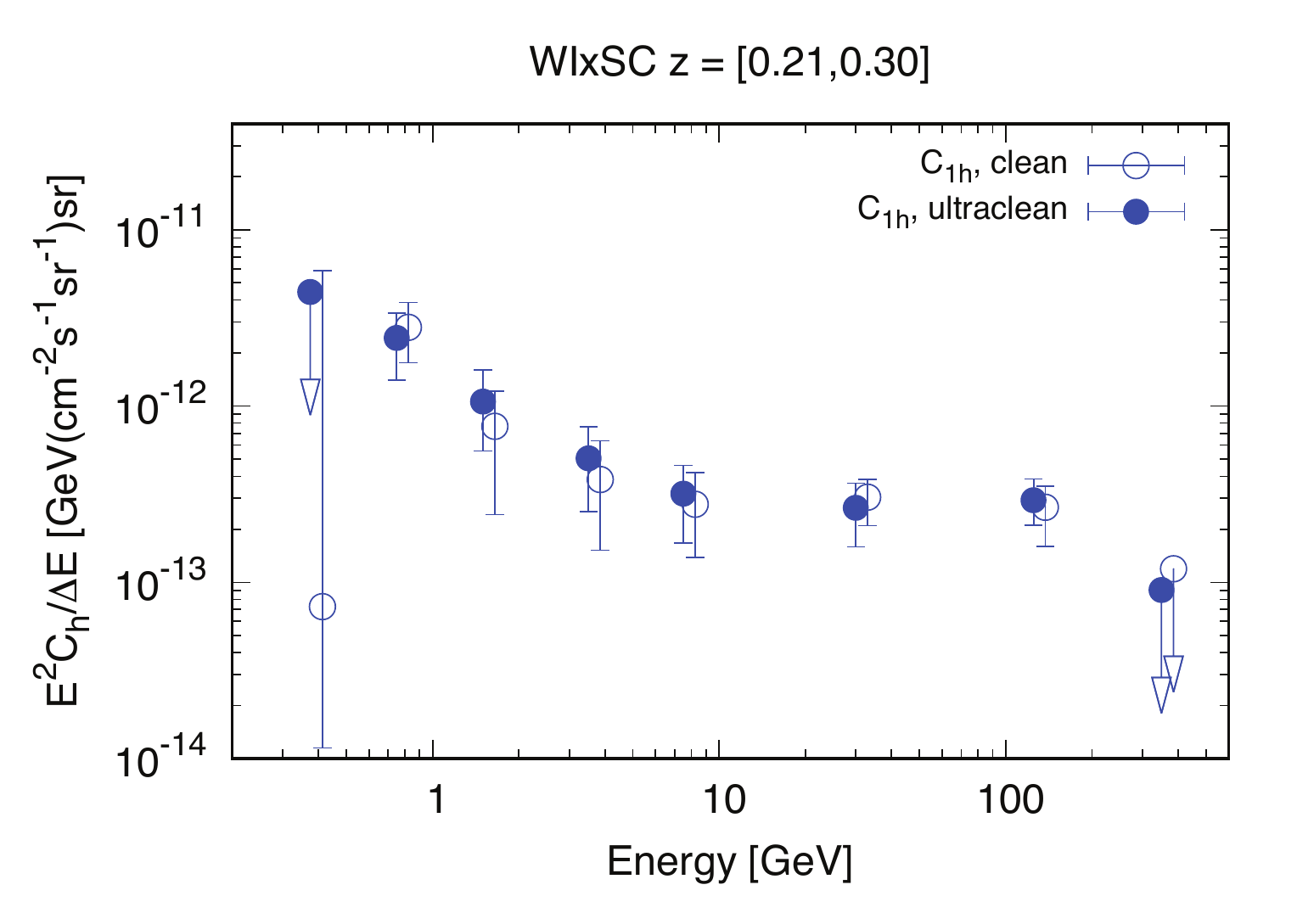, angle=0, width=0.47 \textwidth}
\hspace{-2cm}
\vspace{-0.3cm}
\caption{
The 1-halo term as a function of energy for the \texttt{P8R2\_ULTRACLEANVETO\_V6} and \texttt{P8R2\_CLEAN\_V6}
cases.}
\label{fig:checks-ultraclean}
\end{figure}

{\bf Data class.}
The effect of changing the data selection procedure is illustrated {in the right-column panels of} Fig.~\ref{fig:checks-mask}.
The default  procedure is the  \texttt{P8R2\_CLEAN\_V6} selection with a zenith angle cut of $100^\circ$.
In the figure we compare the standard CCF with the one obtained when $\gamma$-ray data are selected
using the  \texttt{P8R2\_ULTRACLEANVETO\_V6}   with
 a zenith angle cut of $90^\circ$. This alternative selection  has the lowest cosmic-ray contamination
of the $\gamma$-ray sample among the different available classes. Increasing purity comes  at   the price of
reducing the effective area  by $\sim 30\%$ with respect to the \texttt{P8R2\_CLEAN\_V6} case.
The tighter choice of a zenith angle cut of $90^\circ$ instead of $100^\circ$,
removes more aggressively any residual contamination from the bright $\gamma$-ray Earth Limb.
The right-hand panels of Fig.~\ref{fig:checks-mask} shows the CCF  that we measure when the  \texttt{P8R2\_ULTRACLEANVETO\_V6}
is adopted. Fig.~\ref{fig:checks-ultraclean}, instead, compares the energy dependence of the 1-halo terms in the
two cases. The results obtained using the two selection procedures are fully consistent   with each other.

{\bf Data sub-sample.}
 As a further test we have considered only \texttt{FRONT} events.
 The \texttt{FRONT} data amount to about half of the total, with a
better PSF, about $\sim 50\%$ more compact than the global average.
The corresponding CCF, {shown, again,  in  the right-column panels of} Fig.~\ref{fig:checks-mask}, has a slightly larger amplitude in
 the energy bin $E \in [0.5,1.0]$ GeV. This is not surprising. It reflects the convolution effect
 of a PSF that is more compact than the standard one. Once the signal is deconvolved
 from the PSF, like in the CAPS case,  the  correlation signal obtained using the
  \texttt{P8R2\_CLEAN\_V6:FRONT} data is fully consistent with the
  standard  one.

\begin{figure}
\centering \epsfig{file=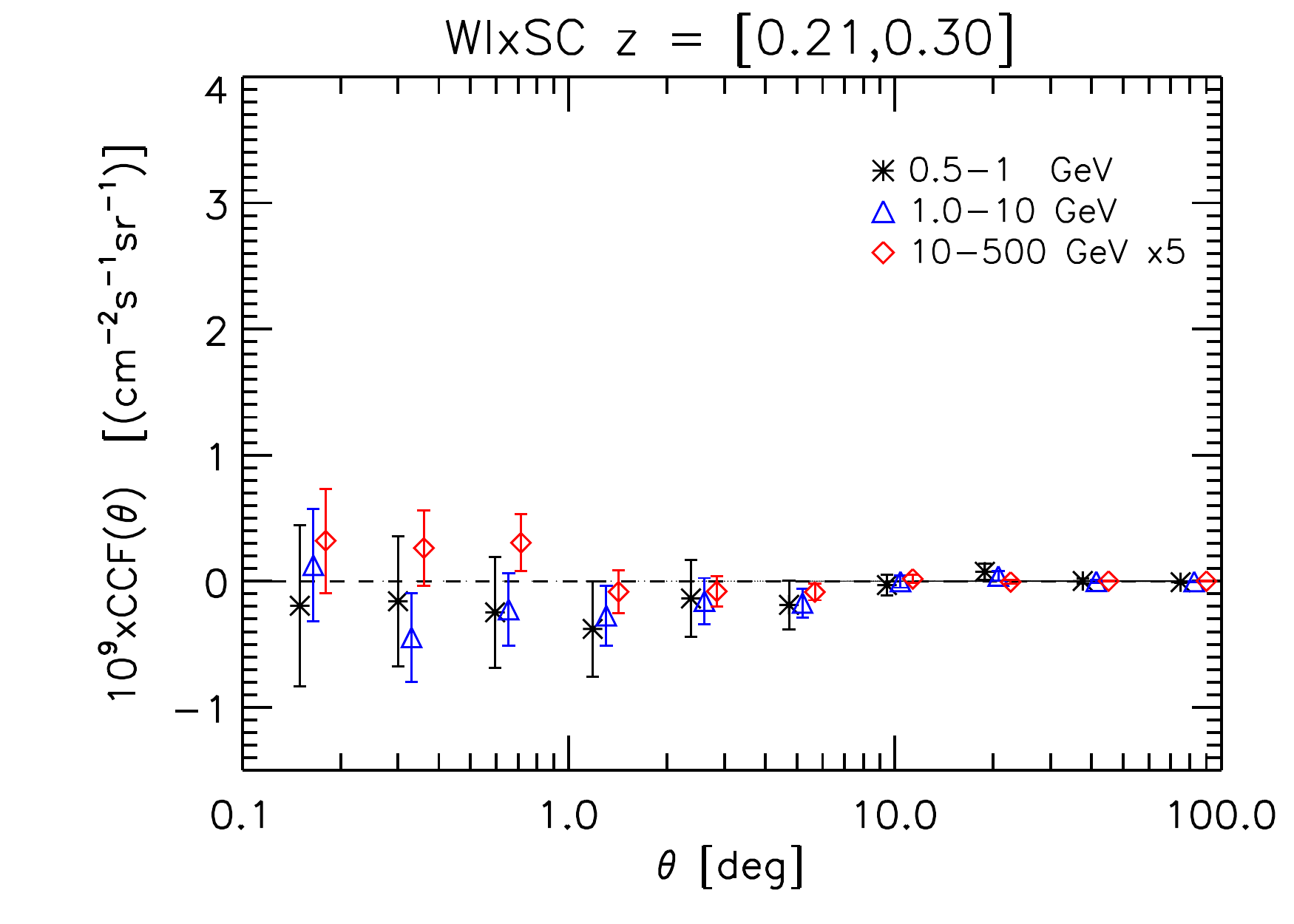, angle=0, width=0.45 \textwidth}
\caption{Same as Fig.~\ref{fig:wixsc_ccf_fermi} except that the angular coordinated of
\WISC\ galaxies have been transformed as $(l,b) \rightarrow (l,-b)$.}
\label{fig:checks-minus}
\end{figure}

{\bf No signal tests.}
Finally, Fig.~\ref{fig:checks-minus} is analogous to Fig.~\ref{fig:wixsc_ccf_fermi} except for the fact that
Galactic latitude $b$ of each object  has been switched to $-b$. The same transformation
has been applied to the angular coordinates of the pixels of the \WISC\  mask.
This transformation is expected to preserve the angular auto-correlation of
\WISC\  galaxies and remove the cross-correlation signal. The figure shows that
this is indeed the case. No spurious cross-correlation is detected.

\section{Additional Results}
\label{apdx:moreplots}

In this section we show all the results of the analyses that were not presented in the main text.
This includes various sets of plots illustrating the CCFs of different catalogs, namely  {\it i)}
three  $z$-shells extracted from the QSO DR6 sample (Fig.~\ref{fig:qso_ccf_fermi_full}),;
{\it ii)} four  $z$-shells extracted from the \WISC\  sample (Fig.~\ref{fig:wixsc_ccf_fermi_full}); and
{\it iii)} seven $z$-shells extracted from the SDSS DR12 sample (Fig.~\ref{fig:sdss_ccf_fermi_full}).
A similar set of plots for the 2MPZ case has already been shown in the main text (Fig.~\ref{fig:2mpz_ccf_fermi}).

In addition we show  two tables that summarize all the results obtained in this work.
Table~\ref{tab:T1apdx} expands (and also includes) Table~\ref{tab:T01} and contains the results of the
best fits to the CAPS of the different catalogs measured in all $z$-bins and in three {wide
energy bins: $E \in [0.5,1]$, $[1,10]$ and $[10,500]$ GeV.

Table~\ref{tab:T2apdx} lists the best-fit parameters of the models that
describe the energy dependence of the 1-halo and 2-halo terms that contribute to the CAPS.
This table shows the results of all the energy models: SPL, DPL, and BPL, whereas Table~\ref{tab:T02}
in the main text only contains the DPL results.

\begin{figure*}
\centering \epsfig{file=QSO_CCF_EAll_zAll.pdf, angle=0, width=0.45 \textwidth}
\centering \epsfig{file=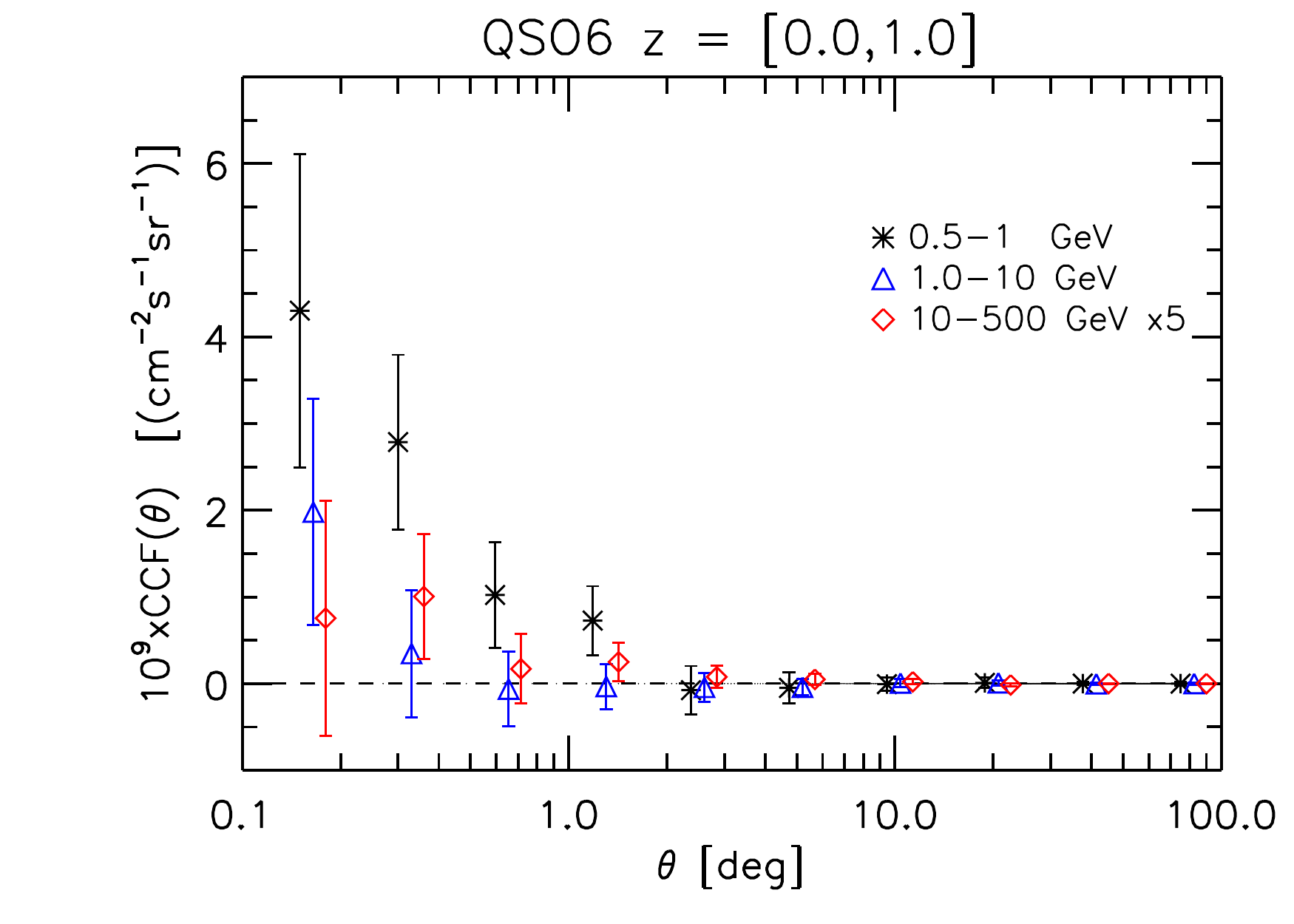, angle=0, width=0.45 \textwidth}
\centering \epsfig{file=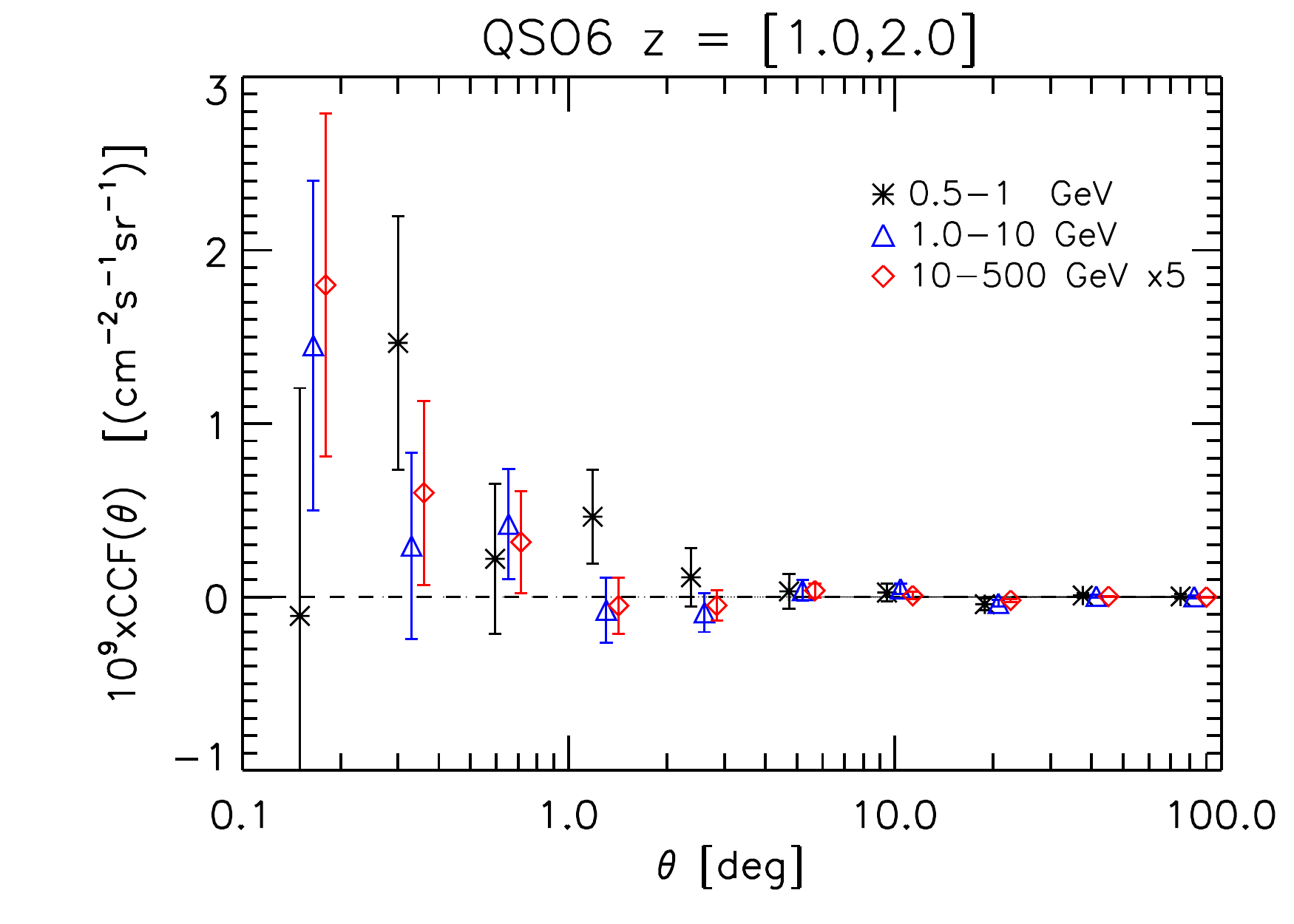, angle=0, width=0.45 \textwidth}
\centering \epsfig{file=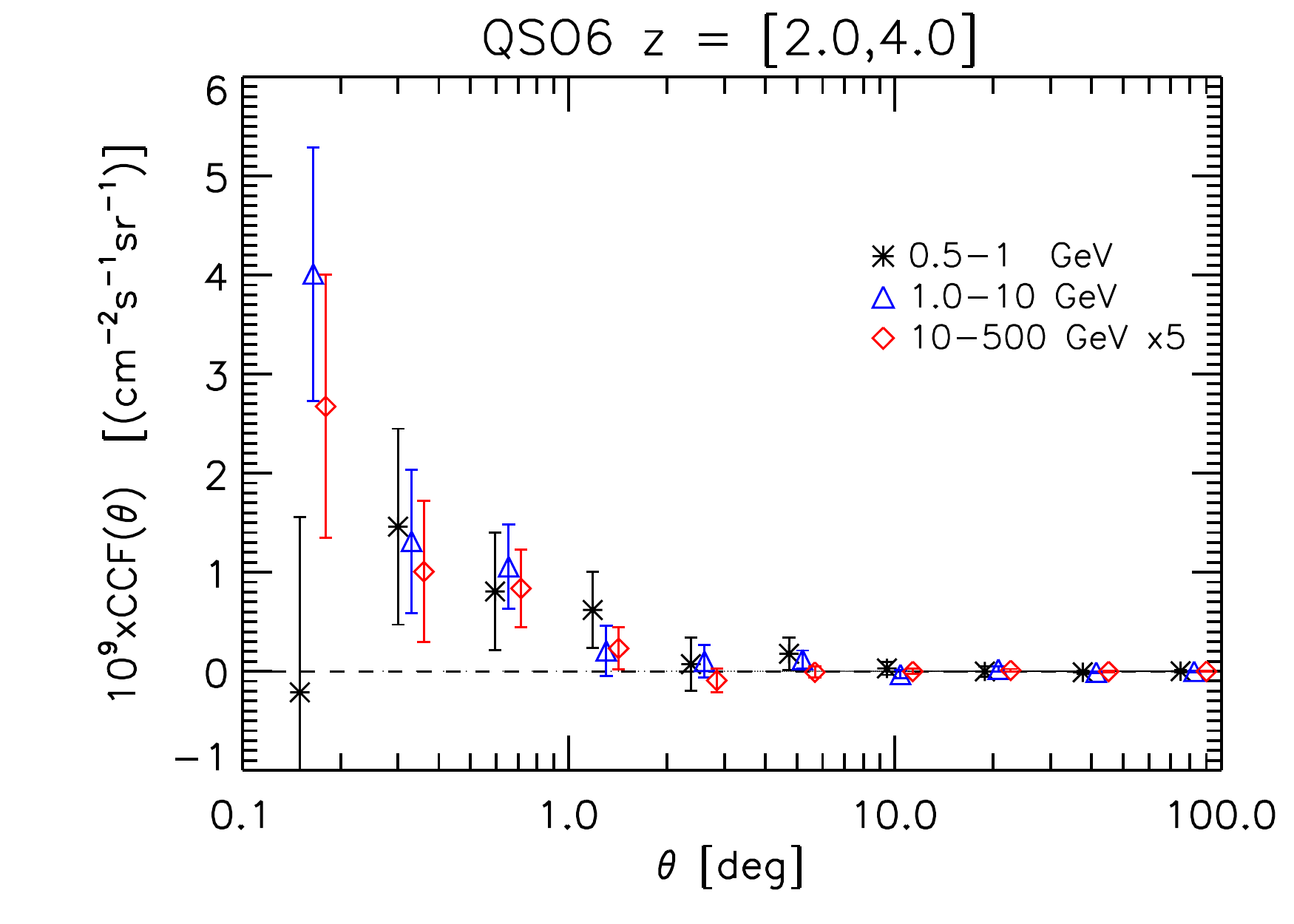, angle=0, width=0.45 \textwidth}
\vspace{-0.4cm}
\caption{Angular CCF for DR6-QSOs  in different redshift and energy bins.
Top left panel is for  $z=$ All, top right for  $z \in [0.0,1.0]$,  bottom left  for  $z \in [1.0,2.0]$, and and bottom right for $z \in [2.0,4.0]$.
Energy bins are $E\in [0.5,1]$ GeV, $[1,10]$ GeV, $[10,500]$~GeV.}
\label{fig:qso_ccf_fermi_full}
\end{figure*}

\begin{figure*}
\centering \epsfig{file=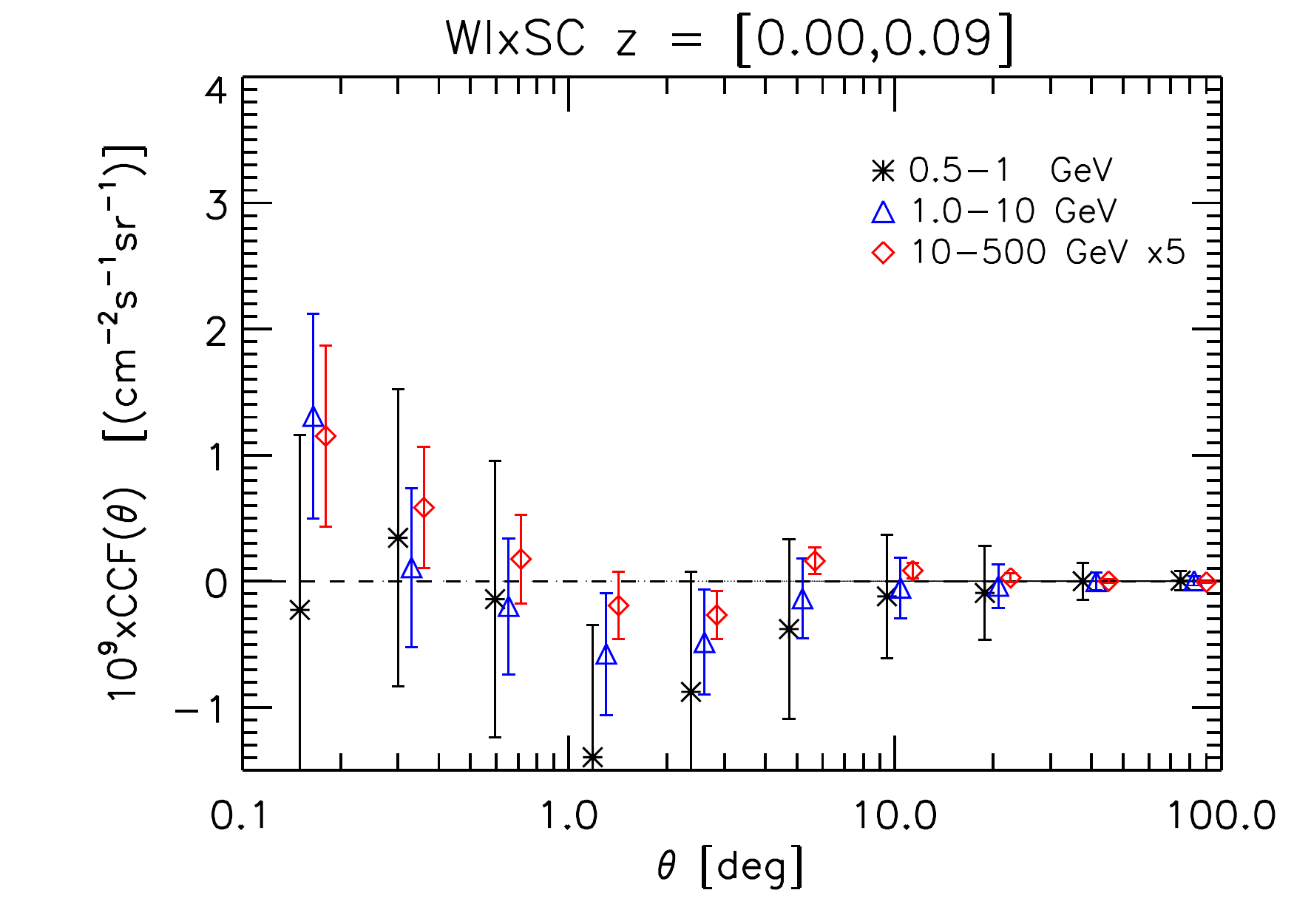, angle=0, width=0.45 \textwidth}
\centering \epsfig{file=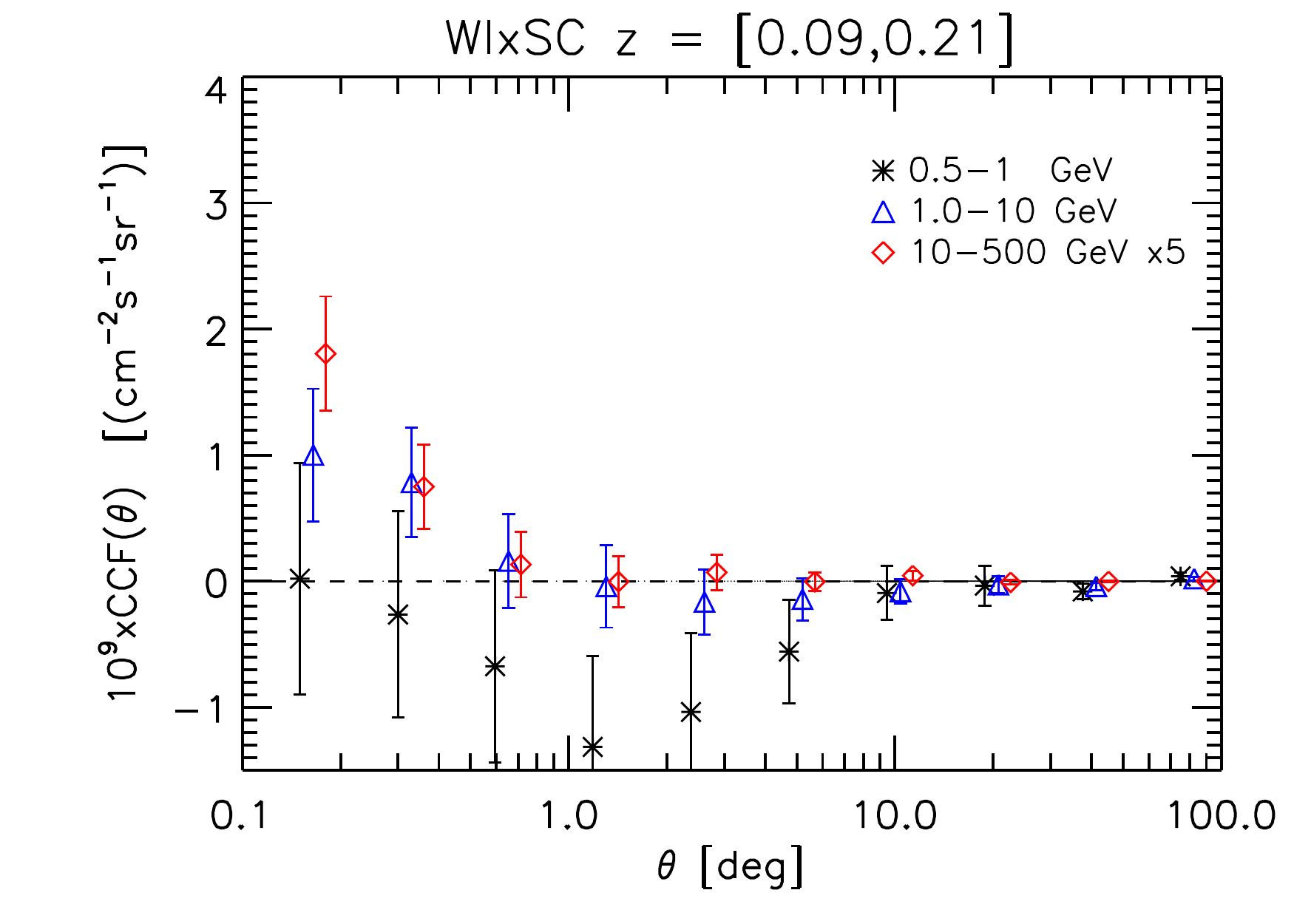, angle=0, width=0.45 \textwidth}
\centering \epsfig{file=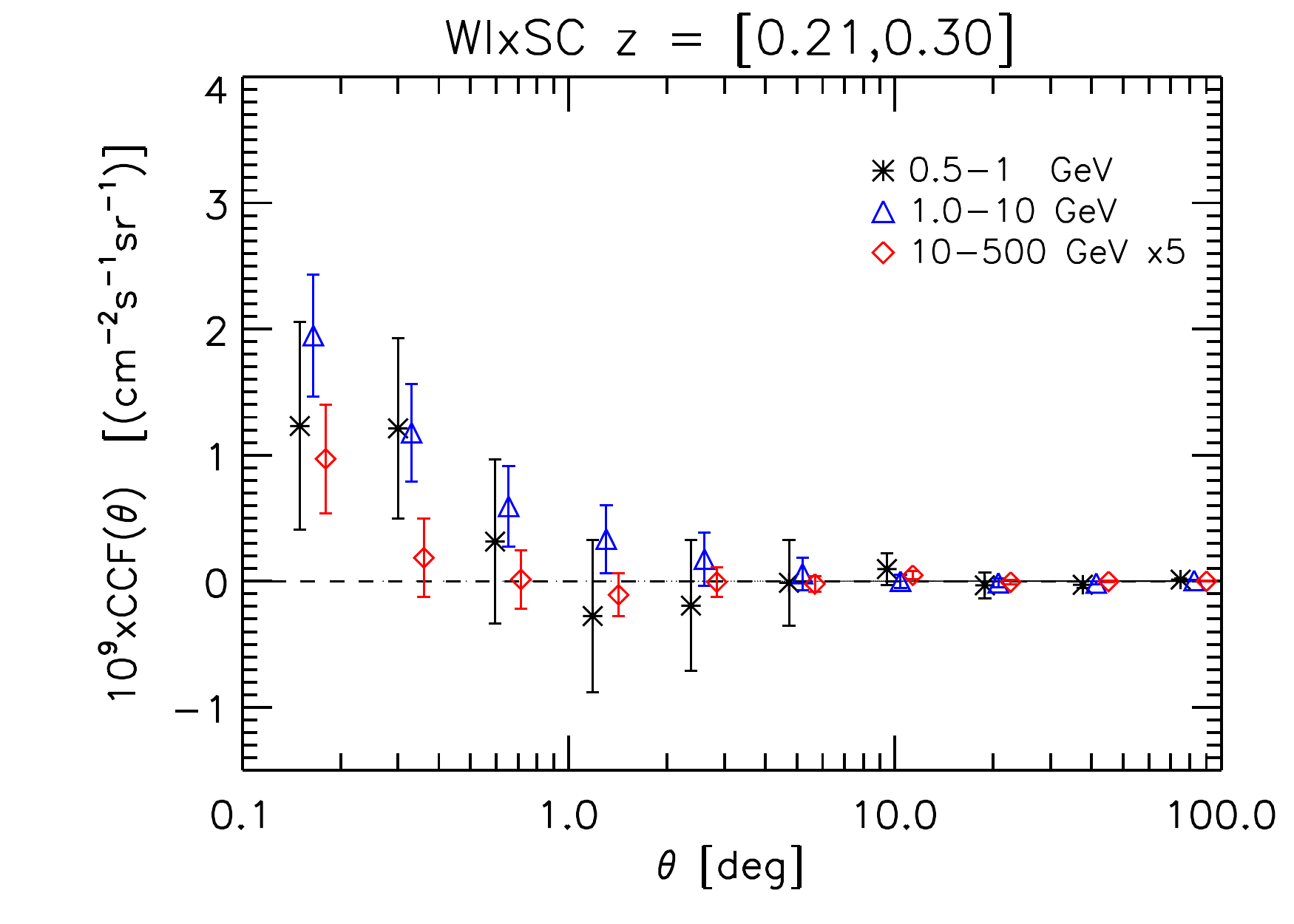, angle=0, width=0.45 \textwidth}
\centering \epsfig{file=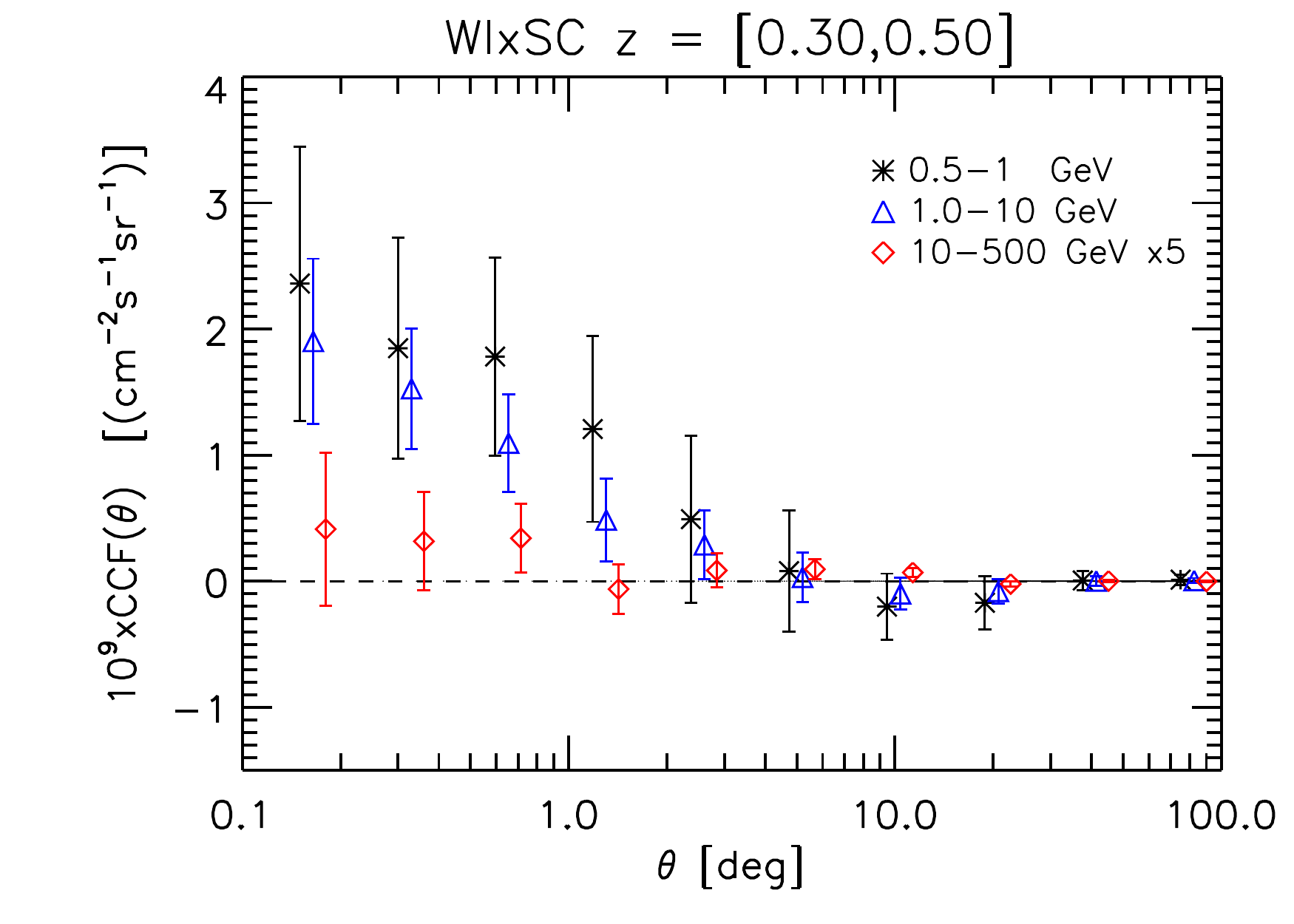, angle=0, width=0.45 \textwidth}
\vspace{-0.4cm}
\caption{Angular CCF for \WISC\ galaxies in different redshift and energy bins.
Top left panel is for  $z \in [0.0,0.09]$,  Top right: $z \in [0.09,0.21]$,  Bottom left: $z \in [0.21,0.30]$, and
Bottom right: $z \in [0.30,0.50]$.
Energy bins  are for $E \in [0.5,1]$ GeV, $[1,10]$ GeV $[10,500]$ GeV.}
\label{fig:wixsc_ccf_fermi_full}
\end{figure*}

\begin{figure*}
\centering \epsfig{file=DR12_CCF_EAll_zAll.pdf, angle=0, width=0.45 \textwidth}
\centering \epsfig{file=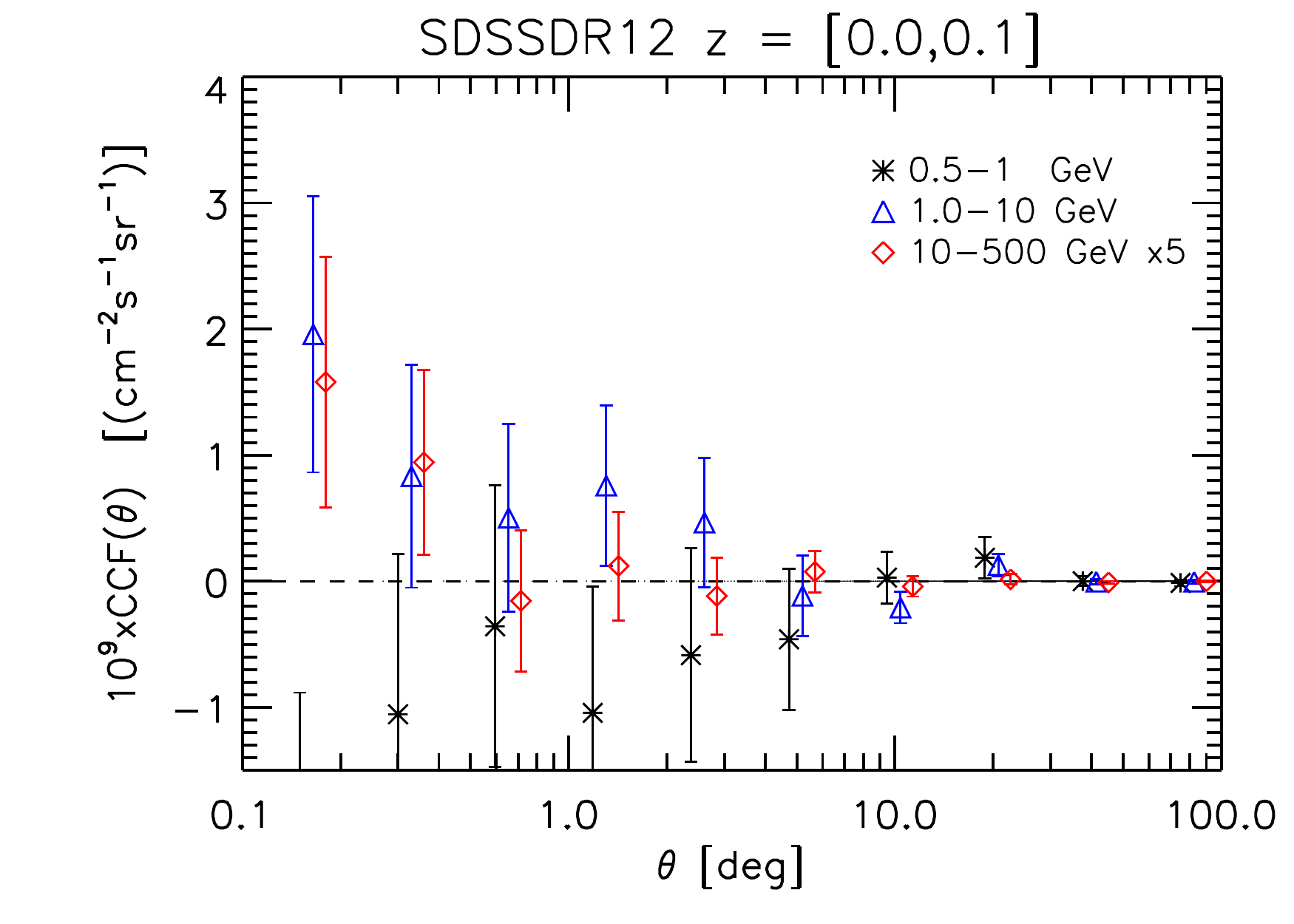, angle=0, width=0.45 \textwidth}
\centering \epsfig{file=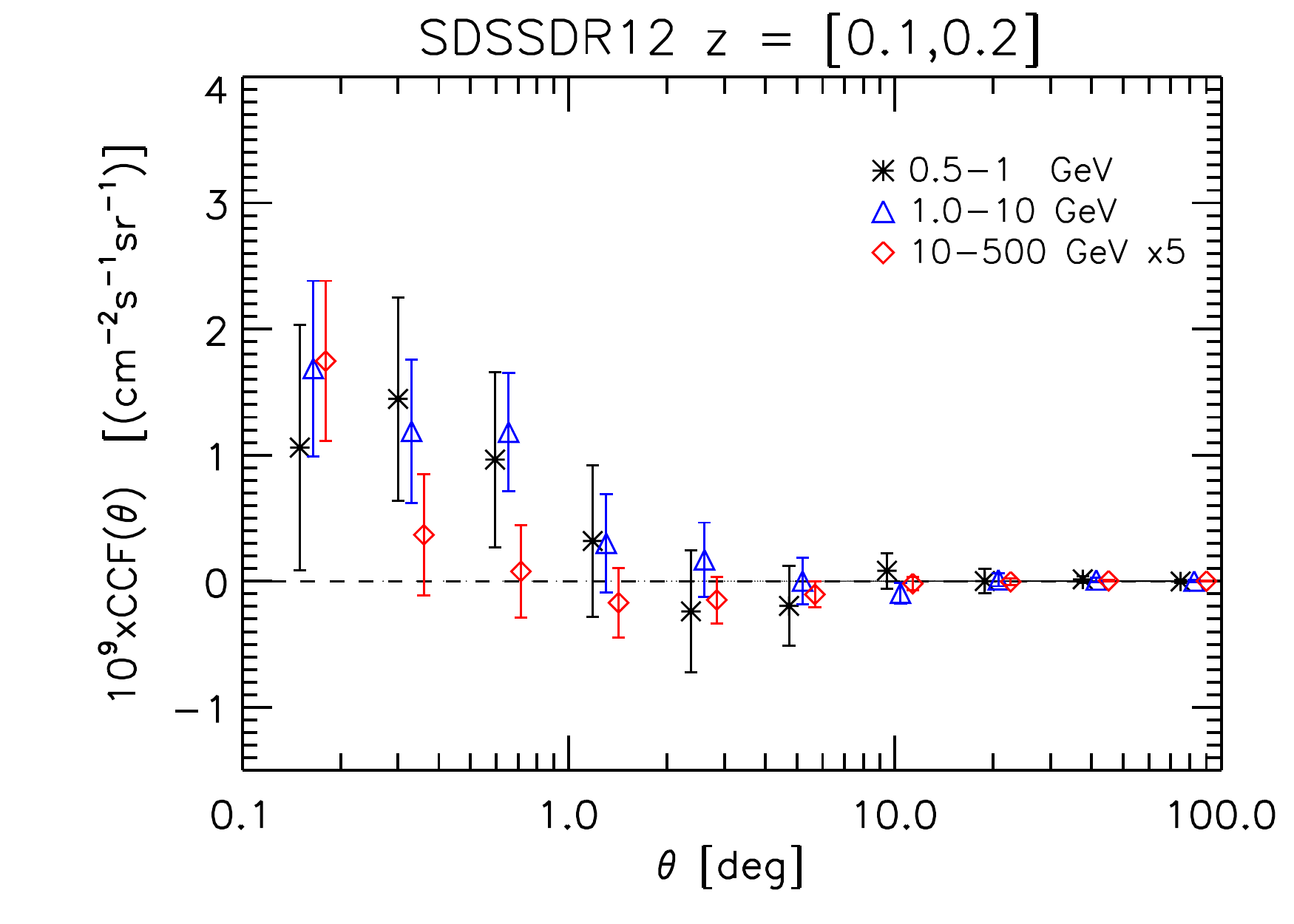, angle=0, width=0.45 \textwidth}
\centering \epsfig{file=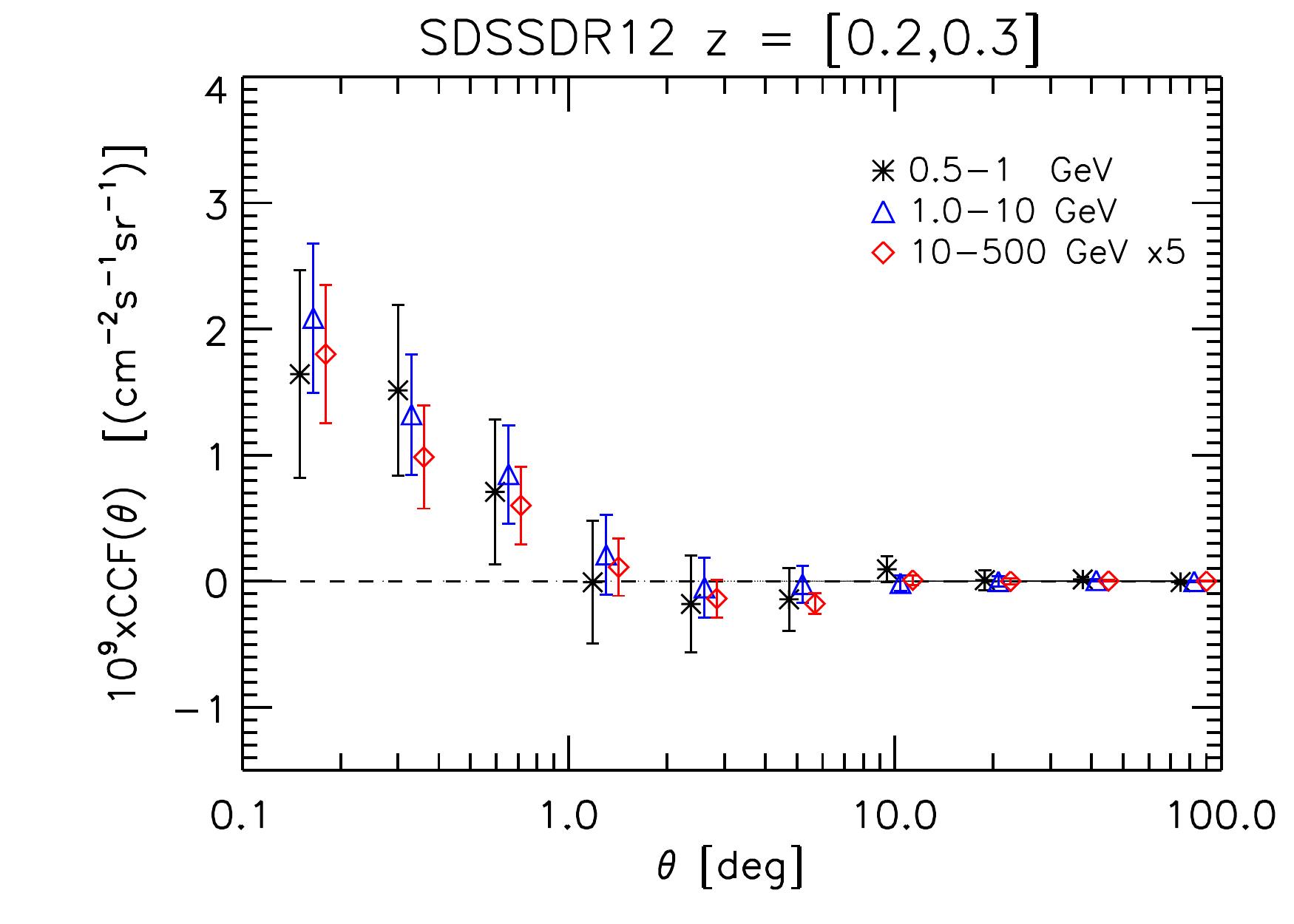, angle=0, width=0.45 \textwidth}
\centering \epsfig{file=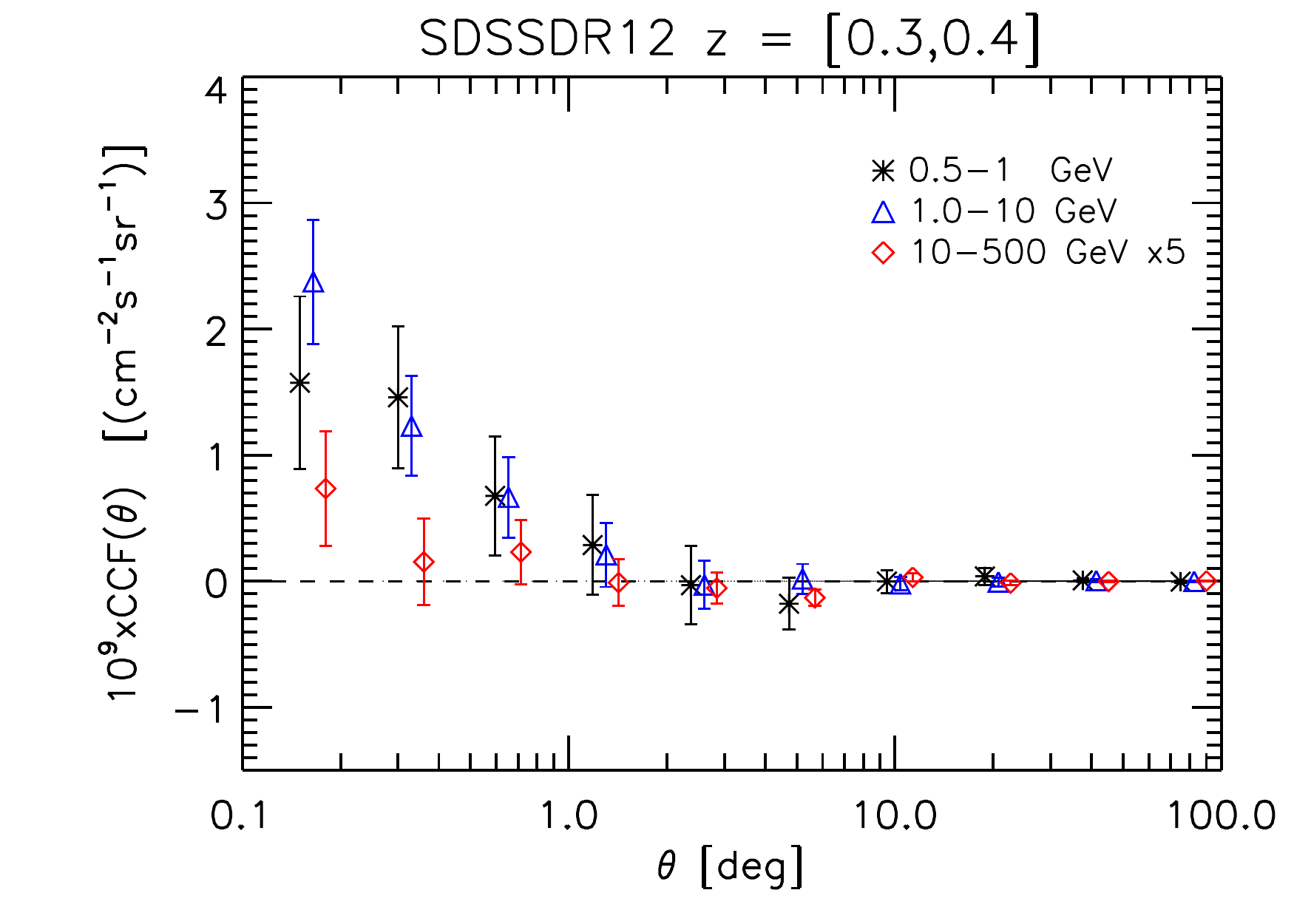, angle=0, width=0.45 \textwidth}
\centering \epsfig{file=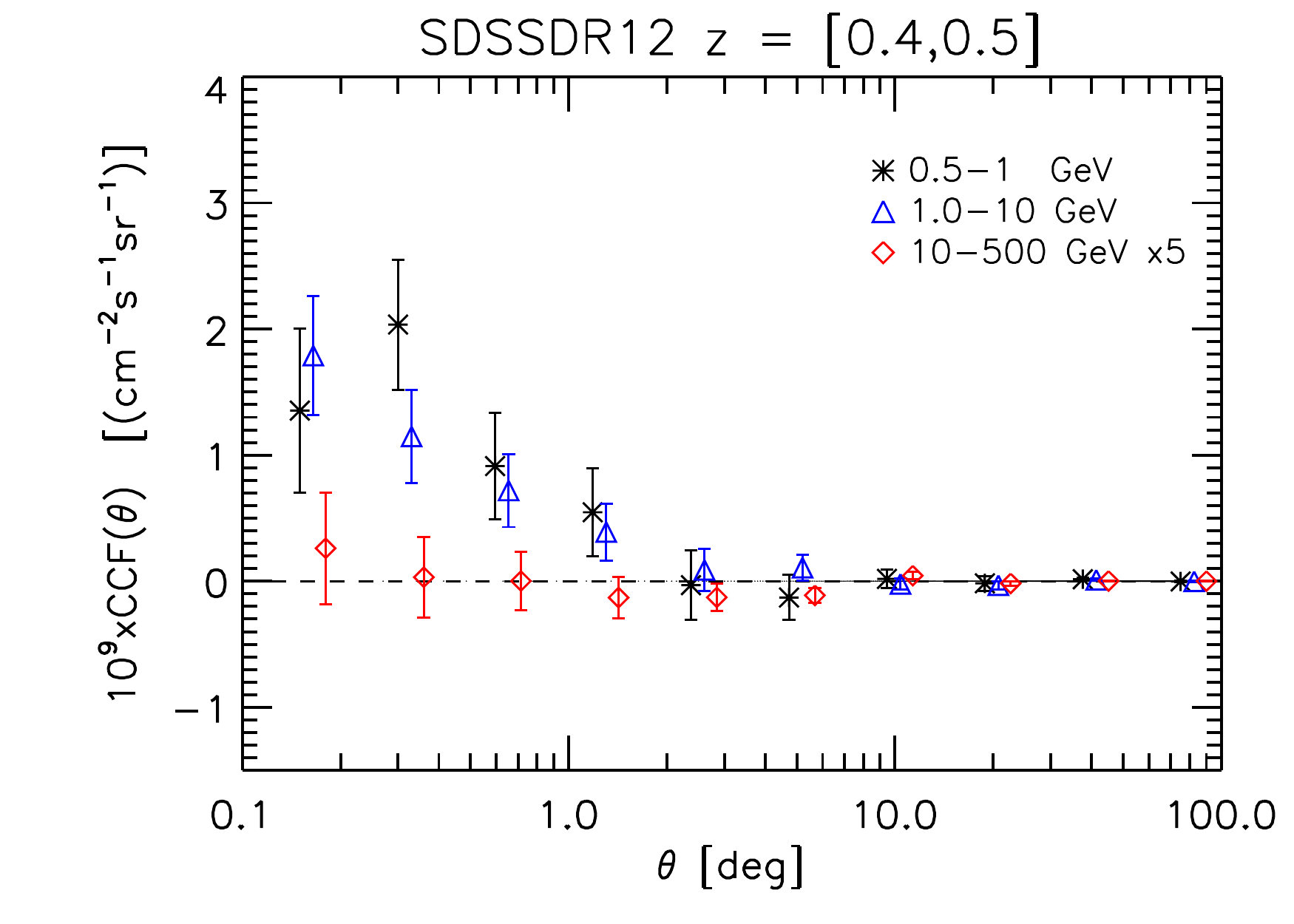, angle=0, width=0.45 \textwidth}
\centering \epsfig{file=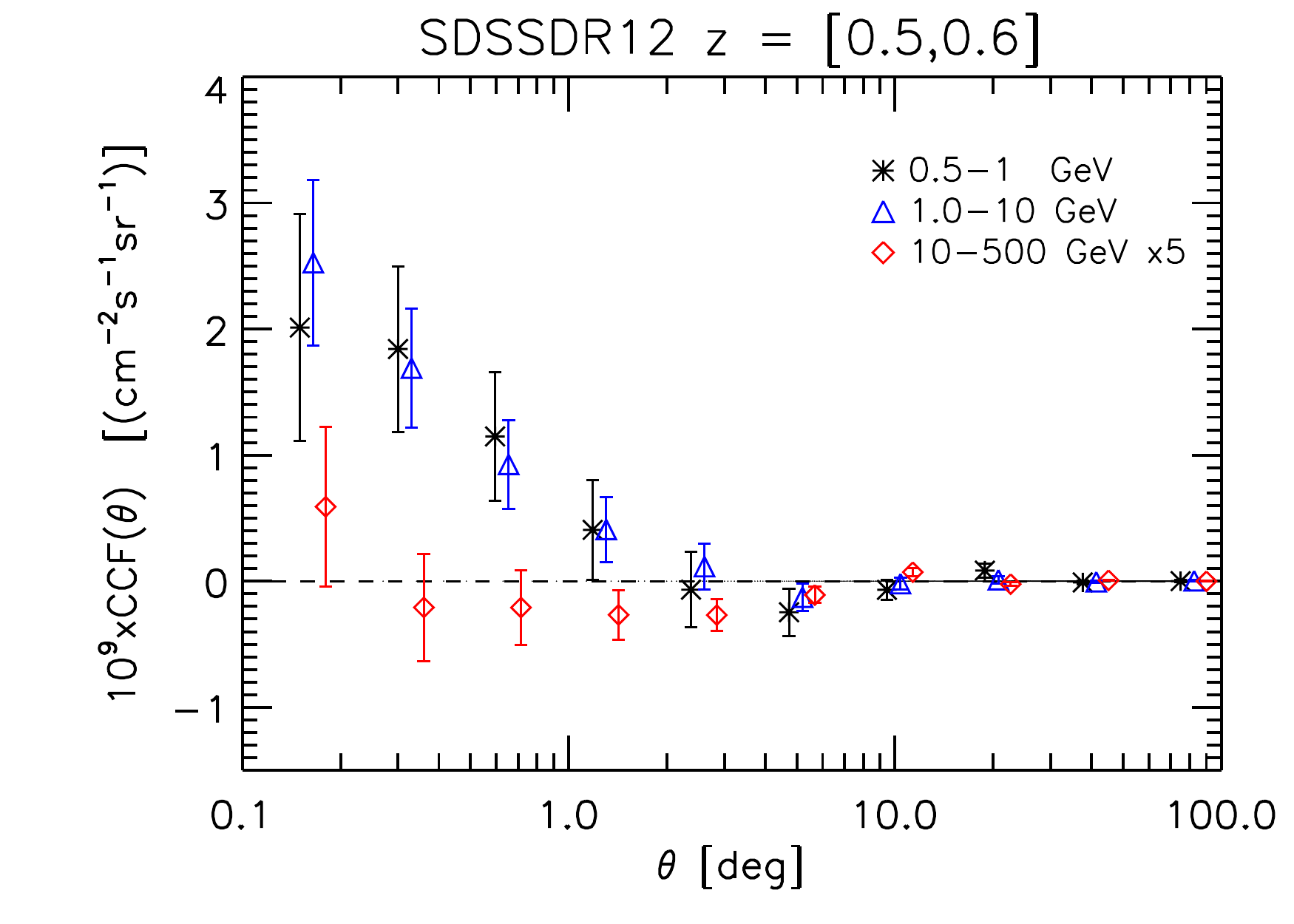, angle=0, width=0.45 \textwidth}
\centering \epsfig{file=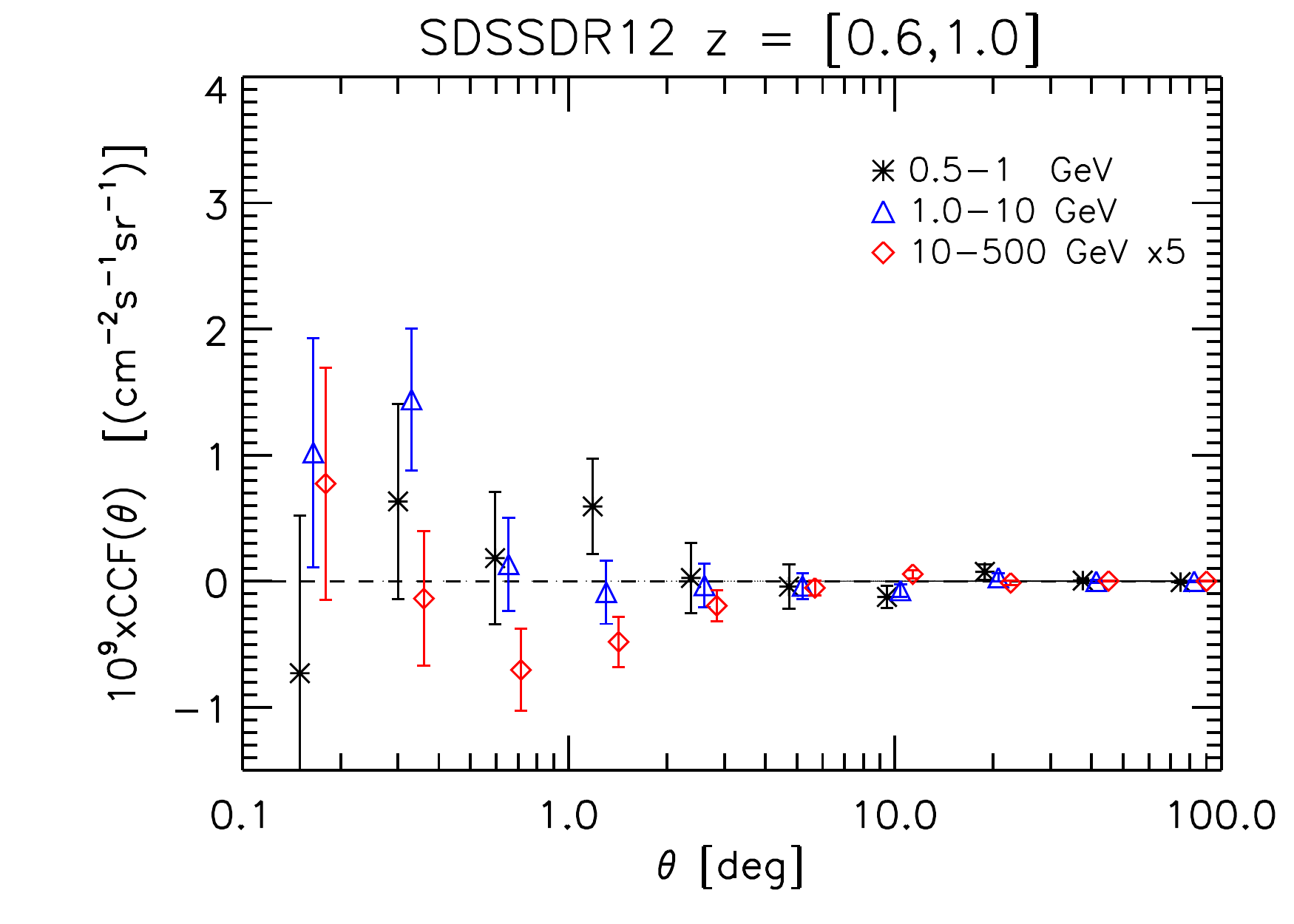, angle=0, width=0.45 \textwidth}
%
\caption{Angular CCF for SDSS-DR12 galaxies in different redshift and energy bins.
Top row is for $z=$  All and $z \in [0.0,0.1]$,  second from top  $z \in [0.1,0.2]$ and   $\in [0.2,0.3]$, 
third from top $z \in [0.3,0.4]$ and  $z \in [0.4,0.5]$, bottom row $z \in [0.5,0.6]$ and $z \in [0.6,1.0]$.
Energy bins are $E \in [0.5,1]$ GeV, $[1,10]$ GeV and $[10,500]$ GeV.}
\label{fig:sdss_ccf_fermi_full}
\end{figure*}

\renewcommand\tabcolsep{2.5pt}
\begin{table*}
\caption{Best fit to CAPS. Col. 1: Subsample name.
Col. 2(7):  minimum $\chi^2$ value.
Cols. 3(8) and 4(9): Values of the tests statistics
TS = ($\chi^2(0)-\chi_{min}^2$) and corresponding statistical significance.
Cols. 5(10) and 6(11):  68\% C.L. constraints on the one-halo term $C_{\rm 1h}$ and on the
two-halo term  $A_{\rm 2h} \times C_{\rm \ell=80}$
both expressed in units of $10^{13} \times \ ($cm$^{-2}$s$^{-1}$sr$^{-1})$ sr.
{The fit in each row is performed using 12 data points and 2 fit parameters,
for a total of 10 degrees of freedom.}}
\label{tab:T1apdx}
\begin{center}
\begin{tabular}{c|c|c|c|c|c|c|c|c|c|c|c|c}
\hline \hline
Sample & $\chi^2_{\rm min}$&TS& $\sigma$ & $C_{\rm 1h}$ & $A_{\rm 2h}C_{80}$  & \ \ \ \ & Sample & $\chi^2_{\rm min}$&TS& $\sigma$ & $C_{\rm 1h}$ & $A_{\rm 2h} C_{80}$ \\
\hline
\hline
NVSS ZA E1 &$20.3$ & $15.0$  & 3.5 & $47^{+12}_{-13}$ & $<15.7$ &&2MPZ Z1 E1 & $4.40$&$0.20$& 0.1 & $<90.7$ & $<59.8$\\
NVSS ZA E2 & $32.7$ & $110$  & 10.3 &  $26.1^{+2.6}_{-2.7}$ & $<4.95$  && 2MPZ Z1 E2 & $7.97$&$4.27$& 1.6 &  $<24.9$ & $<24.8$\\
NVSS ZA E3 & $5.49$ & $64.4$ & 7.7 &  $0.94^{+0.12}_{-0.11}$&$<0.372$ && 2MPZ Z1 E3 & $15.5$&$4.04$& 1.5 &  $<0.780$ & $<1.79$\\
\hline
QSO Z1 E1  &$10.4$&$11.9$ & 3.0 &  $ 41^{+12}_{-21}$ & $<26.1$ && 2MPZ Z2 E1 & $ 8.64$ & $0.168$ & 0.1 &   $<62.7$ & $<49.7$ \\
QSO Z1 E2   &$10.8$& $1.21$ & 0.6 &  $<9.89$ & $<6.83$ & & 2MPZ Z2 E2 & $6.35$ & $1.11$ & 0.6 &  $<12.5$ & $<21.7$ \\
QSO Z1 E3 & $10.3$ & $6.11$ & 2.0 &  $<0.470$ & $ <1.56$ & & 2MPZ Z2 E3 & $9.33$ & $3.25$& 1.3 &  $<0.448$ & $<2.15$ \\
\hline
QSO Z2 E1 &$9.47$&$0.897$ & 0.5 &  $<24.9$ & $<26.1$  && 2MPZ Z3 E1 &$6.89$&$1.88$& 0.9 &  $<94.9$& $<47.5$\\
QSO Z2 E2 &$14.1$&$3.42$& 1.3 &  $<9.01$ & $<10.4$ &&  2MPZ Z3 E2  & $2.44$&$15.4$& 3.5 &  $ 19.8^{+5.1}_{-7.0}$ & $<20.7$\\
QSO Z2 E3 &$5.00$&$3.22$& 1.3 &  $<0.258$ & $<1.24$ & & 2MPZ Z3 E3  & $8.26$ & $17.1$ & 3.7 &  $ 0.71^{+0.21}_{-0.23}$ & $<2.15$\\
\hline
QSO Z3 E1 &$13.9$ &$1.35$& 0.7 & $<36.0$ & $<41.4$ && 2MPZ ZA E1  & $7.85$&$0.911$ & 0.5 &  $<59.8$ & $<37.7$\\
QSO Z3 E2 &$8.60$&$11.9$& 3.0 &  $<17.2$ & $<20.7$ && 2MPZ ZA E2  &$8.18$&$8.18$ & 2.4 &  $ 8.61^{+3.54}_{-4.30}$ & $<18.7$\\
QSO Z3 E3 & $10.7$&$7.93$ & 2.3 &  $<0.310$ & $ 1.08^{+0.36}_{-0.47}$ && 2MPZ ZA E3  & $12.3$&$13.3$ & 3.2 &  $ 0.31^{+0.11}_{-0.13}$ & $<1.63$ \\
\hline
QSO ZA E1 & $5.53$ & $7.25$  & 2.2 &  $<29.9$ & $<23.8$ && MG12 Z1 E1  &$7.70$&$1.40$ & 0.7 &  $<34.4$ & $<24.9$\\
QSO ZA E2 & $11.3$ & $12.0$  & 3.0 &  $ 5.7^{+1.7}_{-2.1}$ & $<5.18$ && MG12 Z1 E2  &$9.60$&$1.76$ & 0.8 &  $<10.4$ & $<16.4$\\
QSO ZA E3 & $11.4$ & $12.3$ & 3.1 &   $<0.224$&$ 0.71^{+0.26}_{-0.29}$ && MG12 Z1 E3   &$4.32$&$6.16$& 2.0 &  $<0.356$ & $<1.79$\\
\hline
WIxSC Z1 E1 & $14.8$&$3.56$& 1.4 &  $ <52.1$ & $<23.8$ && MG12 Z2 E1 &$5.29$&$3.20$& 1.3 &  $<45.4$ & $<27.3$\\
WIxSC Z1 E2 & $16.0$&$8.50$ & 2.4 &  $ 6.2^{+2.1}_{-2.4}$ & $<7.85$ && MG12 Z2 E2 &$8.24$&$5.55$ & 1.9 &  $<5.18$ & $<17.2$\\
WIxSC Z1 E3 & $16.0$&$5.55$ & 1.9 &  $<0.258$ & $<1.36$ && MG12 Z2 E3 &$7.79$&$7.41$& 2.2 &  $<0.178$ & $<1.42$\\
\hline
WIxSC Z2 E1 &  $11.7$&$4.17$ & 1.5 &  $ <39.5$ & $<13.0$&& MG12 Z3 E1 &$9.35$&$5.95$ & 2.0 &  $ <47.5$ & $<21.7$\\
WIxSC Z2 E2 & $3.67$&$9.73$ & 2.7 &  $ 4.3^{+1.6}_{-2.1}$ & $<9.01$ && MG12 Z3 E2  &$5.74$&$13.3$& 3.2 &  $ 4.1^{+2.4}_{-2.1}$ & $<16.4$\\
WIxSC Z2 E3 & $12.4$&$15.6$ & 3.5 &  $<0.141$ & $ 0.71^{+0.29}_{-0.32}$  && MG12 Z3 E3  &$14.0$&$19.1$ & 4.0 &  $ 0.10^{+0.061}_{-0.051}$ & $0.78^{+0.43}_{-0.39}$\\
\hline
WIxSC Z3 E1 &$12.9$&$7.52$& 2.3 &  $ 24.9^{+9.2}_{-9.8}$ & $<15.7$ && MG12 Z4 E1 &$7.04$&$4.93$& 1.7 &   $<37.7$ & $<31.4$\\
WIxSC Z3 E2 & $4.54$&$16.2$& 3.6 &  $ 4.7^{+1.6}_{-1.9}$ & $<10.4$ && MG12 Z4 E2   &$5.52$&$17.5$& 3.8 &  $ 4.3^{+2.0}_{-2.0}$ & $<14.3$\\
WIxSC Z3 E3  & $5.39$&$22.4$& 4.3 &  $ 0.178^{+0.034}_{-0.050}$ & $<0.619$ && MG12 Z4 E3   &$5.00$&$7.86$& 2.3 &   $<0.170$ & $<0.896$ \\
\hline
WIxSC Z4 E1 & $9.94$&$7.98$ & 2.4 &  $<24.9$ & $ 33^{+12}_{-15}$ && MG12 Z5 E1 &$8.12$&$11.0$ & 2.9 &  $<43.3$ & $<37.7$\\
WIxSC Z4 E2 &$12.0$&$9.92$ & 2.7 &  $<5.95$ & $10.4^{+3.8}_{-5.2}$ && MG12 Z5 E2 &$5.73$&$12.9$ & 3.2 &  $<6.83$ & $<15.0$\\
WIxSC Z4 E3 &  $9.67$&$8.02$& 2.4 &  $ 0.155^{+0.064}_{-0.072}$ & $<0.896$ && MG12 Z5 E3  &$10.3$&$5.33$& 1.8 &   $<0.162$ & $<0.711$\\
\hline
WIxSC ZA E1 & $16.2$ & $22.0$ & 4.3 &  $32.8^{+7.3}_{-7.0}$ & $<5.95$ && MG12 Z6 E1 &$6.67$&$7.54$& 2.3 &  $<49.7$ & $<41.4$ \\
WIxSC ZA E2 & $9.32$ & $26.5$ & 4.8 &  $4.1^{+1.5}_{-1.7}$ & $<11.4$ && MG12 Z6 E2  &$6.93$&$15.5$ & 3.5 &  $<9.89$ & $<19.8$\\
WIxSC ZA E3 & $1.99$ & $35.3$ & 5.6 &  $0.098^{+0.040}_{-0.040}$ & $0.56^{+0.26}_{-0.27}$&&MG12 Z6 E3 &$13.7$&$7.00$ & 2.2 &  $ 0.155^{+0.057}_{-0.065}$ & $<0.711$\\
\hline
 &  &  & &  & & & MG12 Z7 E1 &$7.62$&$1.08$& 0.6 &   $<24.9$ & $<29.9$\\
 &  &  & &  & & & MG12 Z7 E2 &$3.94$&$6.21$& 2.0 &  $ <11.4$ & $<9.44$\\
 &  &  & &  & & & MG12 Z7 E3   &$13.9$&$4.80$& 1.7 &   $ <0.390$ & $<0.515$\\
\hline
 &  &  & &  & & & MG12 ZA E1  &$6.90$&$11.5$& 2.9 &  $ 21.7^{+9.8}_{-10.8}$ & $<31.4$\\
 &  &  & &  & & & MG12 ZA E2   &$7.69$&$26.9$& 4.8 &   $ 3.0^{+1.6}_{-1.5}$ & $6.8^{+3.4}_{-3.3}$\\
 &  &  & &  & & & MG12 ZA E3  &$8.73$&$23.5$& 4.5 &  $ 0.098^{+0.032}_{-0.034}$ & $<0.780$\\
\hline
\hline
\end{tabular}
\end{center}
\end{table*}

\renewcommand\tabcolsep{1.5pt}

\begin{table*}
\caption{CAPS energy dependence.
Results of the best fit when the SPL, DPL and BPL models are use. Col. 1: sub-sample considered.
Col 2/7/12:  minimum $\chi^2$ values 
(the $\chi^2$ is calculated as a sum over 8 energy bins and 12 multipole bins, i.e., 96 bins in total.
{The number of fitted parameters is 3 for the SPL fit, 4 for the DPL fit and 5 for the BPL fit,
for a total of 93, 92 and 91 degrees of freedom, respectively}).
Cols 3/8/13 and 4/9/14: Values of the tests statistics
TS = ($\chi^2(0)-\chi_{min}^2$) and corresponding statistical significance.
Col 5/10/15 and 6/11/16: best-fit values of the one-halo term $C_{\rm 1h}$ and
 two-halo term  $A_{\rm 2h} \times C_{\rm \ell=80}$
 both expressed in units of of $10^{13} \times$  (cm$^{-2}$s$^{-1}$sr$^{-1}$GeV$^{-1}$)sr.
}
\label{tab:T2apdx}
\begin{center}
\begin{tabular}{c|c|c|c|c|c|c||c|c|c|c|c|c|c||c|c|c|c|c|c|c|c}
\hline \hline
Sample&$\chi^2_{\rm min}$&TS&   $\sigma$  &$\alpha$ & $C_{\rm 1h}$ & $A_{\rm 2h} C_{80}$&$\chi^2_{\rm min}$&TS& $\sigma$  &$\alpha_{\rm 1h}$&$\beta_{\rm 2h}$ & $C_{\rm 1h}$ & $A_{\rm 2h} C_{80}$ & $\chi^2_{\rm min}$&TS& $\sigma$  & $E_{\rm break}$&$\alpha$&$\beta$ & $C_{\rm 1h}$ & $A_{\rm 2h} C_{80}$\\
\hline
\hline
\multicolumn{7}{c}{SPL} & \multicolumn{7}{c}{DPL}  & \multicolumn{8}{c}{BPL}  \\
\hline
NVSS ZA & $126.$ & $274.0$ & 16.2  & $2.32$ & $44.5$ & $0.035$ & $126.$ & $274.0$ &  16.1 &   $2.32$ & $4.54$ & $44.1$ & $0.047$ & $118.$ & $282.0$ &  16.2 &   $2.33$ & $2.27$ & $8.32$ & $41.9$ & $0.005$\\
\hline
QSO6 Z1 & $110.$ & $16.0$ & 3.3  &  $3.25$ & $19.8$ & $0.055$ & $108.$ & $18.0$ &  3.2 &    $3.59$ & $1.50$ & $17.0$ & $0.769$ & $107.$ & $19.0$ &   3.1 &   $1.40$ & $3.27$ & $2.52$ & $20.0$ & $0.742$\\
QSO6 Z2 & $96.1$ & $5.04$ & 1.4  &  $3.07$ & $8.13$ & $0.573$ & $96.0$ & $5.14$ &  1.1 &    $3.30$ & $2.17$ & $5.72$ & $2.62$ & $95.7$ & $5.44$ &   0.9 &   $0.939$ & $8.05$ & $3.11$ & $96.2$ & $20.4$\\
QSO6 Z3 & $95.6$ & $19.2$ & 3.7  &  $2.19$ & $0.033$ & $18.2$ & $94.3$ & $20.5$ &  3.5 &    $3.19$ & $2.05$ & $8.31$ & $11.8$ & $94.0$ & $20.8$ &  3.3 &    $1.22$ & $2.37$ & $2.17$ & $0.088$ & $23.6$\\
\hline
WIxSC Z1& $93.8$ & $19.8$ & 3.7  &  $2.41$ & $10.4$ & $0.0071$ & $93.7$ & $19.9$ &  3.5 &    $2.48$ & $1.58$ & $11.3$ & $0.036$ & $93.5$ & $20.1$ &  3.2 &    $2.03$ & $2.35$ & $2.81$ & $9.40$ & $0.295$ \\
WIxSC Z2& $96.7$ & $25.3$ & 4.3  &  $2.18$ & $3.26$ & $ 3.97$ & $96.3$ & $25.7$ &  4.1 &    $2.39$ & $1.80$ & $5.39$ & $ 1.85$ & $93.5$ & $28.5$ &  4.2 &    $1.08$ & $3.17$ & $2.51$ & $15.1$ & $0.457$\\
WIxSC Z3& $71.6$ & $62.7$ & 7.3  &  $2.21$ & $5.65$ & $ 6.22$ & $71.4$ & $62.9$ &  7.2 &    $2.30$ & $1.87$ & $7.00$ & $ 3.60$ & $70.5$ & $63.9$ &  7.1 &    $1.25$ & $2.56$ & $2.35$ & $10.6$ & $4.14$\\
WIxSC Z4& $85.7$ & $30.6$ & 4.9  &  $2.24$ & $2.79$ & $ 16.8$ & $82.2$ & $34.1$ &  5.0 &    $1.90$ & $2.67$ & $1.29$ & $ 23.6$ & $83.2$ & $33.2$ &  4.6 &    $1.40$ & $2.65$ & $2.27$ & $5.41$ & $21.1$\\
\hline
2MPZ Z1 & $83.8$ & $7.98$ & 2.0  &  $2.50$ & $22.5$ & $0.486$ & $83.7$ & $8.1$ &  1.7 &    $2.53$ & $3.53$ & $21.7$ & $0.033$ & $81.7$ & $10.1$ &  1.8 &    $1.72$ & $2.12$ & $8.82$ & $10.7$ & $0.798$\\
2MPZ Z2 & $62.6$ & $5.68$ & 1.6  &  $2.00$ & $2.58$ & $ 4.54$ & $61.7$ & $6.58$ &  1.4 &    $1.89$ & $2.51$ & $1.76$ & $ 8.45$ & $62.4$ & $5.88$ &   1.0 &   $0.576$ & $5.21$ & $2.35$ & $10.7$ & $16.3$\\
2MPZ Z3 & $69.7$ & $38.2$ & 5.5  &  $2.22$ & $24.0$ & $0.173$ & $69.6$ & $38.3$ &  5.3 &    $2.22$ & $1.77$ & $22.8$ & $1.35$ & $66.5$ & $41.4$ &  5.4 &    $1.03$ & $2.53$ & $2.28$ & $33.6$ & $2.51$\\
\hline
MG12 Z1 & $56.5$ & $13.8$ & 3.0  &  $2.03$ & $3.16$ & $ 2.97$ & $56.4$ & $13.9$ &  2.7 &    $2.03$ & $1.91$ & $3.11$ & $2.05$ & $55.3$ & $15.0$ &  2.6 &    $0.461$ & $7.43$ & $2.06$ & $3.08$ & $8.95$\\
MG12 Z2 & $83.0$ & $18.3$ & 3.5  &  $2.06$ & $0.084$&$ 13.4$ & $82.1$ & $19.2$ & 3.4 &    $4.47$ & $2.02$ & $2.97$ & $ 12.3$ & $82.8$ & $18.5$ &   3.0 &   $1.34$ & $2.17$ & $2.03$ & $0.0690$ & $14.5$\\
MG12 Z3 & $86.7$ & $46.4$ & 6.2  &  $2.13$ & $3.18$ & $ 12.5$ & $86.4$ & $46.7$ &  6.0 &    $2.23$ & $2.00$ & $3.86$ & $ 10.3$ & $85.4$ & $47.7$ &  5.9 &    $1.07$ & $2.44$ & $2.19$ & $4.85$ & $12.3$\\
MG12 Z4 & $75.5$ & $36.3$ & 5.4  &  $2.32$ & $2.15$ & $ 19.2$ & $69.4$ & $42.4$ &  5.7 &    $3.59$ & $1.95$ & $16.2$ & $ 8.23$ & $73.3$ & $38.5$ &  5.1 &    $1.41$ & $3.02$ & $2.50$ & $9.38$ & $18.2$\\
MG12 Z5 & $86.7$ & $35.2$ & 5.4  &  $2.64$ & $4.05$ & $ 22.7$ & $80.5$ & $41.4$ &  5.6 &    $3.79$ & $2.07$ & $14.1$ & $ 9.99$ & $83.7$ & $38.2$ &  5.1 &    $1.36$ & $2.96$ & $2.53$ & $7.57$ & $17.2$\\
MG12 Z6 & $61.5$ & $27.0$ & 4.5  &  $2.30$ & $5.45$ & $ 11.4$ & $61.4$ & $27.1$ & 4.3 &    $2.36$ & $2.38$ & $6.22$ & $ 11.2$ & $61.0$ & $27.5$ &  4.1 &    $0.902$ & $2.55$ & $2.35$ & $7.07$ & $10.8$\\
MG12 Z7 & $69.8$ & $12.3$ & 2.7  &  $2.30$ & $7.27$ & $ 2.89$ & $69.7$ & $12.4$ &  2.5 &    $2.28$ & $2.34$ & $6.64$ & $ 2.44$ & $68.7$ & $13.4$ &  2.3 &    $2.30$ & $2.39$ & $2.19$ & $8.80$ & $1.14$\\
\hline\hline
\end{tabular}
\end{center}
\end{table*}


\end{document}